\RequirePackage{ifpdf}
\documentclass[12pt]{JHEP3}
\usepackage{mathrsfs}
\usepackage{amsmath,amssymb}
\usepackage{epsfig}
\usepackage{relsize}
\usepackage{graphicx}

\input epsf

\usepackage{enumitem}

\usepackage{color}

\definecolor{grey}{rgb}{0.4,0.4,0.5}
\definecolor{darkgreen}{rgb}{0,0.5,0}
\definecolor{darkred}{rgb}{0.6,0.0,0}
\definecolor{lightbrown}{rgb}{1,0.9,0.8}
\definecolor{brown}{rgb}{0.6,0.3,0.3}
\definecolor{darkblue}{rgb}{0,0,0.8}
\definecolor{darkmagenta}{rgb}{0.5,0,0.5}

\def\x'{\mathaccent 19 x}
\def\y'{\mathaccent 19 y}
\def\n'{\mathaccent 19 n}
\def\u'{\mathaccent 19 u}

\def\et'{\mathaccent 19 \eta}
\def\th'{\mathaccent 19 \theta}
\def\lam'{\mathaccent 19 \lambda}
\def\varet'{\mathaccent 19 \vartheta}
\def\rh'{\mathaccent 19 \rho}
\def\ph'{\mathaccent 19 \phi}
\def\xb'{\mathaccent 19 {\bar{x}}}

\def\l{{\lambda}}

\def\sl(2){\alg{sl}(2)}
\def\sl{\alg{sl}}

\def\det{\hbox{det}}
\def\be{\begin{equation}}
\def\ee{\end{equation}}

\newcommand{\bea}{\begin{eqnarray}}
\newcommand{\eea}{\end{eqnarray}}
\newcommand{\bei}{\begin{itemize}}
\newcommand{\eei}{\end{itemize}}
\newcommand{\bee}{\begin{enumerate}}
\newcommand{\eee}{\end{enumerate}}

\def\r {\rho}
\def\a {\alpha}
\def\b {\beta}

\def\pa {\partial}
\def\P {\mathscr{P}}

\def\g {\gamma}
\def\om {\omega}
\def\Om {\Omega}
\def\p{\phi}
\def\la{\label}
\def\e{\epsilon}
\def\ov{\over}

\def\Tr{{Tr}}
\def\G{\Gamma}
\def\S{\Sigma}

\def\bA{{\mathbb{A}}}

\def\bI{{\mathbb I}}

\def\bR{{\mathbb R}}

\newcommand{\alg}[1]{\mathfrak{#1}}
\newcommand{\su}{\alg{su}}

\newcommand{\AdS}{{\rm  AdS}_5\times {\rm S}^5}
\newcommand{\ads}{${\rm  AdS}_5\times {\rm S}^5\ $}

\newcommand{\bem}{\left (\begin{matrix}}
\newcommand{\eem}{\end{matrix} \right )}

\def\cA{{\cal A}}
\def\S{{\cal S}}
\def\P{{\cal P}}

\def\I{{\cal I}}
\def\cJ{{\cal J}}
\def\cO{{\cal O}}
\def\cV{\,{\cal V}}
\def\cK{\,{\cal K}}
\def\cN{\,{\cal N}}
\def\cF{\,{\cal F}}
\def\cZ{{\cal Z}}

\def\S{{\mathbb S}}

\def\eps{\epsilon}

\def\sh{\sinh}

\def\t{\theta}
\def\Tr{\makebox{Tr}}

\def\F{{\mathbb F}}

\def\J{{\mathbb J}}
\def\R{{\mathbb R}}
\def\bP{{\mathbb P}}
\def\bK{{\mathbb K}}
\def\u{{\alg u}}

\def\rok{\chi_k^+}
\def\lok{\chi_k^-}
\def\rom{\chi_m^+}
\def\lom{\chi_m^-}
\def\roi{\chi_i^+}
\def\loi{\chi_i^-}
\def\rjok{\cJ_k^+}
\def\ljok{\cJ_k^-}

\author{Stephen Britton\footnote{Email: brittons@tcd.ie, frolovs@maths.tcd.ie} 
\  and Sergey Frolov\footnote{Correspondent fellow at Steklov Mathematical Institute, Moscow.}
 \\ {\it Hamilton Mathematics Institute and School of Mathematics,
Trinity College, Dublin 2, Ireland}}

\abstract{The free field representation of the Zamolodchikov-Faddeev algebra for the chiral Gross-Neveu model is analysed in detail, and used to construct an integral representation for form factors of the model.}

\title{Free field representation and form factors of the chiral Gross-Neveu model}

\preprint{
          \smaller{\smaller{\smaller{HMI-13-01}}}\\[-.5ex]
          \smaller{\smaller{\smaller{TCDMATH 13-07}}}}

\begin{document}

\renewcommand{\thefootnote}{\arabic{footnote}}
\setcounter{footnote}{0}

\section{Introduction}

In quantum field theory  the matrix elements, $\langle out|O({x})|in \rangle$,  of  local operators $O$ between $in$- and $out$-states
determine the field theory correlation functions. In a two-dimensional relativistic theory the  crossing invariance allows one to express a generic matrix element in terms of analytically continued form factors which are matrix elements between the vacuum, $|vac \rangle$,  and $in$-states
\be\la{FFdef}
F_{a_1\ldots a_n}(\a_1,\ldots,\a_n)=\langle vac|O(0)|\a_1,\ldots,\a_n\rangle^{in}_{a_1\ldots a_n}\,.
\ee
Here $a_i$ is a flavour index of the $i$-th particle, and $\a_i$ is its usual rapidity variable related to its energy and momentum by $E_i = m_i\cosh \a_i\,,\ p_i = m_i\sinh \a_i$.

 In an integrable two-dimensional relativistic theory the form factors satisfy a set of axioms \cite{KW78}-\cite{Smirnov92}, collected in appendix \ref{axioms}, whose solutions were found and studied for some models, see e.g. \cite{Smirnov92}-\cite{Babujian2009} and references therein. Finding a solution to the axioms is a complicated problem which requires understanding and employing  the form factors' analytic properties. It was observed by Lukyanov \cite{Lukyanov}  (by generalising the ideas in \cite{Lukyanov:1992sc}) that the problem of computing the form factors can be reduced to the problem of constructing a free field representation of the Zamolodchikov-Faddeev (ZF) algebra \cite{Zam,Fad} for the model under consideration. The free field representation approach has been successfully applied to several models \cite{Lukyanov}, \cite{Lukyanov:1997bp}-\cite{HT1996}  including the SU(2) Thirring and sine-Gordon models \cite{Lukyanov}. 
An advantage of this approach is that in principle the construction of form factors does not require  a complete understanding of their analytic properties. Moreover, the analytic properties follow from a free field representation. This might be important for understanding analytic properties of form factors of nonrelativistic models. In particular, an important model to keep in mind is the  \ads superstring sigma model in the light-cone gauge \cite{AFrev}. Even though it is relatively straightforward to generalise most of the form factors axioms to the case \cite{KM}, finding a solution appears to be highly nontrivial   in particular because the analytic properties of  \ads form factors are not known. It is quite possible that Lukyanov's approach will appear to be more efficient in the \ads case.

The goal of this paper is to extend Lukyanov's results for the SU(2) Thirring model  to a more general case of the SU(N) chiral Gross-Neveu (GN) model \cite{GN}. The model has a very rich spectrum of particles.
Its ``elementary'' particles transform  in the rank-1 fundamental representation of  SU(N), and they can form bound states transforming in all the other fundamental representations of  SU(N)  \cite{Kurak:1978su}.  
Anti-particles of  rank-$r$ particles are rank-$(N$-$r)$ particles. In particular,  anti-particles of  elementary particles are at the same time their bound states \cite{Kurak:1978su}.
The exact GN S-matrix was found by combining the SU(N) invariance with the $1/N$ expansion \cite{Berg:1977dp}-\cite{Koberle:1979ne}.
The chiral  GN model was extensively studied in the axiomatic approach in \cite{Smirnov92, Takeyama:2001ui, Babujian2009} where form factors of several local operators were constructed. It is therefore useful for understanding how the free field approach works in case of models containing bound states and invariant under higher-rank symmetry algebras. 
In this paper we construct a free field representation of the GN model ZF algebra for elementary particles and their bound states, and find a large class of operators generating form factors of local operators through Lukyanov's trace formula \cite{Lukyanov}.
A free field representation for elementary particles of the chiral GN model was also constructed in \cite{Kojima}, and it agrees with our findings up to some Klein factors necessary to satisfy the ZF algebra relations.  

The paper is laid out as follows. We begin in section \ref{gener} by recalling some facts about the GN model, particularly the properties of its scattering matrix. Here, we also explain the general free field bosonization process, and how this is used to construct form factors. In section \ref{free} we consider how to apply the bosonization process to the GN model and construct the ZF algebra and Hamiltonian in this case. Next, we consider the construction of bound states in section \ref{bound}. This is done in a general way and the relation to the anti-particles is established. Section \ref{localO} contains the construction of representations of local operators in terms of bosonic fields. A new set of these fields (closely related to the first) is introduced and used to construct a family of local operators. Finally, we construct form factors in section \ref{form}, where general formulae for constructing form factors are established. We use this to construct the form factors of the current operator. In several appendices we collect the necessary functions and present derivations of some results stated in the main text.

\section{Generalities}\la{gener}

\subsection{The S-matrix of the chiral Gross-Neveu model}

The spectrum of particles of the chiral SU(N) GN model consists of $N$ elementary particles of mass $m$ transforming   in the rank-1 fundamental representation of  SU(N), and their $r$-particle bound states of mass $m_r=m \sin {\pi r\ov N}/\sin {\pi\ov N}$ transforming  in the rank $r=2,\ldots,N-1$  fundamental representation of  SU(N). A rank-$r$ particle with rapidity $\t$ is created by a ZF operator ${\cal A}^\dagger_{K}(\t)$, and annihilated by $\cA^{K}(\t)$ where ${K} = (k_1,\ldots, k_r)$ has integer-valued components ordered as $1\le k_1< k_2<\cdots< k_r\le N$. The creation and annihilation operators satisfy the ZF algebra
\bea\la{ZFu}
\cA^\dagger_{K_1}(\t_1)\cA^\dagger_{K_2}(\t_2)&=& \cA^\dagger_{N_2}(\t_2)\cA^\dagger_{N_1}(\t_1)S_{K_1K_2}^{N_1N_2}(\t_{12})\,,\\
\cA^{K_1}(\t_1)\cA^{K_2}(\t_2)&=& S^{K_1K_2}_{N_1N_2}(\t_{12})\cA^{N_2}(\t_2)\cA^{N_1}(\t_1)\,,\\
\cA^{K_1}(\t_1)\cA^\dagger_{K_2}(\t_2)&=& \cA^\dagger_{N_2}(\t_2)S_{N_1K_2}^{K_1N_2}(\t_{21})\cA^{N_1}(\t_1)+2\pi \delta_{K_2}^{K_1}\delta(\t_{12})\,,
\eea
where $\t_{ij}\equiv \t_i-\t_j$ and  $S_{K_1K_2}^{N_1N_2}(\t_{12})$ is the scattering matrix of  particles of ranks $r_1$ and $r_2$ with rapidities $\t_1$ and $\t_2$.  Since higher rank particles are bound states of  elementary particles, their S-matrices are obtained from the GN S-matrix  for elementary particles by the fusion procedure.  It is often convenient to use the matrix
form of the GN S-matrix. 
We introduce $N$-dimensional rows $E^i$ and columns $E_i$ with
all vanishing entries except the one in the $i$-th position which
is equal to the identity, and the matrix unities
$E_i^{~j}=E_i\otimes E^j$ with the only non-vanishing element
on the intersection of the
$i$-th row with the $j$-th column.  Then  the entries of the GN S-matrix for elementary particles can be combined in the
following $N\times N$ matrix \bea \label{matelement}
\S^{GN}(\t)=S^{kl}_{ij}(\t)\, E_k^{~i}\otimes E_l^{~j}\,. \eea
Explicitly  the S-matrix of the chiral GN model for elementary particles is given by
\begin{equation}       
{\mathbb S}^{GN}(\theta) =S(\theta)\, \R(\theta)\,,\qquad S(\theta)=\frac{\Gamma \left( \frac{i \theta}{2\pi} \right)       
\Gamma \left(\frac{N-1}{N}- \frac{i \theta}{2\pi} \right)}       
{\Gamma \left(-\frac{i \theta}{2\pi} \right)       
\Gamma \left(\frac{N-1}{N}+ \frac{i \theta}{2\pi} \right)}\,,
  \label{S-matrix}  
\end{equation}        
where the scalar factor $S(\t)$ does not have any poles in the physical strip $0\le Im(\theta)\le \pi$ and for real $\t$ has the nice integral form
\be\la{so}
S(\theta)=\exp\left(-2\, i\, \int^{\infty}_{0}{dt\ov t}\, {\sinh{(N-1)\pi t \ov N}\ov \sinh \pi t}e^{\pi t\ov N}\sin\t t \right)\, .
\ee
It satisfies the crossing symmetry condition
\bea
\prod
   _{k=-\frac{N-1}{2}}^{\frac{N-1}{2}} S(\theta + \frac{2\pi i}{N}\,k) =(-1)^{N-1}\frac{ \theta - i\pi {N-1\ov N}}{\theta + i\pi {N-1\ov N}} \,,
\eea
and has 
the large $\t$ asymptotics
$
S(\pm\infty)=e^{\mp i\pi {N-1\ov N}}\, .
$
 The matrix and pole structure of the GN S-matrix is given by 
the standard SU(N)-invariant R-matrix 
\be       
\R(\theta)=\frac{\theta\, \bI-\frac{2\pi i}{N}\, \bP}{\theta -\frac{2\pi i}{N}}   \,,
\ee
where $\bI$ is the identity operator and $\bP = E_k^{~i}\otimes E_i^{~k}$ is the permutation operator which exchanges the flavour indices of the scattering  particles. 
Introducing the projection operators 
$$
\bP_s = {1\ov2}(\bI+\bP)\,,\quad \bP_a = {1\ov2}(\bI-\bP)\,,
$$
onto the symmetric and antisymmetric parts of the tensor product of two fundamental representations one gets
\be       
\R(\theta)=\bP_s+ \frac{\theta +\frac{2\pi i}{N}}{\theta -\frac{2\pi i}{N}}\,\bP_a   \,,
\ee
which exhibits the pole at $\theta =\frac{2\pi i}{N}$  in the antisymmetric part.  This leads to the existence of bound states composed of two, three, and up to $N-1$ elementary particles. The $(N$-$1)$-particle bound states are identified with anti-particles of the elementary particles. In general
 a rank-$r$ particle and a  rank-$(N$-$r)$ particle created by $\cA^\dagger_{K}$ and $\cA^\dagger_{\overline K}$ form a particle-antiparticle pair if ${\overline K}$ is such that $K \cup{\overline K} =\P(1,2,\ldots,N)$ where $\P$ is some permutation of  $1,2,\ldots N$. 
In what follows in such a pair we refer to a bound state of smaller rank (that is  $r<N/2$) as a particle. If $N$ is even, $N=2 p$, then a bound state with the label $K = (1,k_2,\ldots, k_p)$ is considered as a particle.
The ZF operators can be normalised in such a way that for a particle $\cA^\dagger_{K}$ and antiparticle $\cA^\dagger_{L}$  the charge conjugation matrix $C_{KL}=\e_{KL}$ where $\e_{KL}\equiv\e_{i_1\ldots i_N}$ is skew-symmetric, and $\e_{1\ldots N}=1$.

\subsection{Form factors}

The $in$- and $out$-bases of asymptotic states are expressed in terms of  the ZF creation operators as follows\bea\nonumber
\hspace{-0.3cm} &&|\t_1,\t_2, \cdots , \t_n
\rangle^{(in)}_{K_1,...,K_n} =\cA^\dagger_{K_n}(\t_n)\cdots  \cA^\dagger_{K_1}(\t_1)|vac \rangle
 \,,\quad\qquad  \t_1<\t_2<\cdots <\t_n\,,\\\nonumber
\hspace{-0.3cm} &&|\t_1,\t_2, \cdots , \t_n
\rangle^{(out)}_{K_1,...,K_n} = \cA^\dagger_{K_1}(\t_1)\cdots  \cA^\dagger_{K_n}(\t_n)|vac \rangle 
   \,,\quad\qquad  \t_1<\t_2<\cdots <\t_n\,,
\eea 
and are related to each other by the scattering S-matrix. The vacuum state 
$|vac \rangle$ is annihilated by $\cA^K(\t)$, and has the unit norm, $\langle vac|vac \rangle=1$.  

Form factors of a local operator $O(x)$ are the matrix elements of  $O(0)$ between $n$-particle $in$-states and the vacuum state
\be\la{defF}
F_{K_1\ldots K_n}(\t_1,\dots,\t_n) = \langle vac| O(0) \cA^\dagger_{K_n}(\t_n)\cdots  \cA^\dagger_{K_1}(\t_1)|vac \rangle\,.
\ee
Being analytically continued to complex $\t_i$ the form factors satisfy a set of axioms, see appendix \ref{axioms}.  An important observation by Lukyanov  \cite{Lukyanov} reduces the problem of computing form factors to the problem of finding a representation of a so-called {\it extended} ZF algebra. It is generated by  vertex operators 
$A_I(\t)$\,,\footnote{They should not be confused with the ZF creation and annihilation operators $\cA^\dagger_I(\t)$ and $\cA^I(\t)$. 
}  the angular Hamiltonian $\bK$, and the central elements $\Omega_I$ 
obeying the defining relations
\bea\la{eZFa}
A_{I}(\t_1)A_{J}(\t_2)&=& A_{L}(\t_2)A_{K}(\t_1)S_{IJ}^{KL}(\t_{12})\,,\\\la{eZFb}
A_{I}(\t_1)A_{J}(\t_2)&=& -{i\, C_{IJ}\ov \t_{12}-i\pi}+\cO(1)\,,\quad \t_{12}\to i\pi\,,\\\la{eZFc}
{d\ov d\t}A_I(\t ) &=&-\left[\, \bK\,, \,A_I(\t) \right] 
- i\Omega_I A_I(\t)
\,.
\eea
The relations \eqref{eZFa} and  \eqref{eZFb} show that one can think of $A_I(\t)$ and 
$
C^{IJ}A_J(\t+{i\pi})$, $C^{IJ}C_{JK}=\delta^I_K$ as representing the ZF creation and annihilation operators, respectively. For some models the relations  \eqref{eZFb} have to be modified by replacing $C_{IJ}$ with $C_{IJ}\G$ where $\G$ is an auxiliary element satisfying $\G^2=id$ which either commutes or anticommutes with $A_I$. In particular this is the case for the SU(2p) chiral GN model.

In addition to the relations above if  the particles $\cA^\dagger_K$ of the same mass with $K\in \cK$ are bound states of particles $\cA^\dagger_I$ and $\cA^\dagger_J$ with 
$I\in \I$ and $J\in \cJ$
then the vertex operators $A_I$ and $A_J$  must satisfy the following bootstrap conditions
\be\la{eZFd}
A_{I}(\t' + i\u_+)A_{J}(\t-i\u_-)= {i\ov \t'-\t}\sum_{K\in\cK}\Gamma^K_{IJ}A_K(\t)+\cO(1)\,,\quad \t'\to \t\,.
\ee 
Here $\Gamma^K_{IJ}$ are some constants, and $\u_\pm$ are found from the equations
\be
\u_++\u_-= \u_{IJ}^K\,,\quad m_I \sin \u_+ = m_J \sin \u_-\,,
\ee
and the scattering matrix of the particles $\cA^\dagger_I$ and $\cA^\dagger_J$ has a pole at $\t = i\u_{IJ}^K$. The mass of the bound state $\cA^\dagger_K$ is equal to $m_K = m_I\cos\u_++m_J\cos\u_-$. The relations \eqref{eZFd} can be inverted and used to derive the vertex operators for the bound states from the vertex operators for elementary particles.  
 
Let us now assume that a representation of the extended ZF algebra is constructed and the vertex operators act in some space $\pi_{A}$. According to Lukyanov a local operator $O$ of the model under consideration corresponds to a linear operator $\Lambda(O)$ acting in $\pi_{A}$ which satisfies the following two conditions 
\be\la{lamO}
e^{\t \,\bK}\Lambda(O) e^{-\t\, \bK}=e^{\t s(O)}\Lambda(O)\,,\quad \Lambda(O) A_I(\t) = e^{2\pi i\,\Omega(O,I)}A_I(\t)\Lambda(O) \,,
\ee
where $s(O)$ is the spin of the local operator $O(x)$, and 
$\Omega(O,I)$ appears if the particle $\cA^\dagger_I$ has nontrivial statistics with respect
to $O(x)$.  
Then, the form factor \eqref{defF} is given by the formula
 \be\la{ffax}
F_{K_1\ldots K_n}(\t_1,\ldots,\t_n)= \cN_O\, {\Tr_{\pi_{A}}\left[e^{2\pi i \,\bK}\Lambda(O)A_{K_n}(\t_n) \cdots A_{K_1}(\t_1) \right]\ov\Tr_{\pi_{A}}\left[e^{2\pi i \,\bK}\right]}\,,
\ee
where the normalization constant $ \cN_O$ depends only on the local operator $O$ and has to be fixed by other means. Assuming that \eqref{ffax} satisfy the necessary analyticity properties, the form factor axioms then follow from the cyclicity of the trace and the defining relations of the extended ZF algebra. 

\subsection{Free field realization of the extended ZF algebra}\la{ffrZ}

\begin{figure}[t]
\begin{center}
\includegraphics[width=0.36\textwidth]{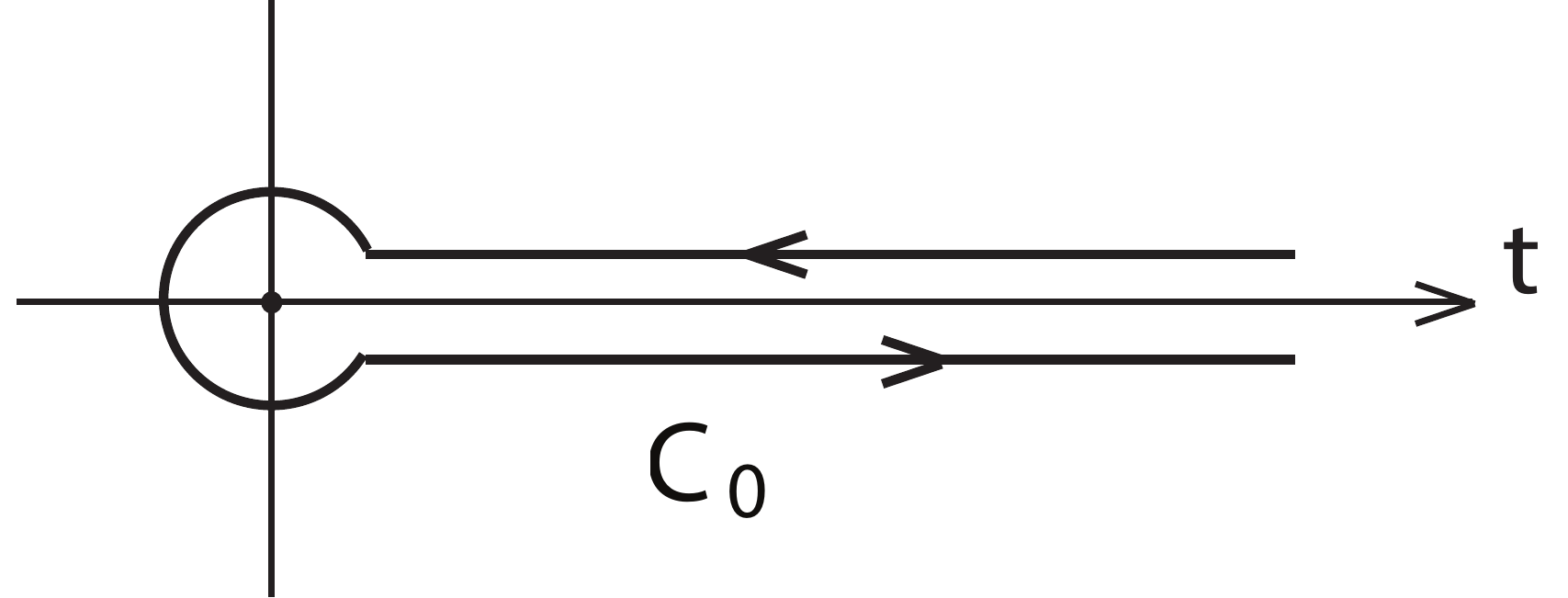}
\caption{\small The integration contour $C_0$ in the integral $\int_{C_0}\, {dt\ov 2\pi i}\, F(t)\, \ln(-t)$.}
\label{C0}
\end{center}
\end{figure}

Another important observation by \cite{Lukyanov} is that for many models the extended ZF algebra can be realized in terms of free bosons. Let us sketch the idea of the construction.  One considers particles of the same mass belonging to a highest weight irreducible representation of the symmetry algebra of the model under study. Then the    
 highest weight vertex operator $A_1$ satisfies the following simple relation
 \be\la{A1A1}
 A_1(\t_1)A_1(\t_2) = S(\t_{12})A_1(\t_2)A_1(\t_1)\,.
 \ee
Here the scattering matrix $S(\t)$ of the two highest weight particles obeys
$S(0)=-1$, and admits the representation
\be
S(\t)={g(-\t)\ov g(\t)}\,,
\ee
where $g(\t)$ is  an analytic function without zeroes and poles in the lower half plane $Im(\t)\le 0$ except a simple zero at $\t=0$, and 
\be
\pa_\t \ln g(\t) = \cO({1\ov\t})\,,\quad \t\to\infty\,,\ Im(\t)\le 0\,.
\ee
These properties of $g(\t)$ imply that for  $Im(\t)< 0$ it admits the following integral representation
\be
g(\t)=\exp\left(-\int_{0}^\infty\, {dt\ov t}\, f(t)e^{-i\theta t}\right)\,,
\ee
where $f(t)$ asymptotes to 1 at large $t$. The function $f(t)$ does not have to vanish at $t=0$, and the integrals of the form 
\be
\int_{0}^\infty\, {dt}\, F(t)
\ee
will be always understood as \cite{JKM}
\be
\int_{0}^\infty\, {dt}\, F(t) \equiv 
\int_{C_0}\, {dt\ov 2\pi i}\, F(t)\, \ln(-t)\,,
\ee
where the integration contour $C_0$ goes from $+\infty+i0$ above the real axis, then around zero, and finally below the real axis to  $+\infty-i0$, see Figure \ref{C0}. 
Let us also mention that for real values of $\t$ the scattering matrix $S(\t)$ has the integral representation
\be
S(\t)=\exp\left(-2i\,\int_{0}^\infty\, {dt\ov t}\, f(t)\sin(\theta t)\right)\,.
\ee
Let us now  introduce the bosonic operators satisfying the commutation relations 
\be
[a(t),a(t')]=t f(t)\delta(t+t')\,,
\ee
where  $a(t)$ and $a(-t)$ for $t>0$ are the annihilation and creation operators, respectively
\be
a(t)|0\rangle = 0 \ \ {\rm for}\ t>0\,.
\ee
We use the operators to define the following free bosonic field
\be
\p(\t)=Q+\int^\infty_{-\infty}{dt\over i t}\,a(t)e^{i\theta t}\,,
\ee
which satisfies the following relations 
\bea\begin{aligned}
\left[\phi(\t_1), \phi(\t_2)\right]&=\ln S(\t_{21})\, ,\quad
\langle\phi(\t_1) \phi(\t_2)\rangle=-\ln g(\t_{21})\, .~~~~
\end{aligned}
\eea
The operator $Q$ is a zero mode coordinate operator which commutes with $a(t)$. It appears in an explicit ultraviolet regularization of the free field \cite{Lukyanov}. In addition a regularized free field also contains the zero mode momentum operator $P$ which annihilates the vacuum $|0\rangle$ and also commutes with $a(t)$. In fact, as will be explained, $P$ is an element of the Cartan subalgebra of the symmetry  algebra of the model.  The regularized free fields for the GN model are discussed in appendix \ref{selrule}.

The field $\p$ is used to construct the basic vertex operator 
\be
V(\t)=: e^{i\phi(\t)}:\,,
\ee
which satisfies the ZF algebra relation \eqref{A1A1} and 
\be
V(\t_1)V(\t_2)
=g(\t_{21}):V(\t_1)V(\t_2):\, .\la{A1A1rel}
\ee
The highest weight vertex operator $A_1$ is then realized as 
\be
A_1(\t)=\om_1 V(\t)\,,
\ee
where $\om_1$ is a ``Klein'' factor which commutes with $V(\t)$ and might be  necessary to satisfy all the ZF algebra relations. 

The remaining vertex operators are then obtained by acting on the highest weight vertex operator $A_1$ by 
the lowering symmetry operators (``screening charges'') $\cJ_{k}^-$.
The action of $\cJ_{k}^-$ on the vertex operators depends on the coproduct of the symmetry algebra. In particular in the case of a quantum group with the deformation parameter $q$ the vertex operators are constructed as
\be\la{AFop}
A_1(\t)=\om_1 V(\t)\, ,\quad
A_{k+1}(\t)=\cJ_{k}^-\,A_{k}(\t)  -q\,A_{k}(\t)\,\cJ_{k}^-\,,\quad k=1,2,\ldots\,.
\ee
A free field representation of $\cJ_{k}^-$ is found by assuming the following ansatz\footnote{This ansatz does not  work if the symmetry algebra is the sum of two algebras.}
\be
\cJ_{k}^-\sim\int_{C} d\a : e^{i\phi_k(\a)}: \,,\la{Jop}
\ee
where the commutation relations of the free fields $\phi_k$ with $\p$ and themselves, and the integration contour $C$ are determined by requiring that \eqref{AFop} satisfies the extended ZF algebra. In the next section we discuss how this works for the chiral GN model.

\section{Free field representation of  the  GN model ZF algebra} \la{free}

It appears that if one omits the Klein factors mentioned in the previous section then Lukyanov's procedure gives a free field representation 
for the ZF algebra of a model  with a twisted S-matrix. It is invariant under  the $\sl(N)$  algebra with a rather unusual coproduct which however implies the action \eqref{AFop} with $q=-1$ of the $\sl(N)$ lowering generators on the vertex operators.

\subsection{Twisted GN S-matrix}

There is a simple generalisation of  the standard SU(N)-invariant R-matrix
$\bR(\theta)$.
One can check that the R-matrix of the form
\be
\R^{\Sigma}(\theta)=\frac{\theta\, \Sigma-\frac{2\pi i}{N}\, \bP}{\theta -\frac{2\pi i}{N}}   \,,
\ee 
where $\Sigma$ is a diagonal matrix satisfies the Yang-Baxter Equation (YBE) and the unitarity condition if and only if  $\Sigma$ is given by
\be
\Sigma = \sum_{i,j=1}^{N} s_{ij} E_{ii}\otimes E_{jj} \,,\quad s_{ij}s_{ji} = 1\quad  \forall\ i,j\,.
\ee
In particular the coefficients $s_{ij}$ satisfy the conditions  $s_{ii} = \pm1$ for any $i$. 
The physical unitary condition requires $\Sigma$ to be unitary and therefore $s_{ij}= e^{i \p_{ij}}$ where $\p_{ij}$ are real and obey $\p_{ij}+\p_{ji}=0 \ mod \ 2\pi$. 

One can use $\R^\Sigma$ to define the twisted GN S-matrix as ${\S}^{\Sigma}=S(\theta)\, \R^\Sigma$. It is unclear if  such a twisted S-matrix corresponds to any local field theory which would be a multi-parameter deformation  of the GN model.  The ZF algebra with the twisted GN S-matrix ${\S}^{\Sigma}$ has the form 
\be
\bA_1^{\Sigma}(\t_1) \bA_2^{\Sigma}(\t_2) =\bA_2^{\Sigma}(\t_2) \bA_1^{\Sigma}(\t_1)S_{12}^\Sigma(\t_{12})\,, 
\ee
where $\bA^{\Sigma}$
is a row $\bA^{\Sigma}=A_i^{\Sigma}\, E^i$ with $A_i^{\Sigma}$ being the ZF vertex operators. The relations can be written explicitly in components
\be\la{aaii}
A_i^{\Sigma}(\t_1) A_i^{\Sigma}(\t_2) =s_{ii}S(\t_{12})A_i^{\Sigma}(\t_2) A_i^{\Sigma}(\t_1)\,,
\ee
\be\la{aaij}
A_i^{\Sigma}(\t_1) A_j^{\Sigma}(\t_2) =S(\t_{12})\left[{s_{ij}\t_{12}\ov \t_{12}-{2\pi i\ov N}} A_j^{\Sigma}(\t_2) A_i^{\Sigma}(\t_1)-{{2\pi i\ov N}\ov \t_{12}-{2\pi i\ov N}} A_i^{\Sigma}(\t_2) A_j^{\Sigma}(\t_1)\right]\,.
\ee
Notice that only the transition amplitudes depend on the twist parameters $s_{ij}$.  From eqs.(\ref{aaii}, \ref{aaij}) one can see that if $s_{ii}=1$ and $s_{ij}=-1$ for $i\neq j$ then for each pair of indices $i,j$ the ZF relations are the same as for the SU(2) Thirring model discussed by Lukyanov. 
It is therefore not surprising that a free field representation for the ZF algebra of the GN model is related to the twisted S-matrix with $\Sigma$ of the form
\be
\Sigma^{(-1)} = \sum_{i=1}^{N} E_{ii}\otimes E_{ii}- \sum_{i\neq j}^{N}  E_{ii}\otimes E_{jj} = 2\sum_{i=1}^N E_{ii}\otimes E_{ii} - I
 \,.
\ee
To simplify the notations in what follows we denote the twisted S-matrix ${\S}^{\Sigma^{(-1)}}$ as ${\S}^{{(-1)}}$.  It is easy to check that ${\S}^{{(-1)}}$ satisfies the invariance conditions
\be
{\S}^{{(-1)}}\,\Delta_{(-1)}(\J_k^-)=\Delta^{op}_{(-1)}(\J_k^-)\,{\S}^{{(-1)}}\,,
\ee
where $\J_k^-=E_{k+1}{}^k$ are the $\sl(N)$ lowering generators in the rank-1 fundamental representation, $\Delta^{op}_{(-1)}(\J)\equiv\bP\,\Delta_{(-1)}(\J)\,\bP$ and
\be\la{copr}
\Delta_{(-1)}(\J_k^-) =\J_k^-\otimes \bI +\bI\otimes \J_k^- -2(E_{k}{}^k+E_{k+1}{}^{k+1})\otimes \J_k^-\,,
\ee
is the coproduct. 
It is defined in the same way on the raising generators, and it is extended to the whole $\sl(N)$ algebra via the commutation relations.

Let us introduce the ZF vertex operators $Z_i$ satisfying the relations (\ref{aaii},  \ref{aaij}) with $s_{ii}=1$ and $s_{ij}=-1$ for $i\neq j$
\be\la{zzii}
Z_i(\t_1) Z_i(\t_2) =S(\t_{12})Z_i(\t_2) Z_i(\t_1)\,,
\ee
\be\la{zzij}
Z_i(\t_1) Z_j(\t_2) =S(\t_{12})\left[-{\t_{12}\ov \t_{12}-{2\pi i\ov N}} Z_j(\t_2) Z_i(\t_1)-{{2\pi i\ov N}\ov \t_{12}-{2\pi i\ov N}} Z_i(\t_2) Z_j(\t_1)\right]\,,
\ee
and assume that a free field representation for $Z_1$ and the lowering operators  $\loi  $ is found. Then the coproduct \eqref{copr} implies that all the other vertex operators are obtained  through the formulas
\be
Z_{i+1}(\t)=\loi  Z_{i}(\t) + Z_{i}(\t)\loi  \,.
\ee
Then, one can construct operators $A_i^{\Sigma}$ 
satisfying the ZF algebra with $\S^\Sigma$ and all $s_{ii}=1$ through the formula
\be\la{As}
A_i^{\Sigma}(\theta)= \omega_i\, Z_i(\theta)\,,
\ee
where the ``Klein'' factors $\omega_i$ commute with $Z_j$ and satisfy the following algebra
\be\la{omom}
\omega_i \omega_j + s_{ij}\omega_j\omega_i=0\,,\quad i\neq j\,.
\ee
In particular for the canonical (untwisted) S-matrix $s_{ij}=1$ and if one also imposes extra conditions $\omega_i^2=\eta_{ii}$ where $\eta_{ii}$ are equal to either 1 or $-1$ then it is just the Clifford algebra. 
In the general case the relations \eqref{omom} can be solved by representing $\om_i$ as  zero mode ``vertex'' operators 
\be
\om_i=\Gamma_i\,e^{i\varphi_i}\,, \quad [\varphi_i,\varphi_j]=i \phi_{ij}\,,
\ee
where $\Gamma_i$ satisfy the Clifford algebra $\Gamma_i\Gamma_j+\Gamma_j\Gamma_i=2\eta_{ij}$ with $\eta_{ij}=\eta_{ii}\delta_{ij}$.
In what follows we find it convenient to choose $\eta_{ii}$ in such a way that 
$\G\equiv \G_1\G_2\cdots \G_N$ satisfies the condition $\G=1$ for odd $N$ which can be achieved by choosing $\eta_{ii}=1$ for $i=1,\cdots N-1$ and $\G_N=\G_{N-1}\cdots \G_1$, while for even $N$ the element $\G$ satisfies the condition $\G^2=1$. 

Thus it is sufficient to find a free field representation for $Z_i$ only. 
In what follows we will be interested only in the untwisted case and we will denote the corresponding vertex operators as $A_i$ without any superscript.

To conclude this discussion let us also mention that 
introducing the twist 
\be
\F_{12}=\sum_{i,j=1}^{N} e^{-i \tau_{ij}} E_{ii}\otimes E_{jj}\,,\quad \F_{21}\equiv \bP\,\F_{12}\,\bP\,,
\ee
where the parameters $\tau_{ij}$ satisfy the conditions 
$\tau_{ij} -\tau_{ji}=\p_{ij} \ mod \ 2\pi$, 
one can easily check that $\R^\Sigma$ is a twisted R-matrix
\be
\R^\Sigma =  \F_{21}\,\R\, \F_{12}^{-1}\,,
\ee
and therefore it satisfies the invariance conditions
\be
\Delta^{op}_\F(\J)\, \R^{\Sigma}=\R^{\Sigma}\,\Delta_\F(\J)\,,
\ee
where $\J$ are $\sl(N)$ generators, $\Delta^{op}_\F(\J)=\bP\,\Delta_\F(\J)\,\bP$ and
\be\la{coprF}
\Delta_\F(\J) = \F_{12}\,(\J_1+\J_2)\,\F_{12}^{-1}\,,
\ee
is the twisted coproduct.\footnote{Strictly speaking one should consider $gl(N)$ (or $u(N)$ if $\F$ is unitary) because for generic $\F$ the coproduct of $\J$ which is from $\sl(N)$ is not in the tensor product of two universal enveloping $\sl(N)$ algebras.} There is however no twist which would lead to the coproduct \eqref{copr}.

\subsection{Free fields}

According to the discussion in subsection \ref{ffrZ}, to construct a free field representation for the vertex operators $Z_k$ of the elementary particles of the twisted ZF algebra (\ref{zzii}, \ref{zzij}) one needs a bosonic operator $a_0(t)$ for the highest weight vertex operator $Z_1$, and  
$N-1$ bosonic operators $a_k(t)$ for the lowering operators $\lok $. 
For any $\mu$ the operators $a_\mu(t)$ and $a_\mu(-t)$ for $t>0$ are the annihilation and creation operators, respectively: $a_\mu(t)|0\rangle = 0 \ {\rm for}\ t>0$. 
Since $(N$-$1)$-particle bound states are antiparticles of the elementary particles, only $N-1$ bosonic operators may be independent. 

The commutation relations of the operators $a_\mu$ can be written in the uniform form
\be\la{comrel}
[a_\mu(t),a_\nu(t')]=t f_{\mu\nu}(t)\delta(t+t')\,,\quad \mu=0,1,\ldots,N-1\,,
\ee
where $f_{\mu\nu}$ must satisfy the relations $f_{\mu\nu}(-t)=f_{\nu\mu}(t)$. In addition we also impose the conditions  $f_{\nu\mu}(t)=f_{\mu\nu}(t)$ which were satisfied in the $N=2$ case, and appear to hold for general $N$ too. We also introduce the zero mode operators $Q_\mu$ and $P_\mu$ such that $P_\mu |0\rangle = 0$. Their  commutation relations are listed in appendix \ref{selrule} but will not be important in this section.

We then define the free fields
\be
\p_\mu(\t)=Q_\mu+\int^\infty_{-\infty}{dt\over i t}\,a_\mu(t)e^{i\theta t}\,,
\ee
which satisfy the following relations 
\bea\begin{aligned}
\left[\phi_\mu(\t_1), \phi_\nu(\t_2)\right]&=\ln S_{\mu\nu}(\t_2-\t_1)\, ,\quad
\langle\phi_\mu(\t_1) \phi_\nu(\t_2)\rangle=-\ln g_{\mu\nu}(\t_2-\t_1)\, .~~~~
\end{aligned}
\eea
Here the S-matrices $S_{\mu\nu}$ and Green's functions $g_{\mu\nu}$ are related to $f_{\mu\nu}$ as follows
\be
S_{\mu\nu}(\t)=\exp\left(-2i\,\int_{0}^\infty\, {dt\ov t}\, f_{\mu\nu}(t)\sin(\theta t)\right)\,, \quad g_{\mu\nu}(\t)=\exp\left(-\int_{0}^\infty\, {dt\ov t}\, f_{\mu\nu}(t)e^{-i\theta t}\right)\,,
\ee
and they are related to each other as
\be
S_{\mu\nu}(\t)={g_{\mu\nu}(-\t)\ov g_{\mu\nu}(\t)}\,.
\ee
The fields $\p_\mu$ are used to construct the basic vertex operators
\be
V_\mu(\t)=: e^{i\phi_\mu(\t)}:\,,
\ee
which obey the following  relations
\be
V_\mu(\t_1)V_\nu(\t_2)
=g_{\mu\nu}(\t_{21}):V_\mu(\t_1)V_\nu(\t_2):\, ,\quad
V_\mu(\t_1)V_\nu(\t_2)
=S_{\mu\nu}(\t_{12})V_\nu(\t_2)V_\mu(\t_1)\, .\la{VVrel}
\ee
The free field realization of the ZF algebra with the twisted S-matrix $\S^{(-1)}$ is constructed as follows
\bea\nonumber
Z_1(\t)&=&\r\, V_0(\t)\, ,\quad \r = e^{i\pi\ov N} e^{\g{N-1\ov 2N}}N^{-{1\ov 2N}} \, ,\quad \lok =\r_\chi\int_{C} d\a V_k(\a) 
 \, ,\quad \r_\chi={e^\g\ov 2\pi}\,,\\
Z_{k+1}(\t)&=&\lok \,Z_{k}(\t)+Z_{k}(\t)\,\lok  \,,\quad k=1,\ldots,N-1\,.\la{ZFop}
\eea
Here $\g$ is Euler's constant and the normalization constants for $Z_1$ and  $\lok $ have been chosen for future convenience. Then  the integration contour $C$ in $\lok $ depends on operators located to the right or to the left of  $\lok $ and is specified for any operator $\chi$ which involves integration as follows  \cite{Lukyanov}.  One first brings the product of all vertex operators in a monomial containing $\chi$ to the normal form which is considered as a regular operator. This produces a product of various Green's functions which may have poles. Then the contour $C$ runs from Re$\,\a=-\infty$ to Re$\,\a=+\infty$ and it lies above all poles due to operators to the right of $\chi$ but below all poles due to operators to the left of $\chi$. 
Note that if one then acts by the resulting monomial operator on other operators the contour $C$ should be additionally deformed according to the procedure described. 
As an example, let us consider the monomial $V_0(\t_L)\,\chi^-_1\, V_0(\t_{R,1})V_2(\t_{R,2})$ and assume for definiteness that $\t_{R,1}<\t_{L}<\t_{R,2}$. One gets
\bea\nonumber
&&V_0(\t_L)\,\chi^-_1\, V_0(\t_{R,1})V_2(\t_{R,2})=\int_C d\a :V_0(\t_L)V_1(\a)V_0(\t_{R,1})V_2(\t_{R,2}): \\\nonumber
&&\times g_{01}(\a-\t_L)g_{10}(\t_{R,1}-\a)g_{12}(\t_{R,2}-\a)g_{00}(\t_{R,1}-\t_L)g_{02}(\t_{R,2}-\t_L)g_{02}(\t_{R,2}-\t_{R,1})\,.
\eea
As will be shown below the Green's functions which depend on $\a$ are equal to 
$$g_{01}(\a)=g_{10}(\a)=g_{12}(\a)= - {ie^{-\g}\ov \a+{i\pi\ov N}}\,,$$
and therefore the poles are at
$$
\a= {\t_L - {i\pi\ov N}}\,,\quad \a= {\t_{R,1}+ {i\pi\ov N}}\,,\quad \a= {\t_{R,2}+{i\pi\ov N}}\,.
$$
\begin{figure}[t]
\begin{center}
\includegraphics[width=0.75\textwidth]{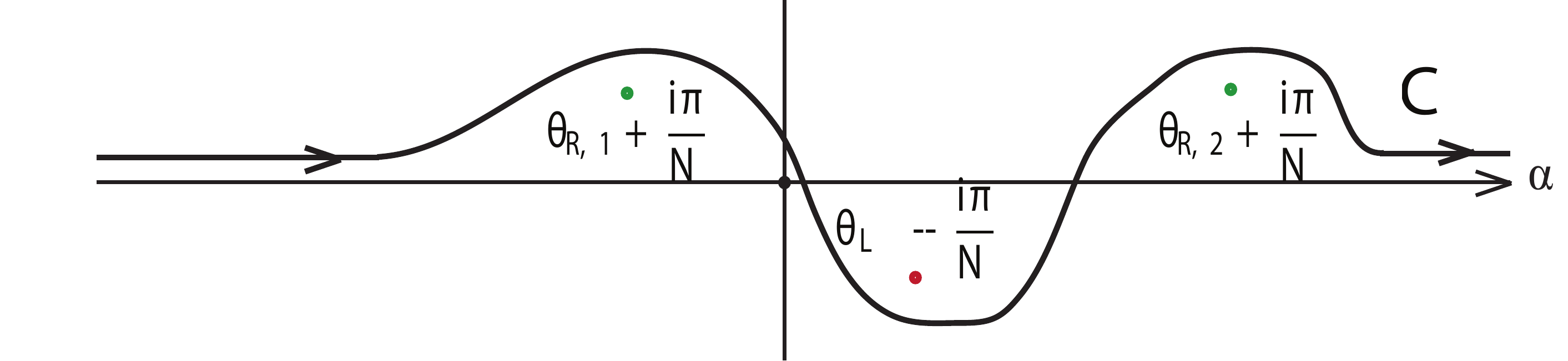}
\caption{\small The integration contour $C$ in the monomial $V_0(\t_L)\,\chi^-_1\, V_0(\t_{R,1})V_2(\t_{R,2})$. }
\label{Contour}
\end{center}
\end{figure}Thus the integration contour $C$ runs above $\t_{R,1}+ {i\pi\ov N}$ and $\t_{R,2}+ {i\pi\ov N}$ but below $\t_{L}-{i\pi\ov N}$, see Figure \ref{Contour}.

Matrix elements of the ZF operators are therefore given by multiple integrals. One sometimes needs to compute integrals of functions which behave as $1/\a$ for large $\a$.  They should be computed with the principle value prescription. In particular $\int_{-\infty}^\infty d\a/(\a\mp i) = \pm\pi i$.

The ZF operators $A_k$ of the GN model are then constructed as
\bea\nonumber
A_k(\t)&=&\G_k Z_k(\t)\, ,\quad  \ljok =\G_{k+1}\G_k^{-1}\lok \,,\\
A_{k+1}(\t)&=&\ljok \,A_{k}(\t)-A_{k}(\t)\,\ljok  \,,\quad k=1,\ldots,N-1\,,\la{AZFop}
\eea
where $\ljok $ are the $\sl(N)$ algebra lowering generators.

Let us now sketch how the commutation relations between $a_\mu$ can be found. From the discussion in subsection \ref{ffrZ} and eq.\eqref{so} we conclude that 
\be
f_{00}(t) = {\sinh{(N-1)\pi t \ov N}\ov \sinh \pi t}e^{\pi |t|\ov N}\,,\quad 
g_{00}(\t)=e^{\frac{(N-1) (\gamma +\log (2 \pi ))}{N}}\frac{ \Gamma
   \left(\frac{i \t}{2 \pi }-\frac{1}{N}+1\right)}{\Gamma \left(\frac{i
   \t}{2 \pi }\right)}\,,
\ee
where $\g$ is Euler's constant.
Then the commutation relations between $a_0$ and $a_1$ can be easily guessed by generalising the $N=2$ relations from \cite{Lukyanov}
\be
f_{01}(t)=- e^{\pi |t| \ov N}\,,\ \ g_{01}(\t)=-\frac{i e^{-\gamma }}{\theta +\frac{i \pi }{N}}\,,\quad f_{11}(t)=1+e^{\frac{2 \pi  |t|}{N}}\,,\ \ g_{11}(\t)=-e^{2 \gamma } \theta  \left(\theta +\frac{2 i \pi }{N}\right)\,.
\ee
It is straightforward to check that $Z_1$ and $Z_2$ indeed obey the ZF algebra relations (\ref{zzii}, \ref{zzij}).  
To find the remaining commutation relations one notices that 
since $\lok $ basically are $\sl(N)$ algebra generators they should have the following properties
\be\la{chikm}
\lok \lom=\lom\lok  \quad {\rm if}\quad |k-m|\neq 1\,,
\ee
and 
\be\la{chiz1}
\lok Z_{m}(\t)=Z_{m}(\t)\lok  \quad {\rm unless}\quad m=k \quad {\rm or}\quad m=k+1\,.
\ee
In addition, for $m=k+1$ one should find
\be\la{chiz2}
\lok Z_{k+1}(\t)=-Z_{k+1}(\t)\lok \, .
\ee
This indeed can be checked for $k=1$ by using the $g_{01}$ and $g_{11}$ functions. To this end one should use \eqref{ZFop} for $Z_2$ and take  the integration contours in $\chi_{1}^-$  to the real line. 
Then the integrand in the double integral can be symmetrized in the integration variables and gives 0, and  the integrands in the single integrals appearing due to the poles in $g_{01}$ sum up to 0 as well. Eq.\eqref{chiz2} can in fact be used to find $g_{11}$.

It is then easy to check that the commutativity of $Z_1$ with $\chi_{2}^-$ guaranties that $Z_1$ and $Z_3$ satisfy the ZF algebra relation \eqref{zzij}.
The relations (\ref{chikm}, \ref{chiz1}) show that only adjacent operators can have nontrivial commutation relations, and therefore only the functions $g_{\mu\mu}$ and $g_{\mu,\mu\pm1}=g_{\mu\pm1,\mu}$ are nontrivial.  
Thus, using the integration contour prescription and the relations \eqref{VVrel} one can represent $Z_{k+1}$ in the following (slightly symbolic) form
\be
Z_{k+1}(\t)=\r\int_{C_k} d\a_k \cdots\int_{C_1} d\a_1 \, \prod_{j=1}^k g_{j-1,j}^s(\a_j-\a_{j-1})\,:V_0(\t)V_1(\a_1)\cdots V_k(\a_k): \,,
\la{Zk}
\ee
where $\a_0\equiv\theta$ and
\be
g_{j-1,j}^s(\a)\equiv \r_\chi\big(g_{j-1,j}(\a)+g_{j-1,j}(-\a)\big)\,.
\ee
The integration contour $C_j$ runs below the poles of $g_{j-1,j}(\a_j-\a_{j-1})$ and above the poles of $g_{j-1,j}(\a_{j-1}-\a_j)$. Strictly speaking \eqref{Zk} is a sum of $2^k$ integrals with contours specified by the product of $g_{j-1,j}$ functions in each integrand.
Only $Z_{2}$ is given exactly by  \eqref{Zk} with $g_{01}^s(\a)=-\frac{1/ N}{\a^2+\pi ^2/N^2}$ and the poles of $g_{01}^s$ at $\a_1= \t-{i\pi\ov N}$ and  $\a_1 = \t+{i\pi\ov N}$ lying above and below  
 $C_1$, respectively.
 
 To find the function $g_{12}$ (and therefore $f_{12}$) one can use the requirement of  the commutativity of $Z_3$ with $\chi_{1}^-$ which also implies that $Z_2$ and $Z_3$ satisfy the ZF algebra relation \eqref{zzij}.
Then one gets
\bea\la{chi12z3a}
\chi_{1}^-Z_{3}\big(\t\big) &=&\r\int_{C_1^{\a}} d\a_1\int_{C_{2}^{\b}} d\b_{2} \int_{C_1^{\b}} d\b_1\,g_{01}^{s}(\b_{1}-\t)g_{12}^{s}(\b_{2}-\b_1)\\\nonumber
&\times&g_{01}(\t-\a_1)g_{12}(\b_{2}-\a_1)g_{11}(\b_{1}-\a_1) \,
:V_0(\t)V_1(\a_1)V_1(\b_1)V_2(\b_2):\,,
\eea
\bea\la{chi12z3b}
Z_{3}\big(\t\big)\chi_{1}^- &=&\r\int_{C_1^{\a}} d\a_1\int_{C_{2}^{\b}} d\b_{2} \int_{C_1^{\b}} d\b_1\,g_{01}^{s}(\b_{1}-\t)g_{12}^{s}(\b_{2}-\b_1)\\\nonumber
&\times&g_{01}(\a_1-\t)g_{12}(\a_1-\b_2)g_{11}(\a_{1}-\b_1) \,
:V_0(\t)V_1(\a_1)V_1(\b_1)V_2(\b_2):\,.
\eea 
If the integration contours $C_1^\a$ and $C_1^\b,\, C_2^\b$ were the same in all the integrals (recall that both \eqref{chi12z3a} and \eqref{chi12z3b} are sums of 4 integrals), e.g. they all would coincide with the real line, then one could symmetrize the integrands with respect to $\a_1$ and $\b_1$. 
Then assuming that $g_{12}$ has only one pole (as it is for $g_{01}$) and imposing the requirement  that the symmetrized integrands in \eqref{chi12z3a} and \eqref{chi12z3b} are equal to each other, one immediately finds that $g_{12} = g_{01}$.
Making the integration contours coincide with the real line produces extra terms due to the poles of the $g_{01}$'s, and one has to check that these extra terms cancel out as well. One can do this, and a lengthy computation indeed shows that if $g_{12} = g_{01}$, then
$\chi_{1}^-Z_{3}(\t)=Z_{3}(\t)\chi_{1}^-$. The function $g_{22}$ is found by imposing the ZF algebra relation \eqref{zzii} on $Z_3$ and appears to be equal to $g_{11}$. The same considerations are used to determine the remaining functions, and one finally reaches the natural conclusion
\be
f_{j,j+1}(t)=f_{01}(t)\,,\ \ g_{j,j+1}(\t)=g_{01}(\t)\,,\quad f_{jj}(t)=f_{11}(t)\,,\ \ g_{jj}(\t)=g_{11}(\t)\,.
\ee
The functions $f_{\mu\nu}$, $g_{\mu\nu}$ and $S_{\mu\nu}$ are listed explicitly in the appendix \ref{appf}.
Thus all the functions $g_{j-1,j}^s$ are given by
\be
g_{j-1,j}^s(\a)=-\frac{1/ N}{\a^2+\pi ^2/N^2}\,,\quad j=1,\ldots, N-1\,,
\ee
and the integration contour $C_j$ in \eqref{Zk} runs below the pole of $g_{j-1,j}(\a_j-\a_{j-1})$ at $\a_j = \a_{j-1}-{i\pi\ov N}$ and above the pole of $g_{j-1,j}(\a_{j-1}-\a_j)$ at $\a_j = \a_{j-1}+{i\pi\ov N}$.

The free field representation for the ZF algebra with the twisted S-matrix  $\S^{(-1)}$  first appeared in \cite{Kojima} where it was obtained by taking a proper limit of the free boson realization of the type-I vertex operators of the $A_{N-1}^{(1)}$ spin chain constructed in \cite{KY}. However it is claimed in  \cite{Kojima} that it is a representation of the ZF algebra with the canonical S-matrix which disagrees with our consideration. 
It is noticed in \cite{Kojima} that the 
 commutation relations for the operators $a_k$ can be written in the  nice form 
\be
[a_i(t),a_j(t')]= t {\sh{a_{ij}\pi t\ov N}\ov \sh{\pi t\ov N}}e^{\pi|t|\ov N}\delta(t+t')\,,\quad i,j=1,2,\ldots, N-1\,,
\ee
where $a_{ij}=2 \delta_{ij} - \delta_{i-1,j} - \delta_{i+1,j}$ is the Cartan matrix of type $A_{N-1}$. Then, 
the operator $a_0$ is expressed as the following linear combination of $a_k$
\be\la{a0}
a_0(t)=-\sum_{k=1}^{N-1}{\sh{(N-k)\pi t\ov N}\ov \sh{\pi t}}a_k(t)\,.
\ee
Finally, another linear combination of $a_k$
\be\la{aN}
a_{N}(t)=-\sum_{k=1}^{N-1}{\sh{k\pi t\ov N}\ov \sh{\pi t}}a_k(t)\,,
\ee
is used in  \cite{Kojima} to construct the vertex operator $V_N(\t)=: e^{i\phi_N(\t)}:$ for the bound state $Z_{12\ldots N-1}$ which is the antiparticle of $Z_N$. Vertex operators for bound states will be discussed in more detail in section \ref{bound}.

\subsection{The angular Hamiltonian}
The next step in constructing a free field representation is to find the angular Hamiltonian.
The most general Hamiltonian we might expect would be of the form
\be
\bK=i\,\int^{\infty}_{0}dt \, \sum^{N-1}_{i, j=1} h_{ij}(t) a_{i}(-t) a_{j}(t)\, ,
\ee
where the (anti-)hermiticity condition for $\bK$ requires the functions $h_{ij}$ to obey $h_{ij}=h_{ji}$, and we assume that $h_{ij}$ are even fuctions of $t$: $h_{ij}(-t)=h_{ij}(t)$.
 We wish to satisfy the relations \eqref{eZFc}
\be
{d\ov d \t} Z_j(\t)=-\left[\,\bK,Z_j(\t)\right]\, ,
\ee
where we set $\Om_I=0$ because as we will see in a moment they do vanish for the representation we consider.
Computing the derivative of \eqref{Zk} with respect to $\t$ it is straightforward to show that 
\bea\nonumber
{d\ov d \t}Z_{k+1}(\t)&=&\r\int_{C_k} d\a_k \cdots\int_{C_1} d\a_1 \, \prod_{j=1}^k g_{j-1,j}^s(\a_j-\a_{j-1})\\
&&\qquad\qquad\times\sum_{j=0}^k\left({\pa\ov \pa\a_j}\right):V_0(\a_0)V_1(\a_1)\cdots V_k(\a_k): \,.~~~~~~~
\la{dZk}
\eea
Thus it is sufficient to find $\bK$ such that 
\be
\left[\,\bK,V_\mu(\t)\right]=- {d\ov d \t} V_\mu(\t)\, ,\quad \mu=0,1,\ldots,N\,.
\ee
We get
\be
-{d\ov d \t} V_\mu(\t)=-i\int_{-\infty}^\infty\, dt\, e^{i\t t}:a_\mu(t)e^{i\p_\mu(\t)}:\,,
\ee
and
\be
\left[\,\bK,V_\mu(\t)\right]=:[\,\bK,i\p_\mu(\t)]e^{i\p_\mu(\t)}:= -i\,\int_{-\infty}^\infty\, dt\, e^{i\t t}:f_{\mu k}(t)h_{kj}(t)a_j(t)e^{i\p_\mu(\t)}:\,.
\ee
Thus $h_{ij}$ can be found from the equations
\be\la{fh}
\sum_{k=1}^{N-1}f_{0 k}(t)h_{kj}(t)=-{\sh{N-j\ov N}\pi t\ov \sh{\pi t}}\,,\quad \sum_{k=1}^{N-1}f_{i k}(t)h_{kj}(t)=\delta_{ij}\,.
\ee
Solving these equations one gets
\be
h_{i j}(t)= {\sinh{i \ov N}\pi t \ov \sinh{ \pi t\ov N}}{\sinh{N-j \ov N}\pi t \ov \sinh \pi t} e^{-{\pi \ov N}t}\, ,\quad h_{j i}(t)=h_{i j}(t)\,,\quad  i\leq j\,  .
\ee
It is worth mentioning that the matrix $h$ with the entries $h_{ij}$ is inverse to  the matrix $f$ with the entries $f_{ij}$ which will be important in computing the form factors.

\section{Bound states} \la{bound}
Let us recall that a rank-$r$ particle
created by a ZF operator ${\cal A}^\dagger_{K}(\t)$, ${K} = (k_1,\ldots, k_r)$, $1\le k_1< k_2<\cdots< k_r\le N$  is a bound state of  elementary particles ${\cal A}^\dagger_{k_j}$. Thus the vertex operators $A_{K}$ can be obtained from the vertex operators $A_{k_j}$ by using the fusion procedure. We first construct  the vertex operators $Z_{K}$ for the bound states of the twisted S-matrix $\S^{(-1)}$ which will be normalised so that they  satisfy
the relations
\be\la{ZaZ}
Z_K(\t'+i\pi)Z_L(\t) = -{i \delta_{K\overline L}\ov \t'-\t} +{\cal O}(1)\,,\quad \t'\to\t\,.
\ee
The vertex operator $A_{K}$ of a rank-$r$ particle is then given by
\be\la{AvZ}
A_K = \G_K Z_K\,,\quad \G_K\equiv \G_{k_1}\G_{k_2}\cdots \G_{k_r}\,,
\ee
and the formula \eqref{ZaZ} for $A_K$ takes the form 
\be\la{AaA}
A_K(\t'+i\pi)A_{L}(\t) = -{i\, \G\,\e_{K{ L}}\ov \t'-\t} +{\cal O}(1)\,,\quad \t'\to\t\,,
\ee
where $\G\equiv \G_{1}\G_{2}\cdots \G_N$.
Since $\G_i\G_j+\G_j\G_i=2\eta_{ij}$ with $\eta_{ij}=\eta_{ii}\delta_{ij}$ then for odd $N$ one can choose the first $N-1$  $\eta_{i}$'s to be 1, and $\G_N=\G_{N-1}\G_{N-2}\cdots \G_1$ so that $\G_N^2=\eta_N=(-1)^{N-1\ov 2}$. Then $\G=1$ and the relation  \eqref{AaA} takes the usual form. On the other hand if $N$ is even then 
 $\G$ is not proportional to the identity matrix but one can choose $\eta_{N}$ so that it obeys 
 $\G^2=1$. To satisfy the form factor axioms one then should insert under the trace in the form factor formula \eqref{ffax} the projection operator 
 ${1\ov 2}(1+\G)$.

\subsection{Fused vertex operators}

It is clear from the ZF algebra (or the S-matrix) that if $i\neq j$ then $Z_i$ and $Z_j$ can form a two-particle bound state because
\be
Z_i(\t_1) Z_j(\t_2) =- {{2\pi i\ov N}S({2\pi i\ov N})\ov \t_{12}-{2\pi i\ov N}}\, \big( Z_i(\t_2) Z_j(\t_1)+Z_j(\t_2) Z_i(\t_1)\big)+\ldots \,,\quad \t_{12}\to {2\pi i\ov N}\,.
\ee
Let us introduce the following fused vertex operators of rank-$r$ 
\be\la{vvr}
\cZ_{k_1\ldots k_r}(\t)\equiv  \lim_{\eps_{i+1,i}\to 0}\prod_{j=2}^{r}(i\eps_{j+1,j} )\,Z_{k_1}\big(\t_1^\eps\big)Z_{k_2}\big(\t_2^\eps\big)\cdots Z_{k_r}\big(\t_r^\eps\big)\,,\quad \t_j^\eps\equiv \t +i\u_{r-2j+1}+\eps_j\,,
\ee
where all indices $k_a$ are different (if two indices coincide the fused operator vanishes), $\u_k\equiv {\pi\ov N}k$, $\eps_1=0$  and all $\eps_{jk}\equiv \eps_{j}-\eps_{k}$ do not vanish until one takes the limits. 
The fused operators satisfy the following relation
\be\la{vvr2}
\cZ_{k_1\ldots k_r}(\t) = \lim_{\eps\to 0}\,i\eps\,\cZ_{k_1\ldots k_p}(\t+i\u_{r-p})\cZ_{k_{p+1}\ldots k_r}(\t-i\u_p+\eps)\,,
\ee
where $p$ is any integer between 1 and $r$.

By using the fused vertex operators we can write
\be
Z_i(\t_1) Z_j(\t_2) ={i\ov \t_{12}-{2\pi i\ov N}}\, \cZ_{ij}(\t) +\ldots \,,\quad \t_{j}\to \t+i\u_{3-2j}\,.
\ee
Note that $\cZ_{ij}=\cZ_{ji}$ because the twisted S-matrix has a pole in the symmetric channel, and moreover the associativity of the ZF algebra implies that a general rank-$r$ fused operator \eqref{vvr} is also symmetric under the exchange of  its indices.

It is clear that a two-particle bound state ZF vertex operator is given by
\be
Z_{ij}(\t) = \cN_{2}\,\cZ_{ij}(\t)\,,
\ee
where $\cN_{2}$ is a normalization constant. 
The mass of the two-particle bound state is equal to $m_2=m\sin\u_2/\sin\u_1$ where $m$ is the mass of elementary particles.  It is not difficult to see that $Z_i$ and $Z_{jk}$ with all indices different can also form a bound state which is a three-particle bound state of the mass $m_3=m\sin\u_3/\sin\u_1$, and the corresponding ZF vertex operator can be defined as
\bea
&&Z_{ijk}(\t) = \cN_{3}\,\cZ_{ijk}(\t)\\\nonumber
&&~~~= \cN_{3}\, \lim_{\eps\to 0}\,i\eps\,Z_i(\t +i\u_2) \cZ_{jk}(\t -i\u_1+\eps) =\cN_{3}\, \lim_{\eps\to 0}\,i\eps\,\cZ_{ij}(\t +i\u_1) Z_{k}(\t -i\u_2+\eps)\,.
\eea
This procedure can be repeated and one introduces the ZF vertex operator for a $r$-particle bound state of  mass $m_r=m\sin\u_r/\sin\u_1$ by the formula
\be
Z_{k_1\ldots k_r}(\t)=\cN_{r}\,\cZ_{k_1\ldots k_r}(\t)\,.
\ee
Since $\cZ_{k_1\ldots k_r}$ is symmetric under the exchange of the indices we can use the canonical ordering $k_1<k_2<\cdots<k_r$ which also shows that the  bound states of the twisted model are indeed in one-to-one correspondence with the bound states of the chiral GN model. 
The normalization constants $\cN_{r}$ have to be chosen so that  the bound state ZF vertex operators satisfy the relations \eqref{ZaZ}
\be\la{ZaZb}
Z_K(\t'+i\pi)Z_{\overline K}(\t) =\cN_{r}\cN_{N-r}\,\cZ_K(\t'+i\pi)\cZ_{\overline K}(\t) = -{i\ov \t'-\t} +{\cal O}(1)\,,\quad \t'\to\t\,,
\ee
where $K=(k_1,\ldots, k_r)$ is a bound state index, and $K \cup{\overline K} =(1,2,\ldots,N)$ (after reordering the indices).
It is not difficult to see that $\cN_r\cN_{N-r}$ is independent of $K$ because according to \eqref{vvr2}
\be\la{ZaZ2}
\lim_{\eps\to 0}\, i\eps\,\cZ_{K}(\t+i\pi)\cZ_{\overline K}(\t+\eps)  =\cZ_{12\ldots N}(\t+i\u_k) \equiv {\cV}_{N}\ \ \Longrightarrow\ \  \cN_{r}\cN_{N-r}=-{1\ov \cV_N}\,,
\ee
where one has to show that $\cV_{N}$ is indeed a constant. 
The computation of $\cV_{N}$ is outlined in appendix \ref{appbound} where it is shown that with our choice of $\r$ and $\r_\chi$ it is equal to 
$-1$, and therefore in what follows we choose $\cN_{r}=1$ for any $r$. 

\subsection{Highest weight bound state vertex operators}

As was discussed in section \ref{free} all vertex operators $Z_k$ for elementary particles can be obtained from the  highest weight vertex operator $Z_1$ by acting on it with the lowering symmetry operators $\lok $. It appears that the same is true for the bound state vertex operators. 
Any rank-$r$ vertex operator is generated from the  highest weight vertex operator $Z_{12\dots r}$. Indeed, one has (dropping $i\e$ and $\t$ for clarity)
\bea\nonumber
Z_{1\dots r-1, r+1}&=&Z_{1\dots r-1}Z_{r+1}=Z_{1\dots r-1}\big(\chi_{r}^- Z_{r}+Z_{r}\chi_{r}^-\big)\\
&=&\chi_{r}^-Z_{1\dots r-1}Z_{r}+Z_{1\dots r-1}Z_{r}\chi_{r}^- = \chi_{r}^-Z_{1\dots r}+Z_{1\dots r}\chi_{r}^-\,.~~~~~
\eea
It is clear then that $Z_{1\dots r-1, j+1}$ for $j\ge r$ is given by 
\be
Z_{1\dots r-1, j+1}(\t)=\chi_{j}^-Z_{1\dots r-1, j}(\t)+Z_{1\dots r-1, j}(\t)\chi_{j}^-\,,
\ee
as desired. To obtain $Z_{1\dots r-2, r,r+1}$ we act on $Z_{1\dots r-1, r+1}$ by $\chi_{r-1}^-$. Having found $Z_{1\dots r-2, r,r+1}$  we then construct 
all $Z_{1\dots r-2, r,j+1}$, and then $Z_{1\dots r-2, r+1,j+2}$, and so on. 

The simplest case is provided by rank--$(N$-$1)$ vertex operators. They are obtained from $\overline{Z}_N\equiv Z_{12\ldots N-1}$ which is the antiparticle of $Z_N$. Acting on $\overline{Z}_N$ with $\chi_{N-1}^-$ one creates $\overline{Z}_{N-1}\equiv Z_{1\ldots N-2,N}$ which is the antiparticle of $Z_{N-1}$. Then, acting on $\overline{Z}_{N-1}$ with $\chi_{N-2}^-$ one creates $\overline{Z}_{N-2}\equiv Z_{1\ldots N-3,N-1,N}$ which is the antiparticle of $Z_{N-2}$, and so on
\be\la{bZ}
\overline{Z}_{k}(\t) = \lok \overline{Z}_{k+1}(\t)+\overline{Z}_{k+1}(\t)\lok \,,\quad k=N-1,\ldots,1\,.
\ee
The resulting formula agrees with the one in \cite{Kojima}.
 
The highest weight vertex operators $Z_{12\dots r}$ can be simplified to an explicit form which contains no integrals at all. The derivation is presented 
 in appendix \ref{appbound} and here we just state the result
 \be\la{v12k}
Z_{12\ldots r}(\t) = C_{N,r}V_{(r)}(\t)\,,\quad V_{(r)}(\t)\equiv\ :\prod_{k=0}^{r-1}\prod_{j=k+1}^{r}V_{k}\big(\t + i\u_{r+k-2j+1}\big):\,.
\ee
Here the normalization constant $C_{N,r}$ is given by
\be\la{CNr}
C_{N,r}=e^{\frac{i\pi r}{N}} e^{\gamma\frac{r (N-r)}{2 N}} N^{-\frac{r}{2 N}} (2 \pi )^{\frac{(r-1) (2 N-r)}{2 N}} \prod
   _{j=1}^{r-1} \frac{1}{\Gamma \left(\frac{j}{N}\right)}\, ,
\ee
and the fused  vertex operator $V_{(r)}$ can be written in the usual form
\be\la{v12k2}
V_{(r)}(\t)=\ :e^{i\p_{(r)}(\t)}:\,,\quad \p_{(r)}(\t)=\ :\sum_{k=0}^{r-1}\sum_{j=k+1}^{r}\p_{k}\big(\t + i\u_{r+k-2j+1}\big):\ = \int^\infty_{-\infty}{dt\over i t}\,a_{(r)}(t)e^{i\theta t}\,,
\ee
where 
\be\la{ar}
a_{(r)}(t)=\sum_{k=0}^{r-1}a_{k}{\sinh{r-k\ov N}\pi t\ov \sinh{\pi t\ov N}}   \,.
\ee
Notice that $V_{(1)}=V_0$ and $V_{(N-1)}=V_N$ as follows from (\ref{a0}, \ref{aN}).
The requirement that the fused vertex operator $Z_{12\ldots N}$ is the constant $C_{N,N}=\cV_N=-1$, or equivalently $V_{(N)}=1$, leads to the relation \eqref{a0} between $a_0$ and $a_k$ which can be imposed because it is consistent with the commutation relations \eqref{comrel} between $a_\mu$. 
For the highest weight vertex operator $\overline{Z}_N$ the formula also simplifies  
\be\la{z12kb}
\overline{Z}_N(\t)\ =C_{N,N-1}V_{N}(\t)\,,
\ee
where 
\be\la{CNNm1}
C_{N,N-1}=-e^{\gamma\frac{  (N-1)}{2 N}} N^{\frac{1}{2 N}} (-2 \pi )^{-1/N} \Gamma \left(\frac{N-1}{N}\right)\,.
\ee

This concludes the construction of the free field representation of the extended ZF algebra, and now we turn to the determination of operators $\Lambda$ representing local operators of the chiral GN model.

\section{Local operators}\la{localO}

In this section we construct a large set of operators which commute with the ZF vertex operators, and can be used to generate form factors of  local operators of the chiral GN model. The consideration generalises the one in \cite{Lukyanov} where the $N=2$ case was discussed in detail.

\subsection{Primed ZF operators}\la{primedZ}

We follow  \cite{Lukyanov} and  introduce auxiliary operators $a_\mu'\,,\, Q_\mu'$ related to $a_\mu\,,\, Q_\mu$ by
\be
a_\mu'(t) = -e^{-{\pi |t|\ov N}} a_\mu(t)\,,\quad  Q_\mu'=- Q_\mu\,.
\ee
The commutation relations of the operators $a_\mu'$ have the form
\be
[a_\mu'(t),a_\nu'(t')]=t f_{\mu\nu}''(t)\delta(t+t')\,,\quad [a_\mu'(t),a_\nu(t')]=t f_{\mu\nu}^{\prime}(t)\delta(t+t')\,,
\ee
where $f_{\mu\nu}''$ and $f_{\mu\nu}'$ satisfy the same symmetry relations as $f_{\mu\nu}(t)$, and are listed explicitly in the appendix \ref{appf}.
We also define the free fields
\be
\p_\mu'(\t)= Q_\mu'+\int^\infty_{-\infty}{dt\over i t}\,a_\mu'(t)e^{i\theta t}\,,\quad \mu=0,1,\ldots,N\,,
\ee
which satisfy the following relations 
\bea\begin{aligned}
\langle\phi_\mu'(\t_1) \phi_\nu'(\t_2)\rangle&=-\ln g_{\mu\nu}''(\t_2-\t_1)\, ,\quad \\
\langle\phi_\mu(\t_1) \phi_\nu'(\t_2)\rangle&=\langle\phi_\mu'(\t_1) \phi_\nu(\t_2)\rangle=-\ln g_{\mu\nu}^{\prime}(\t_2-\t_1)\,,~~~~
\end{aligned}
\eea
where the functions can be found in appendix \ref{appg}.

The fields $\p_\mu'$ are used to construct the primed vertex operators
\be
V_\mu'(\t)=: e^{i\phi_\mu'(\t)}:\,,
\ee
and the primed $Z_{k+1}'$ 
\bea\begin{aligned}
Z_1'(\t)&=\r'\, V_0'(\t)\, ,\quad\r'=e^{\gamma\frac{  (N-1)}{2 N}} N^{\frac{1}{2 N}}\, ,\quad \rok =\r_\chi\int_{C} d\a V_k'(\a) 
\,,\\
Z_{k+1}'(\t)&= Z_{k}'(\t)\,\rok  +\rok \,Z_{k}'(\t)\,,\quad k=1,\ldots,N-1
\,,\la{ZFopp}
\end{aligned}
\eea
where the integration contour $C$ in $\rok$ is determined in the same way as for $\lok $. 
Then one gets the following representation 
\be
Z_{k+1}'(\t)=\r'\int_{C_k} d\a_k \cdots\int_{C_1} d\a_1 \, \prod_{j=1}^k g_{j-1,j}''\hspace{-0.5cm}{}^s~~~(\a_j-\a_{j-1})\,:V_0'(\t)V_1'(\a_1)\cdots V_k'(\a_k): \,,
\la{Zkp}
\ee
where $\a_0\equiv\theta$ and
\be
g_{j-1,j}''\hspace{-0.5cm}{}^s~~~(\a)\equiv \r_\chi\big(g_{j-1,j}''(\a)+g_{j-1,j}''(-\a)\big)=\frac{1/ N}{\a^2+\pi ^2/N^2}\,,\quad j=0,1,\ldots, N-1\,.
\ee
The integration contour $C_j$ runs above the pole of $g_{j-1,j}''(\a_{j-1}-\a_j)$ at $\a_j = \a_{j-1}-{i\pi\ov N}$ and below the pole of $g_{j-1,j}''(\a_j-\a_{j-1})$ at $\a_j = \a_{j-1}+{i\pi\ov N}$ because 
$g_{j,j+1}''(\t)=-\frac{i e^{-\gamma }}{\theta -\frac{i \pi }{N}}$. Thus for $\theta\in {\mathbb R}$ all the contours coincide with the real line.

The primed operators $A_k'$ of the GN model are then constructed as
\bea\nonumber
A_k'(\t)&=&\G_k^{-1} Z_k'(\t)\, ,\quad  \rjok =\G_{k}\G_{k+1}^{-1}\chi_{k}^-\,,\\
A_{k+1}'(\t)&=&A_{k}'(\t)\,\rjok -\rjok \,A_{k}'(\t) \,,\quad k=1,\ldots,N-1\,,\la{Apop}
\eea
where $\rjok $ are the $\sl(N)$ algebra raising generators.

Similar to the $\lok $ operators,  $\rok $ satisfy the following properties
\be\la{chipkm}
\rok \rom=\rom\rok  \quad {\rm if}\quad |k-m|\neq 1\,,
\ee
and 
\be\la{chipzp1}
\rok Z_{m}'(\t)=Z_{m}'(\t)\rok  \quad {\rm unless}\quad m=k \quad {\rm or}\quad m=k+1\,.
\ee
Then, for $m=k+1$ one finds
\be
\rok Z'_{k+1}(\t)=-Z'_{k+1}(\t)\rok \, ,
\ee
because in the primed case the integration contour in $\rok $  runs between the poles, and the integrand can therefore be symmetrized in the integration variables to give the result. In addition to these properties there are a number of similar mixed relations between primed and unprimed operators. The obvious relations are
\be\nonumber
\rok \chi_{k+1}^- = -  \chi_{k+1}^-\rok\,, \quad \lok \chi_{k+1}^+ = -  \chi_{k+1}^+\lok\,, \quad \rok \lom=\lom\rok \,, \quad{\rm if}\quad |k-m|>1\,,
\ee
\be\nonumber
\rok Z_{m}(\t)=Z_{m}(\t)\rok \,, \quad \lok Z'_{m}(\t)=Z'_{m}(\t)\lok \quad{\rm if}\quad m< k\,,
\ee
\be\la{chipz1}
\rok Z_{k}(\t)=-Z_{k}(\t)\rok  \,,\quad \lok Z'_{k}(\t)=-Z'_{k}(\t)\lok \,.
\ee
Then one can show that
\be\la{chipchi}
\big[ \rok \,,\,\lok \big]\simeq P_{k}\, .
\ee
This relation only holds for arbitrary matrix elements of $Z_k$'s and $Z'_k$'s but not in the operator form which is reflected in $\simeq$. This is the most important relation, and together with 
\be
\big[P_k, Z_k\big] = Z_k\,,\quad \big[P_k, Z_k'\big] = -Z_k'\,,
\ee
it can be used to prove that
\be\la{chiZ}
\rok Z_{k+1}(\t)+Z_{k+1}(\t)\rok \simeq Z_{k}(\t)\, ,\quad
\lok Z'_{k+1}(\t)+Z'_{k+1}(\t)\lok \simeq Z'_{k}(\t)\, ,
\ee
\be\nonumber
\rok Z_{m}(\t)\simeq Z_{m}(\t)\rok \,, \quad \lok Z'_{m}(\t)\simeq Z'_{m}(\t)\lok \quad{\rm if}\quad m>k+1\,.
\ee

A straightforward computation then shows that the primed ZF operators satisfy the following relations
\be
Z_i'(\t_1) Z_i'(\t_2) =S''(\t_{12})Z_i'(\t_2) Z_i'(\t_1)\,,
\ee
\be\la{zpzp}
Z_i'(\t_1) Z_j'(\t_2) =S''(\t_{12})\left[-{\t_{12}\ov \t_{12}+{2\pi i\ov N}} Z_j'(\t_2) Z_i'(\t_1)+{{2\pi i\ov N}\ov \t_{12}+{2\pi i\ov N}} Z_i'(\t_2) Z_j'(\t_1)\right]\,,
\ee
where 
\be
S''(\t) = g''(-\t)/g''(\t) = \frac{\Gamma \left(1-\frac{i \theta }{2 \pi }\right) \Gamma \left(\frac{i \theta
   }{2 \pi }+\frac{1}{N}\right)}{\Gamma \left(\frac{i \theta }{2 \pi
   }+1\right) \Gamma \left(\frac{1}{N}-\frac{i \theta }{2 \pi }\right)}\,.
\ee
$S''(\t) $ has a pole at $\t=2\pi i/N$, and a zero at $\t=-2\pi i/N$ which cancels the pole in the brackets of \eqref{zpzp}.  
Let us introduce the following notation
\be
\t^{[\pm]}\equiv \t \pm {i\pi\ov N}\,.
\ee
Since the ZF relations are regular at $\t_{12}=-{2\pi i\ov N}$,  the operator product $Z_i'(\t^{[-]}) Z_j'(\t^{[+]})$ is regular too. On the other hand since $S''(\t) $ has a pole at $\t=2\pi i/N$ the product $Z_i'(\t'^{[+]}) Z_j'(\t^{[-]})$ would have a pole at $\t'=\t$ for any $i,j$  unless 
\be\la{zpzptmtp}
Z_i'(\t'^{[-]}) Z_j'(\t^{[+]})=Z_j'(\t'^{[-]}) Z_i'(\t^{[+]})\quad {\rm
 for}\ \  \t'\sim \t\,,
 \ee
  because then the expression in the brackets in \eqref{zpzp} would have a zero at $\t_{12}=2\pi i/N$ which cancels the pole. To prove this we first notice that 
\be\la{zpzptmtp0}
 Z_i'(\t'^{[-]}) Z_i'(\t^{[+]})=0
\quad {\rm for}\ \  \t'\sim \t
\ee 
because $S''(\t) $ has a zero at $\t=-2\pi i/N$. Then 
\be
Z_{i}'(\t^{[-]}) Z_{i+1}'(\t^{[+]})=Z_{i}'(\t^{[-]})\roi Z_{i}'(\t^{[+]})= Z_{i+1}'(\t^{[-]}) Z_{i}'(\t^{[+]})\,,
\ee
and
\be\nonumber
Z_{i}'(\t^{[-]}) Z_{i+2}'(\t^{[+]})=Z_{i}'(\t^{[-]}) \Big[\chi_{i+1}^+Z_{i+1}'(\t^{[+]})+Z_{i+1}'(\t^{[+]})\chi_{i+1}^+\Big]=Z_{i+2}'(\t^{[-]}) Z_{i}'(\t^{[+]})\,,
\ee
because $Z_{i}'$ commutes with $\chi_{i+2}^+$. The same proof works for $j-i>2$. 

Writing
\be
S''(\t) = {R''\ov \t-{2\pi i\ov N}} +\  {\rm regular}\,,\quad R'' = -\frac{2 i \pi  \Gamma \left(\frac{N+1}{N}\right)}{\Gamma \left(\frac{2}{N}\right)
   \Gamma \left(\frac{N-1}{N}\right)}\,,
\ee
we can now find
 \bea\la{zpzppm0}
Z_i'(\t^{[+]}) Z_i'(\t^{[-]})&=&-R''\, \partial_\t Z_i'(\t^{[-]}) Z_i'(\t^{[+]})=R''\, Z_i'(\t^{[-]})  \partial_\t Z_i'(\t^{[+]})\,,
\\\nonumber
Z_i'(\t ^{[+]}) Z_j'(\t^{[-]})&+&Z_j'(\t ^{[+]}) Z_i'(\t^{[-]})=-{R''}{N\ov 2\pi i}\, Z_j'(\t^{[-]}) Z_i'(\t^{[+]})\\\nonumber
Z_i'(\t ^{[+]}) Z_j'(\t^{[-]})&-&Z_j'(\t ^{[+]}) Z_i'(\t^{[-]})={R''}\big(\partial_\t Z_i'(\t^{[-]}) Z_j'(\t^{[+]})- \partial_\t Z_j'(\t^{[-]}) Z_i'(\t^{[+]})\big)\\\nonumber&~&~~~~~~~~~~~~~~~~~~~~={R''}\big( Z_j'(\t^{[-]}) \partial_\t Z_i'(\t^{[+]})-  Z_i'(\t^{[-]}) \partial_\t Z_j'(\t^{[+]})\big)\,.~~~
 \eea
 
 Eqs.(\ref{zpzptmtp}, \ref{zpzptmtp0}) show that a natural analog of the fused vertex operators $\cZ_{k_1\ldots k_r}$ in the primed case is
\be\la{vvrp}
Z'_{k_1\ldots k_r}(\t)\equiv  \lim_{\eps_{i+1,i}\to 0}\,Z'_{k_1}\big(\t_1^\eps\big)Z'_{k_2}\big(\t_2^\eps\big)\cdots Z'_{k_r}\big(\t_r^\eps\big)\,,\quad \t_j^\eps\equiv \t -i\u_{r-2j+1}+\eps_j\,,
\ee
where the regularization parameters $\e_j$ satisfy $\eps_1=0$  and all $\eps_{jk}\equiv \eps_{j}-\eps_{k}$ do not vanish until one takes the limits. 
The primed fused operators are  symmetric under the exchange of their indices (if two indices coincide it vanishes) and satisfy the relation
\be\la{vvrp2}
Z'_{k_1\ldots k_r}(\t) = \lim_{\eps\to 0}Z'_{k_1\ldots k_p}(\t-i\u_{r-p})Z'_{k_{p+1}\ldots k_r}(\t+i\u_p+\eps)\,,
\ee
where $p$ is any integer between 1 and $r$. Just as it was for $Z_{k_1\ldots k_r}$, any primed fused operator can be obtained from the lowest weight fused operators $Z'_{12\ldots r}$ by acting on them with the symmetry generators $\rok $. 

 It is shown in appendix \ref{appbound2}  that the lowest weight primed operators $Z'_{12\dots r}$ can be also reduced to the following explicit form
 \be\la{v12kp}
Z'_{12\ldots r}(\t)=D_{N,r}V_{(r)}'(\t)\,,\quad
V_{(r)}'(\t)\equiv\ 
:\prod_{k=0}^{r-1}\prod_{j=k+1}^{r}V'_{k}\big(\t - i\u_{r+k-2j+1}\big):\,,
\ee
where the normalization constant $D_{N,r}$ is given by
\be\la{DNr}
D_{N,r}=e^{-\frac{\gamma  (r) (r-N)}{2 N}} N^{\frac{r}{2 N}} (2 \pi )^{-\frac{(r-1) r}{2 N}} \prod _{m=1}^{r-1} \Gamma
   \left(1-\frac{m}{N}\right)\,,
\ee
and the fused primed vertex operator $V_{(r)}'(\t)=\ :e^{i\p_{(r)}'(\t)}:$ is given by
\be\la{v12k2p}
\p_{(r)}'(\t) = \int^\infty_{-\infty}{dt\over i t}\,a_{(r)}'(t)e^{i\theta t}\,,\quad a_{(r)}'(t) = -e^{-{\pi |t|\ov N}} a_{(r)}(t)\,.
\ee
Due to the relation \eqref{a0} between $a_0$ and $a_k$, and our choice of $\r'$ and $\r_\chi$,  the fused primed operator $Z'_{12\ldots N}$ is just the constant $D_{N,N}$ equal to 1
\be\la{DNN0}
Z'_{12\ldots N}=D_{N,N}=1\, .
\ee
For the rank--$(N$-$1)$ lowest weight vertex operator the formula also simplifies to
\be\la{v12kpb}
\overline{Z}_N'\equiv Z'_{12\ldots N-1}(\t)=D_{N,N-1}V'_{(N-1)}\big(\t\big)=D_{N,N-1}V'_{N}\big(\t\big)\,,
\ee
where 
\be\la{DNNm1}
D_{N,N-1}=\frac{e^{\frac{\gamma  (N-1)}{2 N}} N^{-\frac{1}{2 N}} (2 \pi )^{\frac{N-1}{N}}}{\Gamma \left(\frac{1}{N}\right)}\,.
\ee

\subsection{$\Lambda$ and $T$ operators}

To discuss the operators $\Lambda$ which are  used to generate form factors of local operators we need the following algebra of the ZF vertex operators and the primed operators
\be\la{zzp}
Z_i(\t_1) Z_j'(\t_2) =-(-1)^{\delta_{ij}}S'(\t_{12})Z_j'(\t_2) Z_i(\t_1)\,,
\ee
where 
\be
S'(\t) = {g'(-\t)\ov g'(\t)} =-\frac{\sinh \left(\frac{\theta }{2}-\frac{i \pi }{2 N}\right)}{\sinh \left(\frac{\theta }{2}+\frac{i \pi }{2 N}\right)}\,.
\ee
Obviously, $S'(\t) $ has a pole at $\t=-\pi i/N$, and a zero at $\t=\pi i/N$. 
In addition, for any $\t$, $S'(\t)$ satisfies the following important relation 
\be\la{sprel}
 \prod_{k=1}^{N}S'\Big(\t+2k {i\pi\ov N}\Big) =(-1)^{N-1}\,.
\ee
The commutation relations \eqref{zpzptmtp},  \eqref{zpzptmtp0} and  \eqref{zzp} take a simpler form if  one introduces  the primed operators for the GN model 
\be\la{apr}
A_j'(\t)=\G_j^{-1}Z_j'(\t)\,,\quad A'_{k_1\ldots k_r}(\t) = \G_{k_1}^{-1}\cdots \G_{k_r}^{-1} Z'_{k_1\ldots k_r}(\t)\,.
\ee
Then the relations above take the form 
\be\la{apaptmtp}
A_i'(\t^{[-]}) A_j'(\t^{[+]})=-A_j'(\t^{[-]}) A_i'(\t^{[+]})\,,
 \ee
\be\la{aap}
A_i(\t_1) A_j'(\t_2) =S'(\t_{12})A_j'(\t_2) A_i(\t_1)\,.
\ee
Let us introduce the following operators 
\be\la{Lop}
\Lambda_{j_1j_2...j_N}^{\P}(\t)=\G' A_{j_1}'\big(\t_{\P(1)}\big)\cdots A_{j_N}'\big(\t_{\P(N)}\big)  =\G' \prod_{a=1}^{N}A_{j_a}'\big(\t_{\P(a)}\big)\,,
\ee
where $\P$ is any permutation of  $1,2,\ldots N$, $\G'\equiv \G_{N}\cdots \G_{1}$,  and
$\t_k\equiv \t -i\u_{N-2k+1}$.
Then taking into account \eqref{sprel} one concludes that these operators 
commute with $A_i$ for both $N$ odd and even
\be\la{aL}
A_i(\t_1) \Lambda^{\P}_{j_1...j_N}(\t_2)=\Lambda^{\P}_{j_1...j_N}(\t_2) A_i(\t_1)\,.
\ee
It is worth mentioning that the indices $j_a$ are arbitrary and some of them may coincide. If  the  permutation $\P$ is trivial, $\P=id$ then  
\be\la{Lop0}
\Lambda_{j_1...j_N}^{id}(\t)=\e_{j_1...j_N}\,.
\ee
The simplest nontrivial class of the operators $ \Lambda^{\P}_{j_1...j_N}$ is obtained for a cyclic permutation $\P_\l\equiv (N-\l+1,\dots,N,1,\ldots,N-\l)$ and $\P_0=\P_N=id$. Taking into account \eqref{vvrp} and \eqref{apr} one finds
\be
 \Lambda^{\P_\l}_{j_1...j_N}(\t)=\G'  A'_{j_1...j_\l}(\t+i \u_{N-\l})A'_{j_{\l+1}...j_N}(\t-i\u_\l)\,.
\ee
It is clear that the operators $\Lambda^{\P}_{j_1...j_N}$ are not linearly independent. If for a generic set of indices $j_1,\ldots,j_N$ in a permutation $\P=(p_1,\ldots,p_{k-1},p_k,\ldots,p_N)$ one has $|p_k-p_{k-1}|\ge 2$ for some $k$ then by using the ZF algebra for the primed operators one can exchange the positions of the operators 
$A'_{j_{k-1}}$ and $A'_{j_{k}}$ and express 
$\Lambda^{\P}_{j_1...j_N}$ as a linear combination of 
$\Lambda^{\P'}_{j_1...j_N}$ with 
$\P'=(p_1,\ldots,p_{k},p_{k-1},\ldots,p_N)$. 
This allows one to choose a convenient basis of the operators  $\Lambda^{\P}_{j_1...j_N}$. In a given permutation $\P$ one first moves $N$ to the left either to the first position or to $N-1$. Then one moves $N-1,N$ to the left either to the first position or to $N-2$.  Finally one gets the permutation
$\P'=(N-\l_1+1,\dots,N,\P_{N-\l_1})$ where $\P_{N-\l_1}$ is a permutation of  $1,\ldots,N-\l_1$. Repeating the procedure one eventually gets the following permutation
\be\la{canP}
\P_{\vec \l}=\big(\{N-\l_1+1,\dots,N\},\{N-\l_1-\l_2+1,\dots,N-\l_1\},\ldots,\{1,\dots,N-\sum_{b=1}^{r-1}\l_b\}\big)\,,
\ee
which consists of $r$ length $\l_k$ sequences of consecutive integers. The corresponding operators $\Lambda$  are then given by
\be\la{canL}
 \Lambda^{\P_{\vec \l}}_{j_1...j_N}(\t)=\G'  A'_{j_1...j_{\l_1}}(\t_1)A'_{j_{\l_1+1}...j_{\l_1+\l_2}}(\t_2)\cdots A'_{j_{N-\l_r+1}...j_{N}}(\t_r)\,,
\ee
where $\t_k=\t +i\u_{N-\l_k-2\sum_{i=1}^{k-1}\l_i}$, and $\sum_{i=1}^{r}\l_i=N$. These operators are obviously antisymmetric under the exchange of indices in each of the vertex operators $A'_{j_1...j_{k}}$. 

In addition if one considers a linear combination of $\Lambda^{\P}_{j_1...j_N}$ which is antisymmetric under the exchange of  the indices $j_{k},j_{k+1},\ldots, j_{k+n}$ for some $k$ and $n$ then in the permutation $\P$ one can always reorder $p_{k},p_{k+1},\ldots, p_{k+n}$ so that $p_{k}<p_{k+1}<\ldots<p_{k+n}$. As a result if in the operator \eqref{canL} the right boundary of the sequence $\l_j$ and  the left boundary of the sequence $\l_{j+1}$ lie between the positions 
$k$ and $k+n$ then these sequences can be united into one sequence of length $\l_2+\l_3$. In what follows we will always consider operators \eqref{canL}.

The operators $\Lambda^{\P}_{j_1...j_N}$ and their  products form an overcomplete  basis of an algebra of operators commuting with $A_i$ and can be used to generate form factors of many local operators of the chiral GN model. The local operators however would have the trivial index $\Omega(O,I)$ equal to 0.

\medskip

Another set of interesting operators is 
\be\la{Ttr}
T_r(\t)={\G'\ov N!}\,\e^{a_1a_{2}\ldots a_N}A'_{a_1\ldots a_r}(\t-{i\pi\ov 2})\partial_\t A'_{a_{r+1}\ldots a_N}(\t+{i\pi\ov 2})\,,\quad r=1,\ldots,N-1\,.
\ee
Taking into account the identity \eqref{sprel}, and the relations \eqref{aap} and \eqref{Lop0}, one finds
\be\la{aT}
\big[T_r(\t),A_i(\b)\big] =\partial_\t\log s_r(\t-\b) \, A_i(\b)\,,
\ee
where
\be\la{sr}
s_r(\t)\equiv\prod_{j=1}^{r}S'\big(-\t-i\u_{r-2j+1}-{i\pi\ov 2}\big)=\prod_{j=1}^{r}{1\ov S'\big(\t+i\u_{r-2j+1}+{i\pi\ov 2}\big)}\,.
\ee
The functions $s_r$ are not real unless $r=N/2$. It might be convenient to take the following linear combinations of $T_r$
\be
T_r^{+}(\t)=T_r(\t) + T_{N-r}(\t)\,,\quad T_r^{-}(\t)=i\big(T_r(\t) - T_{N-r}(\t)\big)\,,
\ee
which lead to real functions $s_r^\pm$
\be
s_r^+(\t) =(-1)^{N-1}\prod_{j=1}^{r}{S'\big(\t+i\u_{r-2j+1}-{i\pi\ov 2}\big)\ov S'\big(\t+i\u_{r-2j+1}+{i\pi\ov 2}\big)}\,,
\ee
\be
s_r^-(\t) = 
i\,(-1)^{N-1} \prod_{j=1}^{r}{1\ov S'\big(\t+i\u_{r-2j+1}+{i\pi\ov 2}\big)S'\big(\t+i\u_{r-2j+1}-{i\pi\ov 2}\big)}\,.
\ee
The meaning of the operators $T_r$ was discussed at length in \cite{Lukyanov}. They allow one to generate form factors of commutators of  a local operator with local commutative integrals of motion.  
Let us finally mention that as in the $N=2$ case the operators $\Lambda^{\P}_{j_1...j_N}$  and $T_r$ form a quadratic algebra.

\section{Form factors of the chiral GN model}\la{form}

Following \cite{Lukyanov}, we expect that, up to an overall normalization constant, form factors of  a local operator should be generated by the following functions
\be\la{Ffunc}
\begin{aligned}
& \cF^{\cV}_{M_1\ldots M_n}(\t'_1,\ldots,\t'_k|\t_1,\ldots,\t_n) \\
&\qquad\qquad={\Tr_{\pi_{A}}\left[{1\ov2}(1+ \G)e^{2\pi i \,\bK}\Lambda^{\cV} (\t'_k,\ldots , \t'_1) A_{M_n}(\t_{n})\cdots A_{M_1}(\t_{1})\right]\ov \Tr_{\pi_{A}}\left[{1\ov2}(1+ \G)e^{2\pi i \,\bK}\right]}\,.\qquad
\end{aligned}
\ee
Here $\Lambda^{\cV} (\t'_k,\ldots , \t'_1)$ are linear combinations of the products of $k$ operators $\Lambda^{\P}_{{\bf j}}(\t')$ with ${\bf j}=\{j_1...j_N\}$ transforming in an irreducible representation $\cV$ of $\su(N)$
\be
\Lambda^{\cV} (\t'_k,\ldots , \t'_1) =\sum c^{\cV}_{{\bf j}_k\ldots {\bf j}_1}\, \Lambda^{\P_k}_{{\bf j}_k}(\t'_k)\cdots \Lambda^{\P_1}_{{\bf j}_1}(\t'_1)\,,
\ee
and $A_{M_i}(\t_i)$ is a rank-$r_i$ bound state vertex operator. These functions are therefore combinations of traces of the form
\be\la{trApA}
{\Tr_{\pi_{A}}\left[{1\ov2}(1+ \G)e^{2\pi i\, \bK}\Big(\prod_{j=1}^{N}\prod_{a=1}^{m_j'} A'_{j}(\t'_{j,a})\Big) \Big(\prod_{j=1}^{N}\prod_{a=1}^{m_j} A_{j}(\t_{j,a})\Big) \right]\ov \Tr_{\pi_{A}}\left[{1\ov2}(1+ \G)e^{2\pi i \,\bK}\right]}\,,
\ee
where $\sum_{j=1}^N{m_j'}=M'=kN$ and   $\sum_{j=1}^N{m_j}=M=\sum_{j=1}^n{r_j}$
are the total numbers of $A'_j$ and $A_j$ operators.
 Such a trace does not vanish only if $m_j$ and $m_j'$ satisfy the following selection rules
\be\la{selrulesa}
m_j-m_j' =r\,,\quad M-M' = r N {\rm \ for\ some\ integer\ }r\,.
\ee
This formula shows that if $A$ transforms in the fundamental representation of $\su(N)$, then $A'$ transforms in the antifundamental representation, and the form factor vanishes unless a decomposition of the product of all $A$ and $A'$ into irreducible representations  of $\su(N)$ contains a singlet, which is a natural requirement. 
Since $A_k=\G_k Z_k$, one can also see from \eqref{selrulesa} that up to a sign \eqref{trApA} is equal to 
\be\la{trZpZ}
{\Tr_{\pi_{Z}}\left[e^{2\pi i\, \bK}\Big(\prod_{j=1}^{N}\prod_{a=1}^{m_j'} Z'_{j}(\t'_{j,a})\Big) \Big(\prod_{j=1}^{N}\prod_{a=1}^{m_j} Z_{j}(\t_{j,a})\Big) \right]\ov \Tr_{\pi_{Z}}\left[e^{2\pi i \,\bK}\right]}\,.
\ee
The selection rules   \eqref{selrulesa}  just follow from the requirement that the product of all $Z$ and $Z'$ does not depend on the zero mode operators $Q_\mu$. A careful derivation of  \eqref{selrulesa} would employ an explicit ultraviolet regularization of the free fields similar to the one used in \cite{Lukyanov}. This is done in appendix \ref{selrule} where   \eqref{selrulesa} are derived. 

It is thus clear that the functions \eqref{Ffunc} are combinations of multiple integrals with integrands of the form
\be\la{Rfunc}
 R^{\nu_1\ldots\nu_p}_{\mu_1\ldots \mu_q}(\a_1',\ldots,\a_p'|\b_1,\ldots,\b_q)=\langle\langle V'_{\nu_p}(\a_{p}')\cdots V'_{\nu_1}(\a_{1}')V_{\mu_q}(\b_{q})\cdots V_{\mu_1}(\b_{1}) \rangle\rangle\,,
\ee
where the sets $\{\a_1',\ldots,\a_p'\}$ and $\{\b_1,\ldots,\b_q\}$ contain $\t'_j\,$- and $\t_j\,$-related rapidities respectively, and for any operator $W$ acting in $\pi_Z$ we define
\be\la{dbrac}
\langle\langle W \rangle\rangle\equiv{\Tr_{\pi_{Z}}\left[\,e^{2\pi i \,\bK}\,W\,\right]\ov \Tr_{\pi_{Z}}\left[\,e^{2\pi i \,\bK}\,\right]}\,.\qquad
\ee
It is shown in appendix \ref{traces} that for any operator $W$ which is the product of free field exponents
\be
W = U_n(\t_n)\cdots U_1(\t_1)\,,\quad U_j(\t) =\, :e^{i\p_j(\t)}:\, =\, e^{i\p_j^-(\t)}\, e^{i\p_j^+(\t)}\,,\quad 
\ee
one obtains $\langle\langle W \rangle\rangle$ by applying the Wick theorem
\be
\langle\langle U_n(\t_n)\cdots U_1(\t_1) \rangle\rangle =\prod_{j=1}^nC_{U_j}\prod_{j>k}G_{U_jU_k}(\t_k-\t_j)\,, 
\ee
where
\be
C_{U_j}=\langle\langle U_j(\t_j)\rangle\rangle = \exp\big(-\langle\langle \p^-_j(0)\p^+_j(0)\rangle\rangle  \big)\,,
\ee
\be
G_{U_jU_k}(\t_k-\t_j) = \exp\big(-\langle\langle \p_j(\t_j)\p_k(\t_k)\rangle\rangle  \big)\,.
\ee
The constants $C_{V_\mu}$, $C_{V'_\mu}$, and the functions $G_{\mu\nu}\equiv G_{V_\mu V_\nu}$, $G'_{\mu\nu}\equiv G_{V'_\mu V_\nu}$ and $G''_{\mu\nu}\equiv G_{V'_\mu V'_\nu}$ are computed in appendix \ref{traces}.

The integration contours in \eqref{Ffunc} are chosen in the same way as for the vacuum expectation values 
\be\la{Ffunccon}
\langle 0| \Lambda^{\cV} (\t'_k,\ldots , \t'_1) A_{M_n}(\t_{n})\cdots A_{M_1}(\t_{1})|0 \rangle\,,\qquad
\ee
that is the integration contour $C$ in $\chi^\pm_k$ runs from Re$\,\a=-\infty$ to Re$\,\a=+\infty$ and it lies above all poles of $g_{k\mu}$-functions due to operators to the right of $\chi^\pm_k$ but below all poles due to operators to the left of $\chi^\pm_k$. $G_{k\mu}$-functions however have more poles, and in addition to this rule one also requires that the contour $C$ is in the simply-connected region which contains all the poles of $g_{k\mu}$ but no other poles of $G_{k\mu}$.


\subsection*{Form factors of the current operators}

It is clear from the $SU(2)$ result \cite{Lukyanov} that a linear combination of the operators $\Lambda_{j_1j_2...j_N}^{\P}(\a)$ should generate form factors of the current operators $J^\pm_{i}{}^k$. The $\su(N)$ symmetry obviously tells us that it should be proportional  to a linear combination of $\e^{j_1...j_{N-1}k}\Lambda_{j_1...i...j_{N-1}}^{\P}(\a)$ where one inserts  the index $i$ in the sequence $j_1...j_{N-1}$ at some position $\l$.
According to the discussion in section \ref{localO} there are three relevant types of permutations, and therefore operators to be considered are
\be\la{LPl}
\Lambda^{\P_{\l}}{}^{i}_k(\a)={\G'\,\e_{j_1...j_{N-1}k}\ov (\l-1)!(N-\l)!}A'_{ij_1...j_{\l-1}}(\a+i \u_{N-\l})A'_{j_{\l}...j_{N-1}}(\a-i\u_\l)\,,
\ee
where $\P_\l\equiv(N-\l+1,\dots,N,1,\ldots,N-\l)$ and $\l=1,\ldots, N-1$. Then
\be\la{LPlm1}
\Lambda^{\P_{\l-1}}{}^{i}_k(\a)={\G'\,\e_{j_1...j_{N-1}k}\ov (\l-1)!(N-\l)!}A'_{j_1...j_{\l-1}}(\a+i \u_{N-\l+1})A'_{j_{\l}...j_{N-1}i}(\a-i\u_{\l-1})\,,
\ee
where $\P_{\l-1}\equiv(N-\l+2,\dots,N,1,\ldots,N-\l+1)$ and $\l=2,\ldots, N$. Finally
\be\la{LPlm}
\Lambda^{\P_{\l-1,1}}{}^{i}_k(\a)={\G'\,\e_{j_1...j_{N-1}k}\ov (\l-1)!(N-\l)!}A'_{j_1...j_{\l-1}}(\a+i \u_{N-\l+1})A'_{i}(\a+i\u_{N-2\l-1})A'_{j_{\l}...j_{N-1}}(\a-i\u_{\l})\,,
\ee
where $\P_{\l-1,1}\equiv(N-\l+2,\dots,N,N-\l+1,1,\ldots,N-\l)$
and $\l=2,\ldots, N-1$.
Strictly speaking one should consider only the traceless parts of  the operators. 

We propose that form factors of  the current operators $J^\pm_{k}{}^i$ are generated either by
\be\la{LP1}
\Lambda^{\P_{1}}{}^{i}_k(\a)={\G'\,\e_{j_1...j_{N-1}k}\ov (N-1)!}A'_{i}(\a+i \u_{N-1})A'_{j_{1}...j_{N-1}}(\a-i\u_1)\,,
\ee
or by
\be\la{LPN}
\Lambda^{\P_{N-1}}{}^{i}_k(\a)={\G'\,\e_{j_1...j_{N-1}k}\ov (N-1)!}A'_{j_1...j_{N-1}}(\a+i \u_{1})A'_{i}(\a-i\u_{N-1})\,,
\ee
All operators of these types can be obtained from the relevant highest weight operators
\be\la{LP1hw}
\Lambda^{\P_{1}}{}^{1}_N(\a)={\G'}A'_{1}(\a+i \u_{N-1})A'_{1\ldots N-1}(\a-i\u_1)=D_{N,N-1}{\G'}A'_{1}(\a+i \u_{N-1})\bar A'_{N}(\a-i\u_1)\,,
\ee
and
\be\la{LPNhw}
\Lambda^{\P_{N-1}}{}^{1}_N(\a)={\G'}A'_{1...N-1}(\a+i \u_{1})A'_{1}(\a-i\u_{N-1})=D_{N,N-1}{\G'}\bar A'_{N}(\a+i \u_{1})A'_{1}(\a-i\u_{N-1})\,,
\ee
by acting on them with the lowering $\ljok $ 
operators.

Computing the simplest nontrivial form factors generated by these  operators 
one finds 
\bea\label{FF1a}
\cF^{\P_{1}}(\a|\t_1,\t_2) &\equiv&\langle\langle\Lambda^{\P_{1}}{}^{1}_N(\a)\,\bar A_N(\t_2)A_1(\t_1)\rangle\rangle\\\nonumber
&=&(-1)^{N-1}D_{N,N-1}C_{N,N-1}\langle\langle V'_{0}(\a+i \u_{N-1})V'_{N}(\a-i\u_1)\,V_N(\t_2)V_0(\t_1)\rangle\rangle\\\nonumber
&=&(-1)^{N-1}D_{N,N-1}C_{N,N-1}C_0'C_N'C_0C_NG_{0N}''(-i\pi)G_{0N}(\t_1-\t_2)\\\nonumber
&~&\times\, G_{00}'(\t_1-\a-i \u_{N-1})G_{0N}'(\t_1-\a+i \u_{1})\\\nonumber
&~&\times\, G_{00}'(\t_2-\a+i \u_{1})G_{0N}'(\t_2-\a-i \u_{N-1})\, .
\eea
Taking into account the identities 
(\ref{idG1}-\ref{idG3}) and \eqref{Gp01},  one obtains
\be\label{FF1b}
\cF^{\P_{1}}(\a|\t_1,\t_2) =(-1)^{N}\cN_{\cF_1}{G_{0N}(\t_1-\t_2)\ov 4\sinh{1\ov2}(\t_1-\a-i \u_{N-2})\sinh{1\ov2}(\t_2-\a)}\, ,
\ee
where
\be\label{K1}
\cN_{\cF_1}
\equiv D_{N,N-1}C_{N,N-1}C_0'C_N'C_0C_NG_{0N}''(-i\pi)\, .
\ee
According to \cite{Lukyanov} form factors are  generated by expanding $\cF^{\P_{1}}$ in powers of
$e^{\pm\a}$
\be\la{expanF}
\cF^{\P_{1}}(\a|\t_1,\t_2) =\sum_{s=1}^\infty \, e^{-s\a}\,\F_{s}^{\P_{1}}(\t_1,\t_2) \,,\quad \cF^{\P_{1}}(\a|\t_1,\t_2) =\sum_{s=-\infty}^{-1} \, e^{-s\a}\,\F_{s}^{\P_{1}}(\t_1,\t_2)\,.
\ee
Then the form factors of  the components $J^\pm_{N}{}^1$ of the current operators  are proportional to $\F_{\pm1}^{\P_{1}}(\t_1,\t_2)$, respectively. Explicitly one finds
\be\la{FF1c}
\F_{s}^{\P_{1}}(\t_1,\t_2) = (-1)^{N}\cN_{\cF_1}e^{-s\frac{i \pi (N-2)}{2N}}e^{s\frac{\t_1+\t_2}{2}}\,G_{0N}(\t_1-\t_2)\, ,\quad s=\pm1\, ,
\ee
which up to a constant is the same as in \cite{Smirnov92,Babujian2009}.

On the other hand the operator $\Lambda^{\P_{N-1}}{}^{1}_N(\a)$ leads to
\bea\label{FF2a}
\cF^{\P_{N-1}}(\a|\t_1,\t_2) &\equiv&\langle\langle\Lambda^{\P_{N-1}}{}^{1}_N(\a)\,\bar A_N(\t_2)A_1(\t_1)\rangle\rangle\\\nonumber
&=&\cN_{\cF_1}{G_{0N}(\t_1-\t_2)\ov 4\sinh{1\ov2}(\t_1-\a+i \u_{N-2})\sinh{1\ov2}(\t_2-\a)}\, ,
\eea
and its expansion in powers of
$e^{\pm\a}$ produces
\be\la{FF2c}
\F_{s}^{\P_{N-1}}(\t_1,\t_2) = \cN_{\cF_1}e^{s\frac{i \pi (N-2)}{2N}}e^{s\frac{\t_1+\t_2}{2}}\,G_{0N}(\t_1-\t_2)\, ,\quad s=\pm1\, ,
\ee
which up to a constant agrees with \eqref{FF1c}.
It is thus reasonable to expect that both $\Lambda^{\P_{1}}{}^{1}_N$
and $\Lambda^{\P_{N-1}}{}^{1}_N$ generate form factors of the current operators $J^\pm_{N}{}^1$. 

 Let us also mention that the SU(N) symmetry of the model allows one to express the traces of any operator $\Lambda^{\P_1}{}^{i}_k(\a)$ (or $\Lambda^{\P_{N-1}}{}^{i}_k(\a)$) in terms of those of the highest weight operator $\Lambda^{\P_{1}}{}^{1}_N(\a)$. In particular one gets the following formula for the functions generating the particle-antiparticle form factors  
\be\label{FF1ik}
\langle\langle\Lambda^{\P_{1}}{}^{i}_k(\a)\,\bar A_l(\t_2)A_j(\t_1)\rangle\rangle
=\delta_{ij}\delta_{kl}\cF^{\P_{1}}(\a|\t_1,\t_2) \,.
\ee
The SU(N) symmetry of  the traces of operators $\Lambda^{\cV} (\t'_k,\ldots , \t'_1)$ with the ZF operators follows from the fact that the identities \eqref{chiZ} also hold under the traces of products of these operators. Indeed, concentrating for definiteness on the first identity in \eqref{chiZ}, one can see that it is sufficient to show that under the trace 
\be
\la{chipchichi}
\big[\big[ \chi_{j+1}^+ \,,\,\chi_{j+1} ^-\big], V_{j}(\theta)\big]\simeq V_{j}(\theta)\, .
\ee
The relevant part in the trace  comes from 
\be
\langle\langle e^{2\pi i K}\Big(\prod_{\mu=j}^{j+2}\prod_{k=1}^{n_\mu'} V'_{\mu}(\t'_{\mu,k})\Big) \Big(\prod_{\mu=j}^{j+2}\prod_{k=1}^{n_\mu-\delta_{\mu j}} V_{\mu}(\t_{\mu,k})\Big)V_j(\theta)\rangle\rangle\,,
\ee
where we assume without loss of generality that $V_j(\theta)$ is located to the right of all the other $V_k$'s. Then, replacing it with the l.h.s. of \eqref{chipchichi}, one gets that the following identity should hold
\be
I_++I_-=1\,,
\ee
where
\be
I_\pm = \int_{C_\pm} d\a\, \I_\pm\,,
\ee
with
\bea\nonumber
\I_\pm&=&\pm\r_\chi^2 C_1 C_1' R_\pm\\\nonumber
&\times&\big( G'_{j,j+1}(\t-\a\mp {i\pi\ov N})G_{j,j+1}(\t-\a)-G'_{j,j+1}(\a\pm{i\pi\ov N}-\t)G_{j,j+1}(\a-\t)\big)\\\nonumber
&\times& \prod_{\mu=j}^{j+2}\prod_{k=1}^{n_\mu'} G''_{\mu,j+1}(\a\pm{i\pi\ov N}-\t'_{\mu,k})G'_{\mu,j+1}(\a-\t'_{\mu,k})\\
&\times&\prod_{\mu=j}^{j+2}\prod_{k=1}^{n_\mu-\delta_{\mu j}} G'_{\mu,j+1}(\a\pm{i\pi\ov N}-\t_{\mu,k})G_{\mu,j+1}(\a-\t_{\mu,k})\,,
\eea
and
\be
R_\pm\equiv  \mp 2\pi i\, {\rm Res}\, G'_{j+1,j+1}(\a-\a')|_{\a'=\a\pm {i\pi\ov N}} = {\pi\ov \sin{\pi\ov N}}\,. 
\ee
The constants $C_1$ and $C_1'$ are given by \eqref{cj} and \eqref{cjp}. The integration contour $C_+$ runs above $\t + {i\pi\ov N}$ and below $\t_{\mu,k} - {i\pi\ov N}$, while $C_-$ runs below $\t - {i\pi\ov N}$ and  $\t_{\mu,k} - {i\pi\ov N}$.
It is not difficult to check that all the poles of the integrand $\I_+$ lie below $C_+$, while all the poles in $\I_-$ lie above $C_-$. Moreover, if $n_\mu'$ and $n_\mu$ satisfy the selection rules, then one finds that at large $\a$ 
\be
\I_\pm \to \mp {1\ov 2\pi i\, \a}\,,\quad \pm {\rm Im}(\a) >0\,,\quad \a\to\infty\,,
\ee
and therefore the principal value prescription gives
\be
I_\pm = {1\ov 2}\,,
\ee
as required. This completes the proof of the SU(N) symmetry of the traces of operators $\Lambda^{\cV} (\t'_k,\ldots , \t'_1)$.



\section{Conclusion}

In this paper, the free field representation for the ZF algebra of the chiral SU(N) GN model was developed. This was done by constructing vertex operators of the fundamental particles and bound states. These are written in terms of up to $N-1$ integrals, but can be shown to satisfy the ZF algebra in spite of this. The approach we used should be applicable to any  two dimensional integrable model invariant  under a simple Lie algebra.

In addition we also  constructed a large class of operators, $\Lambda$, 
commuting with the ZF operators in terms of the free field representation. 
The representations of the particles, bound states and operators are then used to construct generating functions of the form factors of local operators through the trace formula.  In particular, we proposed two operators $\Lambda$ which generate the form factors of the current operator. Finding the correct operator representation that will give rise to generating functions that contain form factors of the stress-energy tensor remains an open problem. It would also be interesting to try to establish how the free field approach is related to the off-shell Bethe ansatz approach as appears in \cite{Babujian:2006md} and \cite{Babujian2009}. Obviously, the resulting form factors are the same (as they must be), so there should be some way to understand the correlation between the methods.

In light of the free field  approach that was advocated here, further areas of interest might include finding the free field representation of the SU(N) Principal Chiral Field (PCF) model, which is closely related to the GN model. The main complication is that the model is invariant under the direct sum of two Lie algebras and the ansatz \eqref{Jop} for the lowering symmetry operators used in the paper should be modified: a problem that is not immediately obvious how to resolve. In fact, the only known bosonization for a model invariant under the direct sum of two Lie algebras is for the two-parameter family of integrable models (the SS model), see \cite{Fateev:1996} for details of the model. This bosonization was found by Fateev and Lashkevich in \cite{FL}. It is unclear how to generalise their results to the PCF model. As is the case for the GN model, we would like to be able to identify the form factors of the current operator and the stress-energy tensor. 

In addition, similar methods should allow the free field ZF algebra representation of the \ads superstring sigma model in the light-cone gauge to be developed. In this case, we would want to identify the operators $\Lambda$ corresponding to the target space fields. As mentioned in the introduction, finding form factors for this model is complicated by the fact that their analytic properties are not known. Since the free field bosonization does not require a full understanding of these properties, it is hoped that this approach may be able to shed some light on these form factors and their properties.

\section*{Acknowledgements}
We  thank G. Arutyunov, 
S. Lukyanov, 
T. McLoughlin and  
F. Smirnov 
for useful discussions at various stages of the project.
This work was supported in part by the Science Foundation Ireland under Grant 09/RFP/PHY2142. 


\appendix
\renewcommand{\theequation}{\thesubsection.\arabic{equation}}
\numberwithin{equation}{section} 
\section{Form factors axioms}\la{axioms}

We give here the axioms for form factors as appeared in \cite{Smirnov92}, the first four of  which we present in a slightly generalised form similar to \cite{KM} to cover nonrelativistic models possessing the crossing symmetry invariance. First, we define a form factor for an operator $O$ with {\it in}-states as
in \eqref{FFdef} or, equivalently, \eqref{defF}
\be\la{app:defF}
F_{a_1\ldots a_n}(\a_1,\dots,\a_n) = \langle vac| O(0) \cA^\dagger_{a_n}(\a_n)\cdots  \cA^\dagger_{a_1}(\a_1)|vac \rangle
\ee
from which we construct all matrix elements through the crossing symmetry\footnote{In a nonrelativistic model with the crossing symmetry invariance, e.g. the  \ads superstring, the rapidity variable $\a$ should be chosen so that the energy and momentum of the corresponding particle are meromorphic functions on the rapidity plane, and the crossing symmetry transformation is realised as in any relativistic theory as the shift of $\a$ by $i\pi$: $\a\to \a + i\pi$.}
\bea\begin{aligned}
{}^{b_1\ldots b_m}_{\quad \,out}\langle \b_1,\ldots,&\b_m|O(0)|\a_1,\ldots,\a_n\rangle^{in}_{a_1\ldots a_n}~~~
\\
&=C^{b_1c_1}\cdots C^{b_mc_m} F_{a_1\ldots a_nc_1\ldots c_m}(\a_1,\ldots,\a_n,\b_1+i\pi ,\ldots,\b_m+i\pi)\, ,~~~~~
\end{aligned}\eea
where $C^{ab}$ is the charge conjugation matrix.

Then, these form factors must satisfy the following axioms:
\bee
\item Permutation symmetry (Watson's theorem):
\bea\begin{aligned}
&F_{a_1...a_{j+1}a_j...a_n}
(\a_1,...,\a_{j+1},\a_j,...,\a_n)=\\
&\qquad=S^{c_jc_{j+1}}_{a_ja_{j+1}}(\a_j,\a_{j+1})
F_{a_1...c_jc_{j+1},...a_n}(\a_1,...,\a_j,\a_{j+1},...,\a_n)\, .~~~
\label{axiom3}
\end{aligned}\eea
Here $S^{c_jc_{j+1}}_{a_ja_{j+1}}(\a_j,\a_{j+1})$ is the S-matrix which for relativistic models depends only on the difference $\a_j-\a_{j+1}$.
\item Double-crossing or quasi-periodicity condition:
\be
F_{a_1...a_n}(\a_1,...,\a_{n-1},\a_n+2\pi i)=
e^{2\pi i \Omega (O,\mit a_n)} F_{a_n a_1...a_{n-1}}(\a_n,\a_1,...,\a_{n-1})\, .
\ee
The quantity $\Omega(O,\mit a_n)$ appears if the $n$-th particle $\cA^\dagger_{a_n}$ has nontrivial statistics with respect
to operator $O(x)$. 

\item Simple poles: 
The form factors have simple poles at the points $\a_j=\a_i+ i \pi$. Due to the property \eqref{axiom3} it is sufficient to consider only $j=n$, and $i=n-1$. Then the form factors must have the expansion
\bea\nonumber
&&\hspace{-0.3cm} i\, F_{a_1...a_n}(\a_1,...,\a_{n-1},\a_n)=
C_{a_na'_{n-1}}{F_{a'_1...a'_{n-2}}(\a_1,...,\a_{n-2})\ov \a_n-\a_{n-1}-\pi i}\Big( \delta_{a_1}^{a_1'}\cdots \delta_{a_{n-1}}^{a_{n-1}'}~~\\\nonumber
&& \hspace{-0.3cm} - e^{2\pi i \Omega (O,\mit a_{n-1})} S^{a_{n-1}'a_{1}'}_{c_{1}a_{1}}(\a_{n-1},\a_{1})\cdots S^{c_{n-4}a_{n-3}'}_{c_{n-3}a_{n-3}}(\a_{n-1},\a_{n-3})S^{c_{n-3}a_{n-2}'}_{a_{n-1}a_{n-2}}(\a_{n-1},\a_{n-2})\Big)\\
&&+\,\cO(1)
\eea
at $\a_n\to \a_{n-1}+i \pi$.

\item Bound state poles: 
Let particles $\cA^\dagger_K$ with $K\in \cK$ be bound states of particles $\cA^\dagger_I$ and $\cA^\dagger_J$ with 
$I\in \I$ and $J\in \cJ$. The rapidities $\a_I$ and $\a_J$ of $\cA^\dagger_I$ and $\cA^\dagger_J$ are known functions $f_{IJ}^K$ and $f_{JI}^K$ of the rapidity $\a_K$ of the bound states $\cA^\dagger_K$, and the scattering matrix $S_{IJ}^{ab}(\a_I,\a_J^\eps)$ of  $\cA^\dagger_I$ and $\cA^\dagger_J$ with $\a_I=f_{IJ}^K(\a_K)$  and $\a_J^\eps=f_{JI}^K(\a_K+\eps)$  has a pole at $\eps=0$. 
Then the form factors with $\cA^\dagger_I$ and $\cA^\dagger_J$
as external particles are related to those with $\cA^\dagger_K$
as external particles through the small $\eps$ expansion 
\be\la{formbs}
F_{JIa_3...a_n}(\a_J^\eps,\a_I,\a_3,...,\a_n)=
 {i\ov \eps}\sum_{K\in\cK}\Gamma^K_{IJ}F_{Ka_3...a_n}(\a_K,\a_3,...,\a_n)+\cO(1)\,,
\ee 
where $\Gamma^K_{IJ}$ are some constants determined by the consistency of \eqref{formbs} with itself and  the previous form factor axioms. The relations \eqref{formbs} can be inverted and used to express the form factors of bound states through the form factors of the elementary particles.  
 
 For a relativistic theory $\a_I= \a_K + i \u_+$,  $\a_J= \a_K - i \u_-$ ($\u_\pm$ depend on the indices $I,J,K$), the scattering matrix $S_{IJ}^{ab}(\a)$  of $\cA^\dagger_I$ and $\cA^\dagger_J$ has a pole at $\a=i\u_{IJ}^K$, and $\u_\pm$ are found from the equations
\be
\u_++\u_-= \u_{IJ}^K\,,\quad m_I \sin \u_+ = m_J \sin \u_-\,,
\ee
where $m_I$ and $m_J$ are masses of $\cA^\dagger_I$ and $\cA^\dagger_J$, and the mass of the bound state $\cA^\dagger_K$ is equal to $m_K = m_I\cos\u_++m_J\cos\u_-$. 

\bigskip

The last two axioms are valid only for relativistic models and to stress this we use the letter $\t$ for the rapidity variable.

\item Due to relativistic invariance, form factors should satisfy the equation
\be
F_{a_1...a_n}(\theta_1+\zeta,\theta_2+\zeta,...,\theta_n+\zeta)=
\exp(\zeta s(O))F_{a_1...a_n}(\theta_1,...,\theta_n)\, ,
\ee
where\ $s(O)$\ is the spin of the local operator $O(x)$.

\item Form factors $F_{a_1...a_n}(\theta_1,...,\theta_n)$ must be analytic in each variable $\theta_i-\theta_j$ in the strip $0\leq {\rm Im}  \, \theta \leq 2 \pi$ except for simple poles.

\eee

\section{Various functions}\la{appf}
\subsection*{$f_{AB}$-functions}
Since $f_{AB}$,  $f_{AB}''$ and  $f_{AB}'$ satisfy the relations ${\rm f}_{AB}(t)={\rm f}_{BA}(t)={\rm f}_{BA}(-t)$ where ${\rm f}_{AB}$ is any of the three functions and the indices $A,B$ are either $\mu,\nu=0,\ldots,N$ or $(r),(s)=(1)\ldots,(N-1)$, 
we list only nonvanishing functions $f_{\mu\nu}$ with $\mu\le\nu$, $f_{(r)\mu}$ and  $f_{(r)(s)}$  with $r\le s$  and $t>0$
\be
f_{00}(t)=f_{NN}(t)={\sh{(N-1)\pi t\ov N}\ov \sh{\pi t}}e^{\pi t \ov N}\,,\quad f_{jj}(t)=1+e^{\frac{2 \pi  t}{N}}\,,\quad j=1,\ldots, N-1\,,
\ee
\be
f_{j,j+1}(t)=- e^{\pi t \ov N}\,,\quad j=0,1,\ldots, N-1\,,\quad f_{0N}(t)={\sh{\pi t\ov N}\ov \sh{\pi t}}e^{\pi t\ov N}\,.
\ee
\be
f_{(r)0}(t)=\frac{\sinh\frac{\pi  t (N-r)}{N}}{\sinh \pi  t}e^{\frac{\pi  t}{N}} \,,\quad f_{(r)j}(t)=-\delta_{rj}e^{\frac{ \pi  t}{N}}\,,\quad j=1,\ldots, N-1\,,
\ee
\be
f_{(r)(s)}(t)={\sinh {\pi t\, r\ov N}\sinh {\pi t (N-s)\ov N}\ov \sinh\pi t \sinh{\pi t\ov N}}e^{{\pi t\ov N}}\,,
\ee
\be
f_{ab}''(t)=e^{-{2\pi t \ov N}}f_{ab}(t)\,,\quad f_{ab}'(t)=-e^{-{\pi t \ov N}}f_{ab}(t)\,.
\ee

\subsection*{$g_{\mu\nu}$- and $S_{\mu\nu}$-functions}\la{appg}
Since $f_{\mu\nu}(t)=f_{\mu\nu}(t)$ the functions $g_{\mu\nu}$ and $S_{\mu\nu}$ satisfy the same relations $g_{\mu\nu}(\t)=g_{\nu\mu}(\t)$, $S_{\mu\nu}(\t)=S_{\nu\mu}(\t)$, and we again list only nontrivial (not equal to 1) functions 
\be
g_{00}(\t)=g_{NN}(\t)=e^{\frac{(N-1) (\gamma +\log (2 \pi ))}{N}}\frac{ \Gamma
   \left(\frac{i \t}{2 \pi }-\frac{1}{N}+1\right)}{\Gamma \left(\frac{i
   \t}{2 \pi }\right)}\,,
\ee
\be
g_{jj}(\t)=-e^{2 \gamma } \theta  \left(\theta +\frac{2 i \pi }{N}\right)\,,\quad  j=1,\ldots, N-1\,,
\ee
\be
g_{j,j+1}(\t)=-\frac{i e^{-\gamma }}{\theta +\frac{i \pi }{N}}\,,\quad j=0,1,\ldots, N-1\,,\quad g_{0N}(\t)=\frac{e^{\frac{\gamma +\log (2 \pi )}{N}} \Gamma \left(\frac{i \theta }{2 \pi
   }+\frac{1}{2}\right)}{\Gamma \left(\frac{i \theta }{2 \pi }-\frac{1}{N}+\frac{1}{2}\right)}\,,
\ee

\be
S_{00}(\t)=S_{NN}(\t)= S(\t)=\frac{\Gamma \left( \frac{i \theta}{2\pi} \right)       
\Gamma \left(\frac{N-1}{N}- \frac{i \theta}{2\pi} \right)}       
{\Gamma \left(-\frac{i \theta}{2\pi} \right)       
\Gamma \left(\frac{N-1}{N}+ \frac{i \theta}{2\pi} \right)}\,,
\ee
\be
 S_{jj}(\t)=\frac{\theta -\frac{2 i \pi }{N}}{\theta +\frac{2 i \pi }{N}}\,,\quad j=1,\ldots, N-1\,,
\ee
\be
 S_{j,j+1}(\t)=\frac{\theta +\frac{i \pi }{N}}{-\theta +\frac{i \pi }{N}}\,,\quad j=0,1,\ldots, N-1\,,\quad S_{0N}(\t)=\frac{\Gamma \left(-\frac{i \theta }{2 \pi }+\frac{1}{2}\right) \Gamma \left(\frac{i
   \theta }{2 \pi }-\frac{1}{N}+\frac{1}{2}\right)}{\Gamma \left(\frac{i \theta}{2 \pi }+\frac{1}{2}\right) \Gamma \left(-\frac{i \theta }{2 \pi
   }-\frac{1}{N}+\frac{1}{2}\right)} \,,
\ee

The same is true for $g_{\mu\nu}''$
\be
g_{00}''(\t)=g_{NN}''(\t)=\frac{e^{\frac{(N-1) (\gamma +\log (2 \pi ))}{N}} \Gamma \left(\frac{i \theta }{2
   \pi }+1\right)}{\Gamma \left(\frac{i \theta }{2 \pi }+\frac{1}{N}\right)}\,,
\ee
\be
g_{jj}''(\t)=-e^{2 \gamma } \theta  \left(\theta -\frac{2 i \pi }{N}\right)\,,\quad  j=1,\ldots, N-1\,,
\ee
\be
g_{j,j+1}''(\t)=-\frac{i e^{-\gamma }}{\theta -\frac{i \pi }{N}}\,,\quad j=0,1,\ldots, N-1\,,\quad g_{0N}''(\t)=\frac{e^{\frac{\gamma +\log (2 \pi )}{N}} \Gamma \left(\frac{i \theta }{2 \pi
   }+\frac{1}{N}+\frac{1}{2}\right)}{\Gamma \left(\frac{i \theta }{2 \pi
   }+\frac{1}{2}\right)}\,,
\ee
and for $g_{\mu\nu}'$
\be
g_{00}'(\t)=g_{NN}'(\t)=\frac{e^{\gamma  \left(\frac{1}{N}-1\right)} (2 \pi )^{\frac{1}{N}-1} \Gamma
   \left(\frac{1}{2} \left(\frac{i \theta }{\pi
   }+\frac{1}{N}\right)\right)}{\Gamma \left(\frac{i \theta }{2 \pi }-\frac{1}{2
   N}+1\right)}\,,
\ee
\be
g_{jj}'(\t)=-\frac{e^{-2 \gamma }}{\left(\theta -\frac{i \pi }{N}\right) \left(\theta +\frac{i
   \pi }{N}\right)}\,,\quad  j=1,\ldots, N-1\,,
\ee
\be
g_{j,j+1}'(\t)=i e^{\gamma } \theta\,,\quad j=0,1,\ldots, N-1\,,\quad g_{0N}'(\t)=\frac{e^{-\frac{\gamma }{N}} (2 \pi )^{-1/N} \Gamma \left(\frac{1}{2}
   \left(\frac{i \theta }{\pi }-\frac{1}{N}+1\right)\right)}{\Gamma
   \left(\frac{1}{2} \left(\frac{i \theta }{\pi }+\frac{1}{N}+1\right)\right)}\,,
\ee

\subsection*{Regularized integrals}\la{rint}
Let us introduce the following functions
\bea\nonumber
F_2(z,a)&=&\frac{z^2}{2}-\frac{3
   z}{2}-\gamma\left(\frac{z^2}{2}-z\right) +(z-1) \text{log$\Gamma $}(2-z)+\psi ^{(-2)}(2-z)~~~~~~~~~~~\\
   &-&\left(\frac{z^2}{2}-z\right) \log a\,,
\\\nonumber
F_1(z,a) &=& F_2(z+1,a)-F_2(z,a)~~~~~~~~~~~~~~~~~~\\
&=&
-\gamma ( z-\frac{1}{2})+\text{log$\Gamma $}(1-z) -
z \log a+\frac{1}{2} \log \left(\frac{a}{2 \pi
   }\right)\,,~~~~~~~~
\eea
where $\psi ^{(-2)}\left(z\right)$ is given by
\be
\psi ^{(-2)}\left(z\right) = \int_0^z\, dt\,  \text{log$\Gamma
   $}\left(t\right)\,.
\ee
The following integrals are useful in computing  $G_{\mu\nu}$ functions and can be expressed in terms of  $F_i(z,a)$
\be
\int_0^\infty\, {dt\ov t}\,{e^{w a t}-e^{z a t}\ov (e^{a t}-1)^2} = F_2(w,a)-F_2(z,a)\,,
\ee
\be
\int_0^\infty\, {dt\ov t}\,{e^{z a t}\ov e^{a t}-1} =F_1(z,a)\,.
\ee
\be
\int_0^\infty\, {dt\ov t}\,{e^{z t}} =F_1(z+1,1)-F_1(z,1) = -\g -\log(-z)\,.
\ee

\section{Bound state vertex operators} \la{appbound}

In this appendix we derive the expressions for the highest weight bound state vertex operators $Z_{12\ldots r}$ and the normalization constants  $\cN_r$ and $C_{N,r}$.
For convenience we will perform the computations for the fused vertex operators $\cZ_{k_1\ldots k_r}$.
 
\subsection*{Rank-2 fused vertex operators $\cZ_{1b}$}

Consider first $\cZ_{ab}\big(\t\big)\equiv i\e\, Z_{a}\big(\t_1\big)Z_{b}\big(\t_2\big)$, $a<b$,
in the limit $\eps\to 0$ where $\t_1 = \t +{i\pi\ov N}$ and $\t_2 =\t -{i\pi\ov N}+\eps= \t_1 -{2i\pi\ov N}+\eps$ and $\t$ is arbitrary. We have
\bea\nonumber
&&\cZ_{ab}\big(\t\big)=i \e\, Z_{a}\big(\t_1\big)Z_{b}\big(\t_2\big) =i \e \r^2\int_{C_{a-1}^{\a}} d\a_{a-1}\cdots \int_{C_1^{\a}} d\a_1\int_{C_{b-1}^{\b}} d\b_{b-1}\cdots \int_{C_1^{\b}} d\b_1\\
\nonumber
&&\prod_{m=1}^{a-1}g_{m-1,m}^{s}(\a_{m,m-1})\prod_{n=1}^{b-1}g_{n-1,n}^{s}(\b_{n,n-1})\prod_{j=0}^{a-1} g_{jj}(\b_{j}-\a_j) g_{j,j-1}(\b_{j-1}-\a_j)g_{j,j+1}(\b_{j+1}-\a_j)\\\la{zzab}
&&\times
:\prod_{m=0}^{a-1}V_m(\a_m)\prod_{n=0}^{b-1}V_n(\b_n):\,,
\eea
where $\a_0=\t_1\,,\ \b_0=\t_2$ and $\a_{ij}=\a_i-\a_j\,,\ \b_{ij}=\b_i-\b_j$. Since $g_{j,j}(-2\pi i/N)=0$ for $j>1$ and we have a factor of $\e$, naively \eqref{zzab} vanishes in the limit $\eps\to 0$. The only way this would not happen is if in the limit $\eps\to 0$ two poles of $g_{k,k\pm1}$-functions pinch one of the integration contours in \eqref{zzab}. Let us recall that the integration contour $C_m^{\a}$ runs above the pole of $g_{m-1,m}^{s}(\a_{m,m-1})$ at $\a_m=\a_{m-1}+{i\pi\ov N}$, and below the pole of $g_{m-1,m}^{s}(\a_{m,m-1})$ at $\a_m=\a_{m-1}-{i\pi\ov N}$ and above the pole of $g_{m,m-1}(\b_{m-1}-\a_{m})$ at $\a_m=\b_{m-1}+{i\pi\ov N}$. Then the integration contour $C_m^{\b}$ runs above the pole of $g_{m-1,m}^{s}(\b_{m,m-1})$ at $\b_m=\b_{m-1}+{i\pi\ov N}$, below the pole of $g_{m-1,m}^{s}(\b_{m,m-1})$ at $\b_m=\b_{m-1}-{i\pi\ov N}$, and below the pole of $g_{m-1,m}(\b_{m}-\a_{m-1})$ at $\b_m=\a_{m-1}-{i\pi\ov N}$. Thus for the contour
$C_1^{\a}$ one gets the poles
\bea\nonumber
{\rm below}\  C_1^{\a}:&& \a_1=\t_1+{i\pi\ov N}\,,\quad {\rm Res\,}g_{0,1}^{s}(\a_{1,0})={e^\g\ov 2\pi}{\rm Res\,}g_{0,1}(\a_{0,1})= {i\ov 2\pi}\\
&& \a_1=\t_2+{i\pi\ov N}= \t_1-{i\pi\ov N}+\eps\,,\quad {\rm Res\,}g_{0,1}(\b_{0}-\a_{1})=i e^{-\g}\,,
\\\nonumber
{\rm above}\  C_1^{\a}:&& \a_1=\t_1-{i\pi\ov N}\,,\quad {\rm Res\,}g_{0,1}^{s}(\a_{1,0})={e^\g\ov 2\pi}{\rm Res\,}g_{0,1}(\a_{1,0})= {1\ov 2\pi i}\,,\quad
\eea
and one sees that the poles at $ \t_1-{i\pi\ov N}$ and $ \t_1-{i\pi\ov N}+\eps$ do pinch the integration contour $C_1^{\a}$ in the limit $\eps\to 0$.
Then for the contour $C_1^{\b}$ one gets the poles at 
\bea\nonumber
{\rm below}\  C_1^{\b}:&& \b_1=\t_2+{i\pi\ov N}=\t_1-{i \pi \ov N}+\eps\,,\quad {\rm Res\,}g_{0,1}^{s}(\b_{1,0})={e^\g\ov 2\pi}{\rm Res\,}g_{0,1}(\b_{0,1})= {i\ov 2\pi}\\\nonumber
{\rm above}\  C_1^{\b}:&& \b_1=\t_2-{i\pi\ov N}\,,\quad {\rm Res\,}g_{0,1}^{s}(\b_{1,0})={e^\g\ov 2\pi}{\rm Res\,}g_{0,1}(\b_{1}-\t_{2})= {1\ov 2\pi i}\\
&& \b_1=\t_1-{i\pi\ov N}\,,\quad {\rm Res\,}g_{0,1}(\b_{1}-\t_{1})= -i e^{-\g}\, ,\eea
and the poles at $ \t_1-{i\pi\ov N}$ and $ \t_1-{i\pi\ov N}+\eps$ again pinch the integration contour $C_1^{\b}$.
Since one of the poles is below and the other is above the integration contour only the contribution from one of them should be taken into account. We will always choose the one which is the pole of $g_{m-1,m}^{s}$ to get rid of one of these functions in \eqref{zzab}. If $a>1,\, b>1$ then two integration contours are pinched at the same time and one has to sum the contributions coming from each of the contours. 

Let us now consider $a=1$ and compute the integral over $\b_1$. If $a=1$ then there are no integrals over $\a_j$ and we get
\bea\nonumber
&&\cZ_{1b}(\t)=i\e\, Z_{1}\big(\t_1\big)Z_{b}\big(\t_2\big) =i\e\r^2\int_{C_{b-1}^{\b}} d\b_{b-1}\cdots \int_{C_1^{\b}} d\b_1
g_{00}(\t_{21})g_{01}(\b_{1}-\t_1)g_{01}^{s}(\b_{1,0}) 
\\
\nonumber
&&\qquad\qquad\qquad\qquad\qquad\times \prod_{n=2}^{b-1}g_{n-1,n}^{s}(\b_{n,n-1}) \, :V_0(\t_1)\prod_{n=0}^{b-1}V_n(\b_n): 
\\\la{zz1b2}
&&\qquad=\r^2C_{N,2}^{1}\int_{C_{b-1}^{\b}} d\b_{b-1}\cdots \int_{C_2^{\b}} d\b_2
\prod_{n=2}^{b-1}g_{n-1,n}^{s}(\b_{n,n-1}) 
\,  :V_0(\t_1)\prod_{n=0}^{b-1}V_n(\b_n):\,,
\eea
where we take into account that 
$i\e\, g_{0,0}(\t_{21})g_{0,1}(\t_{21}+{i\pi\ov N})=(2\pi)^{{N-1\ov N}}e^{-{\gamma\ov N} }/\Gamma \left(\frac{1}{N}\right)$ and $\b_1=\t_2+{i\pi\ov N}= \t_1-{i\pi\ov N}=\t$ in \eqref{zz1b2}
 in the limit $\eps\to 0$, and  introduce the constant
\be
C_{N,2}^{1}\equiv (2\pi)^{N-1\ov N}e^{-{\gamma\ov N}} {1\ov \Gamma \left(\frac{1}{N}\right)} = \lim_{\eps\to 0}\int_C\, d\b_1 \,i\e\, g_{0,0}(\t_{21})g_{0,1}(\b_{1}-\t_1)g_{0,1}^{s}(\b_{1,0}) \,,
\ee
with the integration contour specified above. 

In particular for the highest weight fused operator one gets
\be\la{v12}
\cZ_{12}(\t)=C_{N,2} :V_0(\t+{i\pi\ov N})V_0(\t-{i\pi\ov N})V_1(\t):=C_{N,2}V_{(2)}(\t)\,.
\ee
where  $C_{N,2}=\r^2C_{N,2}^{1} = (-1)^{2\ov N}\frac{e^{\gamma\frac{N-2}{N}} N^{-{1\ov N}} (2 \pi )^{\frac{N-1}{N}}}{\Gamma \left(\frac{1}{N}\right)}$.
 
\subsection*{Rank-3 fused vertex operators $\cZ_{12c}$}

 Let us now consider $\cZ_{12c}\big(\t\big)=i\e_1\,i\e\,Z_{1}\big(\t_1\big)Z_{2}\big(\t_2\big)Z_{c}\big(\t_3\big)$ for $2<c$ with $\t_1=\t +{2i\pi\ov N}$, $\t_2 = \t+\eps_1$, $\t_3=\t-{2i\pi\ov N}+\eps$, and take the limit $\e_1\to 0$ after using \eqref{v12} for $i\e_1Z_{1}\big(\t_1\big)Z_{b}\big(\t_2\big)$
 \bea\nonumber
\cZ_{12c}\big(\t\big)&&=i\e\, \cZ_{12}\big(\b_1\big)Z_{c}\big(\t_3\big)=
\r^3C_{N,2}^{1}\int_{C_{c-1}^{\g}} d\g_{c-1}\cdots \int_{C_1^{\g}} d\g_1\,i\e\,\prod_{r=1}^{c-1}g_{r-1,r}^{s}(\g_{r,r-1})
\\
\nonumber
&&\times\, g_{00}(\t_{31})g_{00}(\t_{32})g_{01}(\g_{1}-\t_1)g_{01}(\g_{1}-\t_2)g_{01}(\t_3-\b_1)g_{11}(\g_{1}-\b_1)g_{01}(\g_{2}-\b_1)
\\
\la{zabc}
&&\times\, 
:V_0(\t_1)V_0(\t_2)V_1(\b_1)\prod_{r=0}^{c-1}V_r(\g_r):\,,
\eea
where $\b_1\equiv \t_2+{i\pi\ov N}=\t+{i\pi\ov N}$.
 We want to integrate over $\g_1$ and the relevant terms in the integrand are 
 \bea\nonumber
 &&i\e\, g_{01}^{s}(\g_{01}) g_{00}(\t_{31})g_{00}(\t_{32})g_{01}(\g_{1}-\t_1)g_{01}(\g_{1}-\t_2)g_{01}(\t_3-\b_1)g_{11}(\g_{1}-\b_1)\\\nonumber
  &&= g_{00}(-{4i\pi\ov N})g_{00}(-{2i\pi\ov N})g_{01}(-{3i\pi\ov N})g_{01}^{s}(\g_{01}) g_{01}(\g_{1}-\t_1)g_{01}(\g_{1}-\t_2)g_{11}(\g_{1}-\b_1)\,.
 \eea
The integral over $\g_1$ is taken in the same way as the one over $\b_1$ in the previous subsection, and we get that the expression above becomes
\be
C_{N,3}^{1}g_{1,1}(-{2i\pi\ov N}+\eps)
\ee
where we introduce the constant
\bea
C_{N,3}^{1}&\equiv& C_{N,2}^{1}g_{0,0}(-{4i\pi\ov N})g_{1,0}(-{3i\pi\ov N})^2 \\\nonumber
&=& \lim_{\eps\to 0}\int_C\, d\g_1\,i\e\, g_{0,0}(-{4i\pi\ov N})g_{1,0}(-{3i\pi\ov N})g_{0,0}(-{2i\pi\ov N})g_{0,1}(\g_{1}-\t_1)g_{0,1}^{s}(\g_{0,1}) g_{0,1}(\g_{1}-\t_2)
\eea
and use that the pole is located at $\g_1 = \t_3+{i\pi\ov N}=\t_2-{i\pi\ov N}+\eps$. Since $g_{1,1}(-{2i\pi\ov N})=0$ we have to take the integral over $\g_2$ to get a finite result in the limit $\eps\to 0$.
 The relevant terms in the integrand now  are 
 \be
g_{1,1}(-{2i\pi\ov N}+\eps)g_{1,2}^{s}(\g_{1,2})g_{1,2}(\g_{2}-\b_1)
 \ee
 and the poles are at $\g_2=\g_1\pm{i\pi\ov N}=\t_2-{i\pi\ov N}+\eps\pm{i\pi\ov N}$ and 
 $\g_2=\b_1-{i\pi\ov N}=\t_2$. Thus the contour is pinched at $\g_2=\t_2$, and we get the extra factor 
 \be
C_{N,3}^{2}\equiv \frac{2 \pi e^\g }{N}= \lim_{\eps\to 0}\int_C\, d\g_2\, g_{1,1}(-{2i\pi\ov N}+\eps)g_{1,2}^{s}(\g_{1,2})g_{1,2}(\g_{2}-\b_1)\,,
 \ee
 where one uses that
 $g_{11}(-{2i\pi\ov N}+\eps)g_{12}(\eps-{i\pi\ov N})=\frac{2 e^{\gamma } \pi }{N}$. Thus one gets
 \bea\nonumber
\cZ_{12c}\big(\t\big)&&=\r^3
C_{N,2}^{1}C_{N,3}^{1}C_{N,3}^{2}\int_{C_{c-1}^{\g}} d\g_{c-1}\cdots \int_{C_3^{\g}} d\g_3\prod_{r=3}^{c-1}g_{r-1,r}^{s}(\g_{r,r-1})
\\
\la{z12c}
&&\times\, 
:V_0(\t_1)V_0(\t_2)V_0(\t_3)V_1(\b_1)\prod_{r=1}^{c-1}V_r(\g_r):\,,
\eea
where  $\g_1=\t_3+{i\pi\ov N}=\t-{i\pi\ov N}$, $\g_2=\t_2=\t$.
 In particular for the highest weight fused operator one gets
 \be\la{zz123}
 \cZ_{123}(\t) 
=C_{N,3}:V_0(\t+{2i\pi\ov N})V_0(\t)V_0(\t-{2i\pi\ov N})V_1(\t+{i\pi\ov N})V_1(\t-{i\pi\ov N})V_2(\t):=C_{N,3}V_{(3)}(\t)\,,~~~~
\ee
where $C_{N,3}=\r^3C_{N,2}^{1}C_{N,3}^{1}C_{N,3}^{2}=(-1)^{3\ov N}e^{\frac{3 \gamma (N-3)}{2N}}N^{-\frac{3}{2N}} (2 \pi )^{2-\frac{3}{N}} {1\ov\Gamma
   \left(\frac{2}{N}\right) \Gamma \left(\frac{1}{N}\right)}$.

\subsection*{Rank-$k$ fused highest weight vertex operator $\cZ_{1\ldots k}$}
 
 This pattern continues for the rank-$k$ fused highest weight vertex operator $ \cZ_{1\ldots k}$ which contains no integrals at all. So, we consider 
\bea
 \cZ_{12\ldots k}(\t)&\equiv&\lim_{\e_a\to 0}\, i\e_1\cdots i\e_{k-1}\,Z_{1}\big(\t_1\big)Z_{2}\big(\t_2\big)\cdots Z_{k}\big(\t_k\big)\,,\\  \nonumber
\t_j&\equiv& \t +{i\pi\ov N}(k-2j+1)+\e_j\,, \la{z12k0}
\eea
and we should find that it is equal to 
\be\la{z12k}
 \cZ_{12\ldots k}(\t)=C_{N,k}:\prod_{j=1}^{k}\prod_{a_j=0}^{j-1}V_{a_j}\big(\t_j + {i\pi\ov N}a_j\big): \, =C_{N,k}:\prod_{r=0}^{k-1}\prod_{j=r+1}^{k}V_{r}\big(\t_j + {i\pi\ov N}r\big):\,.
\ee
Note that the $V_{k-1}$ operator only enters as $V_{k-1}(\t)$. 

\bigskip

To find $C_{N,k}$ we use induction. Introducing the notation
\be
\t_j^{[\pm r]}\equiv \t_j \pm {i\pi\ov N}r\,,
\ee
We have ($\a_0=\t_{k+1}=\t_k-{2i\pi\ov N}$)
\bea\la{z12kp1}
&& \cZ_{12\ldots k,k+1}(\t)=i\e\, \r C_{N,k}
\int_{C_{k}^{\a}} d\a_{k}\cdots \int_{C_1^{\a}} d\a_1\prod_{m=1}^{k}g_{m-1,m}^{s}(\a_{m,m-1})
\\
\nonumber
&&\quad\times
\prod_{r=0}^{k-1}\prod_{j=r+1}^{k}g_{r,r-1}(\a_{r-1}-\t_j^{[+r]})g_{r,r}(\a_{r}-\t_j^{[+r]})g_{r,r+1}(\a_{r+1}-\t_j^{[+r]})
\\
\nonumber
&&\quad\times
:\prod_{r=0}^{k-1}\prod_{j=r+1}^{k}V_{r}\big(\t_j^{[+r]}\big)\prod_{m=0}^{k}V_m(\a_m):\,.
\eea
To integrate over $\a_j$ it is better to rearrange
\bea
&&\prod_{r=0}^{k-1}\prod_{j=r+1}^{k}g_{r,r-1}(\a_{r-1}-\t_j^{[+r]})g_{r,r}(\a_{r}-\t_j^{[+r]})g_{r,r+1}(\a_{r+1}-\t_j^{[+r]})
\\
\nonumber
&&=
\prod_{j=1}^{k-1}g_{00}\big(-{2i\pi\ov N}(k-j+1)\big)
\prod_{j=2}^{k}g_{10}\big(-{2i\pi\ov N}(k-j+{3\ov2})\big)g_{00}\big(-{2i\pi\ov N}\big)
\\
\nonumber
&&\times\prod_{m=1}^{k}\prod_{j=m}^{k}g_{m-1,m}(\a_{m}-\t_j^{[+(m-1)]})
\prod_{j=m+1}^{k}g_{m,m}(\a_{m}-\t_j^{[m]})
\prod_{j=m+2}^{k}g_{m+1,m}(\a_{m}-\t_j^{[+(m+1)]})
\eea
Thus the integrand for $\a_1$ is
\bea
&&\prod_{j=1}^{k-1}g_{00}\big(-{2i\pi\ov N}(k-j+1)\big)
\prod_{j=2}^{k}g_{10}\big(-{2i\pi\ov N}(k-j+{3\ov2})\big)g_{00}\big(-{2i\pi\ov N}\big)
\\
\nonumber
&&\times g_{01}^{s}(\a_{10})\prod_{j=1}^{k}g_{01}(\a_{1}-\t_j)
\prod_{j=2}^{k}g_{11}(\a_{1}-\t_j^{[+1]})
\prod_{j=3}^{k}g_{21}(\a_{1}-\t_j^{[+2]})
\eea
We see that the two poles which pinch the contour are at
$\a_1=\t_{k+1}+{i\pi\ov N}=\t_{k}-{i\pi\ov N}+\eps=\t_k^{[-1]}+\eps$
and $\a_1=\t_k^{[-1]}$ giving the following contribution
\be
C_{N,k+1}^1\,g_{1,1}\big(-{2i\pi\ov N}+\eps\big)
\ee
where
\bea
&&C_{N,k+1}^1\equiv C_{N,2}^1\, \prod_{j=1}^{k-1}g_{00}\big(-{2i\pi\ov N}(k-j+1)\big)
\prod_{j=2}^{k}g_{01}\big(-{2i\pi\ov N}(k-j+{3\ov2})\big)
\\
\nonumber
&&\times \prod_{j=1}^{k-1}g_{01}(-{2i\pi\ov N}(k-j+{1\ov2}))
\prod_{j=2}^{k-1}g_{11}(-{2i\pi\ov N}(k-j+1))
\prod_{j=3}^{k}g_{01}(-{2i\pi\ov N}(k-j+{3\ov2}))
\\
\nonumber
&& =C_{N,2}^1\,g_{01}\big(-{i\pi\ov N}(2k-1)\big)^2 \prod_{j=2}^{k}g_{00}\big(-{2i\pi\ov N}j\big)
\prod_{j=2}^{k-1}g_{01}\big(-{i\pi\ov N}(2j-1)\big)^3
\prod_{j=2}^{k-1}g_{11}(-{2i\pi\ov N}j)\,.
\eea
Then the integrand for $\a_2$ is
\bea
g_{12}^{s}(\a_{21})\prod_{j=2}^{k}g_{12}(\a_{2}-\t_j^{[+1]})
\prod_{j=3}^{k}g_{22}(\a_{2}-\t_j^{[+2]})
\prod_{j=4}^{k}g_{32}(\a_{2}-\t_j^{[+3]})
\eea
We see that the two poles which pinch the contour are at
$\a_2=\a_{1}+{i\pi\ov N}=\t_{k+1}^{[+2]}=\t_k+\eps$
and $\a_0=\t_k$ giving the following contribution
\be
C_{N,k+1}^2\,g_{2,2}\big(-{2i\pi\ov N}+\eps\big)=C_{N,k+1}^2\,g_{11}\big(-{2i\pi\ov N}+\eps\big)
\ee
where
\bea\nonumber
&&C_{N,k+1}^2\equiv C_{N,3}^2\, \prod_{j=2}^{k-1}g_{12}(-{2i\pi\ov N}(k-j+{1\ov2}))
\prod_{j=3}^{k-1}g_{22}(-{2i\pi\ov N}(k-j+1))
\prod_{j=4}^{k}g_{32}(-{2i\pi\ov N}(k-j+{3\ov2}))
\\
\nonumber
&&\qquad =C_{N,3}^2\,g_{01}\big(-{i\pi\ov N}(2k-3)\big)
\prod_{j=2}^{k-2}g_{01}\big(-{i\pi\ov N}(2j-1)\big)^2
\prod_{j=2}^{k-2}g_{11}(-{2i\pi\ov N}j)\,.
\eea

The integral over $\a_m$ for $m< k-1$ is computed in the same way with the pole at
$\a_m= \a_{m-1}+{i\pi\ov N}=\t_{k+1}^{[+m]}=\t_k^{[+(m-2)]}+\eps$ and gives
\be
C_{N,k+1}^m\,g_{1,1}\big(-{2i\pi\ov N}+\eps\big)
\ee
where
\bea\nonumber
&&C_{N,k+1}^m\equiv C_{N,3}^2\, \prod_{j=m}^{k-1}g_{01}(-{2i\pi\ov N}(k-j+{1\ov2}))
\prod_{j=m+1}^{k-1}g_{11}(-{2i\pi\ov N}(k-j+1))
\prod_{j=m+2}^{k}g_{01}(-{2i\pi\ov N}(k-j+{3\ov2}))
\\
\nonumber
&&\qquad =C_{N,3}^2\,g_{01}\big(-{i\pi\ov N}(2k-2m+1)\big)
\prod_{j=2}^{k-m}g_{01}\big(-{i\pi\ov N}(2j-1)\big)^2
\prod_{j=2}^{k-m}g_{11}(-{2i\pi\ov N}j)\,.
\eea
For $m=k-1$ and $m=k$ we get
\bea\nonumber
C_{N,k+1}^{k-1}=C_{N,3}^2\,g_{0,1}\big(-{3i\pi\ov N}\big)
\,,\quad C_{N,k+1}^{k}=C_{N,3}^2\,.
\eea
Thus, the result is
\be
C_{N,k+1}=\r C_{N,k}\prod_{j=1}^k C_{N,k+1}^j\,.
\ee
Computing the product one finds
\be
\prod_{j=1}^k C_{N,k+1}^j=2\pi\,e^{-{k\g \ov N}}(2\pi)^{-{k\ov N}} {1\ov \Gamma\left(k\ov N\right)}\, .
\ee
The relation  above can be easily solved giving the final result
\bea\nonumber
C_{N,k+1}&=&\r^{k+1}(2\pi)^{k}\left(e^{- {\g\ov N}}(2\pi)^{-{1\ov N}}\right)^{{k(k+1)\ov 2}}\prod_{j=1}^k {1\ov \Gamma\left(j\ov N\right)}\\\la{CNkp1}
&=&(-1)^{\frac{k+1}{N}} e^{\gamma\frac{(k+1) (N-k-1)}{2 N}} N^{-\frac{k+1}{2 N}} (2 \pi )^{\frac{k (2 N-k-1)}{2 N}} \prod
   _{j=1}^{k} \frac{1}{\Gamma \left(\frac{j}{N}\right)}\, .
\eea
By using the identity
\be
\prod_{j=1}^{N-1} {1\ov\Gamma\left(j\ov N\right)} = \sqrt N(2\pi)^{-{N-1\ov 2}}\,,
\ee
one gets  $C_{N,N}=\cV_N=-1$, and \eqref{CNNm1} for $C_{N,N-1}$.

\section{Primed fused operators} \la{appbound2}

In this appendix we derive the expressions for the highest weight fused primed vertex operators $Z'_{12\ldots r}$. The derivation almost repeats the one for the the highest weight bound state vertex operators $ \cZ_{12\ldots r}$ considered in appendix \ref{appbound}. It is therefore unnecessary to repeat the details and only an outline is given.

\subsection*{Rank-2  fused primed vertex operator  $ Z'_{1b}$}

Consider first $Z'_{ab}\big(\t\big)\equiv  Z'_{a}\big(\t_1\big)Z'_{b}\big(\t_2\big)$, $a<b$,
in the limit $\eps\to 0$ where $\t_1 = \t -{i\pi\ov N}$ and $\t_2 =\t +{i\pi\ov N}+\eps= \t_1 +{2i\pi\ov N}+\eps$ and $\t$ is arbitrary. We notice that the rapidities here are shifted in the opposite direction when compared to those in appendix \ref{appbound}. In addition, all Green's functions will be of the type $g''_{\mu \nu}$. Therefore we will see $g''_{00}\left({2 \pi i \ov N} +\e \right)$ instead of $g_{00}\left(-{2 \pi i \ov N} +\e \right)$, etc. The other major difference is that we don't have an initial factor of $i \e$, but instead we have that $g''_{00}\left({2 \pi i \ov N} \right)=0$. Other than these differences, the derivation is essentially identical.

With this in mind, let us now calculate  the  fused primed vertex operator $Z'_{1b}\big(\t\big)$.
Since $a=1$ there are no integrals over $\a_j$, and computing the integral over $\b_1$ one gets
\be\la{zpzp1b2}
 Z'_{1b}\big(\t\big) =\r'^2D_{N,2}^{1}\int_{C_{b-1}^{\b}} d\b_{b-1}\cdots \int_{C_2^{\b}} d\b_2
\prod_{n=2}^{b-1}g_{n-1,n}^{''s}(\b_{n,n-1}) 
 :V_0'(\t_1)\prod_{n=0}^{b-1}V_n'(\b_n):\,,
\ee
where we take into account that 
$g''_{0,0}(\t_{21})g''_{0,1}(\t_{21}-{i\pi\ov N})=e^{-\frac{\gamma }{N}} (2 \pi )^{-1/N} \Gamma \left(\frac{N-1}{N}\right)$ and $\b_1=\t_2-{i\pi\ov N}= \t_1+{i\pi\ov N}$ in \eqref{zpzp1b2}
 in the limit $\eps\to 0$,  
and introduce the constant
\be
D_{N,2}^{1}\equiv (2\pi)^{-{1\ov N}}e^{-{\gamma\ov N}} \Gamma \left(\frac{N-1}{N}\right) = \lim_{\eps\to 0}\int_C\, d\b_1 \, g''_{0,0}(\t_{21})g''_{0,1}(\b_{1}-\t_1)g_{0,1}^{''s}(\b_{1,0}) \,,
\ee
with the integration contour specified in the usual way.

In particular for the highest weight fused primed operator one gets
\be\la{vp12}
 Z'_{12}(\t)=D_{N,2} :V'_0(\t-{i\pi\ov N})V'_0(\t+{i\pi\ov N})V'_1(\t):=D_{N,2}V'_{(2)}(\t)\,.
\ee
where  $D_{N,2}=\r'^2D_{N,2}^{1}= e^{\gamma\frac{N-2}{N}} N^{{1\ov N}} (2 \pi )^{-\frac{1}{N}}\Gamma \left(\frac{N-1}{N}\right)$.

\subsection*{Rank-3 fused primed vertex operators  $Z_{12c}'$}
Let us now consider 
$Z'_{12c}\big(\t\big)=Z'_{1}\big(\t_1\big)Z'_{2}\big(\t_2\big)Z'_{c}\big(\t_3\big)$, $2<c$, $\t_1=\t -{2i\pi\ov N}$, $\t_2 = \t+\eps_1$, $\t_3=\t+{2i\pi\ov N}+\eps$, take the limit $\e_1\to 0$ and use \eqref{vp12} for $Z'_{1}\big(\t_1\big)Z'_{2}\big(\t_2\big)$. The relevant poles are at $ \b_1\equiv \t_2-{i\pi\ov N}=\t-{i\pi\ov N}$, $\g_1 = \t_3-{i\pi\ov N}=\t_2+{i\pi\ov N}+\eps$ and  $\g_2=\b_1+{i\pi\ov N}=\t_2$, which can be used to reduce the expression to
\bea\nonumber
Z'_{12c}\big(\t\big)&&=\r'^3
D_{N,2}^{1}D_{N,3}^{1}D_{N,3}^{2}\int_{C_{c-1}^{\g}} d\g_{c-1}\cdots \int_{C_3^{\g}} d\g_3\,\prod_{r=3}^{c-1}g_{r-1,r}^{''s}(\g_{r,r-1})
\\
\la{zp12c}
&&\times\, 
:V'_0(\t_1)V'_0(\t_2)V'_0(\t_3)V'_1(\b_1)\prod_{r=1}^{c-1}V'_r(\g_r):\,,
\eea
with the new constants defined as
\bea
D_{N,3}^{1}&\equiv& D_{N,2}^{1}g''_{00}({4i\pi\ov N})g''_{10}({3i\pi\ov N})^2 \\\nonumber
&=& \lim_{\eps\to 0}\int_C\, d\g_1\, g''_{00}({4i\pi\ov N})g''_{10}({3i\pi\ov N})g''_{00}({2i\pi\ov N}+\eps)g''_{01}(\g_{1}-\t_1)g_{01}^{''s}(\g_{01}) g''_{01}(\g_{1}-\t_2)\,,
\eea
and
 \be
D_{N,3}^{2}\equiv -\frac{2 \pi e^{\g} }{N}= \lim_{\eps\to 0}\int_C\, d\g_2\, g''_{1,1}({2i\pi\ov N}+\eps)g_{1,2}^{''s}(\g_{1,2})g''_{1,2}(\g_{2}-\b_1)\,.
 \ee
 In particular for the highest weight fused primed operator one gets
 \bea\nonumber
 Z'_{123}(\t) 
&=&D_{N,3}:V'_0(\t-{2i\pi\ov N})V'_0(\t)V'_0(\t+{2i\pi\ov N})V'_1(\t-{i\pi\ov N})V'_1(\t+{i\pi\ov N})V_2'(\t):\\
\la{zzp123}
&=&D_{N,3}:V'_{(3)}(\t)\,,
\eea
where $D_{N,3}=\r'^3D_{N,2}^{1}D_{N,3}^{1}D_{N,3}^{2}=e^{\frac{3 \gamma  (N-3)}{2 N}} N^{\frac{3}{2 N}} (2 \pi )^{-3/N} \Gamma \left(\frac{N-2}{N}\right) \Gamma
   \left(\frac{N-1}{N}\right)$.
   
\subsection*{Rank-$k$ fused highest weight primed vertex operator $Z'_{1\ldots k}$}

 This pattern continues for the rank-$k$ fused highest weight primed vertex operator $Z_{1\ldots k}$ which contains no integrals at all. So, considering 
\bea
Z'_{12\ldots k}(\t)&\equiv&\lim_{\e_a\to 0}\, Z'_{1}\big(\t_1\big)Z'_{2}\big(\t_2\big)\cdots Z'_{k}\big(\t_k\big)\,,\\  \nonumber
\t_j&\equiv& \t -{i\pi\ov N}(k-2j+1)+\e_j\,, \la{zp12k}
\eea
one should find that it is equal to 
\be\la{zp12kb}
Z^{\prime}_{12\ldots k}(\t)=\r'^kD_{N,k}:\prod_{j=1}^{k}\prod_{r_j=0}^{j-1}V_{r_j}'\big(\t_j - {i\pi\ov N}r_j\big): \, =D_{N,k}:\prod_{r=0}^{k-1}\prod_{j=r+1}^{k}V_{r}'\big(\t_j - {i\pi\ov N}r\big):\,.
\ee
To find $D_{N,k}$ one can use induction. The results are the following:
\be
D_{N,k+1}^1\equiv D_{N,2}^1\,g''_{01}\big({i\pi\ov N}(2k-1)\big)^2 \prod_{j=2}^{k}g''_{00}\big({2i\pi\ov N}j\big)
\prod_{j=2}^{k-1}g''_{01}\big({i\pi\ov N}(2j-1)\big)^3
\prod_{j=2}^{k-1}g''_{11}({2i\pi\ov N}j)\,,
\ee
and
\be\nonumber
D_{N,k+1}^2\equiv D_{N,3}^2\,g''_{0,1}\big({i\pi\ov N}(2k-3)\big)
\prod_{j=2}^{k-2}g''_{0,1}\big({i\pi\ov N}(2j-1)\big)^2
\prod_{j=2}^{k-2}g''_{1,1}({2i\pi\ov N}j)\,.
\ee
The general case is
\be\nonumber
D_{N,k+1}^m\equiv D_{N,3}^2\,g''_{0,1}\big({i\pi\ov N}(2k-2m+1)\big)
\prod_{j=2}^{k-m}g''_{0,1}\big({i\pi\ov N}(2j-1)\big)^2
\prod_{j=2}^{k-m}g''_{1,1}({2i\pi\ov N}j)\,.
\ee
For $m=k-1$ and $m=k$ we get
\bea\nonumber
D_{N,k+1}^{k-1}=D_{N,3}^2\,g''_{0,1}\big({3i\pi\ov N}\big)
\,,\quad D_{N,k+1}^{k}=D_{N,3}^2\,.
\eea
Thus, the result is
\be
D_{N,k+1}=\r'D_{N,k}\prod_{j=1}^kD_{N,k+1}^j\,.
\ee
Computing the product one finds
\be
\prod_{j=1}^kD_{N,k+1}^j=e^{-\g {k\ov N}}(2\pi)^{-{k\ov N}} \Gamma\left(N-k\ov N\right)\, .
\ee
The relation  above can be easily solved giving the final result
\bea\nonumber
D_{N,k+1}&=&\r'^{k+1}\left(e^{- {\g\ov N}}(2\pi)^{-{1\ov N}}\right)^{{k(k+1)\ov 2}}\prod_{j=1}^k \Gamma\left(N-j\ov N\right)\\
\la{DNkp1}
&=&e^{-\frac{\gamma  (k+1) (k-N+1)}{2 N}} N^{\frac{k+1}{2 N}} (2 \pi )^{-\frac{k (k+1)}{2 N}} \prod _{m=1}^k \Gamma
   \left(1-\frac{m}{N}\right)\, .
\eea
By using the identity
\be
\prod_{j=1}^{N-1} \Gamma\left(N-j\ov N\right) = {1\ov \sqrt N}(2\pi)^{{N-1\ov 2}}\,,
\ee
one gets $D_{N,N}=1$, and \eqref{DNNm1} for $D_{N,N-1}$.

\section{Regularized free fields and selection rules}\la{selrule}

We assume that $\p_\mu^{(\e)}(\t)$ are defined on the finite interval
\be
-{\pi\ov\e}\le\t\le{\pi \ov\e}\,,
\ee
and satisfy the commutation relations
\be
\left[\phi_\mu^{(\e)}(\t_1), \phi_\nu^{(\e)}(\t_2)\right]=\ln S_{\mu\nu}^{(\e)}(\t_2-\t_1)\, ,
\ee
where $S_{\mu\nu}^{(\e)}(\t)$ goes to  $S_{\mu\nu}(\t)$ as $\e\to 0$ for finite $\t$ and
\be
S_{\mu\nu}^{(\e)}\big(-{\pi\ov\e}\big) = S_{\mu\nu}(-\infty)\,.
\ee
Taking into account the formulae from appendix \ref{appg} one finds that 
\be
\p_\mu^{(\e)}(\t) = Q_\mu - \e\t P_\mu + \p_\mu^{(\e, osc)}(\t)\,,
\ee
where $ \p_\mu^{(\e, osc)}(\t)$ is periodic on $[-{\pi\ov\e},{\pi \ov\e}]$, and the zero modes $P_\mu$, $Q_\mu$ commute with $ \p_\mu^{(\e, osc)}(\t)$ and  satisfy the algebra
\be
\left[P_\mu, Q_\nu\right]=i a_{\mu\nu}\,, \quad \left[Q_\mu, Q_\nu\right]=\left[P_\mu, P_\nu\right]=0\,,\quad \mu,\nu=0,1,\ldots,N\,,
\ee
where $a_{ij}=2 \delta_{ij} - \delta_{i-1,j} - \delta_{i+1,j}$ is the Cartan matrix of type $A_{N-1}$ for $i,j=1,2,\ldots,N-1$, and 
$Q_0, P_0,Q_N,P_N$ are expressed in terms of $Q_j, P_j$ as 
\be
Q_0=-\sum_{k=1}^{N-1}{N-k\ov N} Q_k\,,\ P_0=-\sum_{k=1}^{N-1}{N-k\ov N} P_k\,,\quad Q_N=-\sum_{k=1}^{N-1}{k\ov N} Q_k\,,\ P_N=-\sum_{k=1}^{N-1}{k\ov N} P_k\,,
\ee
and therefore
$a_{00}=a_{NN}={N-1\ov N}=s$, $a_{01}=a_{N-1,N}=-1$, and all the remaining $a_{\mu\nu}=0$. 

The oscillatory part can be expanded in a Fourier series
\be
\p_\mu^{(\e, osc)}(\t) = \sum_{m\neq 0}{1\ov i m}a_{\mu}(\e m)\exp(im\e\t)\,,
\ee
where $a_{\mu}(\e m)$ satisfy the following commutation relations 
\be
[a_{\mu}(\e m),a_{\nu}(\e n)]= m f_{\mu\nu}(\e m)\delta_{m+n,0}\,,\quad i,j=1,2,\ldots, N-1\,,
\ee
where $f_{\mu\nu}(\e m)$ are given in appendix \ref{appf}. In the limit $\e\to 0$ with  $\e m=t$ and $\e n=t'$ kept fixed, $\delta_{m+n,0}/\e$ goes to $\delta(t+t')$ and one recovers the previous formulae.

The primed fields are defined in the same way
\be
\p_\mu^{'(\e)}(\t) = -Q_\mu + \e\t P_\mu + \p_\mu^{'(\e, osc)}(\t)\,.
\ee
Notice that since 
\be
[P_j, \lok ] = -a_{jk} \lok \,,\quad [P_j, \rok ] = a_{jk} \rok \,,
\ee
$\lok $ is a lowering operator and $\rok $ is a raising operator. 

Thus one can use the formulae from the main text and use the zero modes  only to get $N-1$ selection rules from the requirement that no dependence of $Q_j$ should appear in the trace formula. Assuming that we have the trace of the form
\be
\Tr_{\pi_{Z}}\left[e^{2\pi i K}\Big(\prod_{\mu=0}^{N-1}\prod_{k=1}^{n_\mu'} V'_{\mu}(\t'_{\mu,k})\Big) \Big(\prod_{\mu=0}^{N-1}\prod_{k=1}^{n_\mu} V_{\mu}(\t_{\mu,k})\Big) \right]\,,
\ee
one gets 
\be\la{selr1}
{N-j\ov N} n_0 - n_j -{N-j\ov N} n'_0 + n'_j = 0\,,\quad j=1,\ldots,N-1\,,
\ee
or equivalently
\be\la{selr2}
(N-j)(n_0-n_0') - N( n_j  -  n'_j) = 0\,,\quad j=1,\ldots,N-1\,,
\ee
For $N=2$ one gets
\be
n_0-n_0' - 2( n_1  -  n'_1) = 0\,,
\ee
which agrees with \cite{Lukyanov}.

If one considers $Z_1Z_2\cdots Z_N$ one finds that for this operator
\be
n_0=N\,,\ n_1 = N-1, \ldots,\ n_j = N-j\,,
\ee
which implies that there is no $Q_j$ in this operator because
from \eqref{selr1} one gets $n_j = {N-j\ov N} n_0$. In fact from \eqref{selr1} one sees that for the identity operator $n_0$ must be an integer multiple of $N$. For an arbitrary operator $n_0-n_0'$ must be an integer multiple of $N$.

The selection rules take a simpler form if one uses $Z_j$ and $Z_j'$ operators. Then assuming that we have the trace of the form
\be\la{trZpZa}
\Tr_{\pi_{Z}}\left[e^{2\pi i K}\Big(\prod_{j=1}^{N}\prod_{a=1}^{m_j'} Z'_{j}(\t'_{j,a})\Big) \Big(\prod_{j=1}^{N}\prod_{a=1}^{m_j} Z_{j}(\t_{j,a})\Big) \right]\,,
\ee
one gets
\be
n_\mu=\sum_{k=\mu+1}^N m_k\,,\quad n_\mu'=\sum_{k=\mu+1}^N m_k'\,,\quad \mu=0,\ldots,N-1\,.
\ee 
Thus the selection rules \eqref{selr1} take the form
\be\la{selr1a}
\sum_{k=1}^j (m_k-m_k') ={j\ov N}(M-M')\,,\quad M = \sum_{k=1}^N m_k\,,\quad M' = \sum_{k=1}^N m_k'\,,
\ee 
where $M$ and $M'$ are the total numbers of $Z$ and $Z'$ operators.
The selection rules \eqref{selr1a} then immediately imply
\be\la{selr1b}
m_j-m_j' ={1\ov N}(M-M')\,,\quad j=1,\ldots,N-1\,,
\ee 
and therefore 
\be
m_j-m_j' =k\,,\quad M-M' = k N {\rm \ for\ some\ integer\ }k\,.
\ee
This formula shows that if $Z$ transforms in the fundamental irrep of $\su(N)$ then $Z'$ transforms in the antifundamental irrep (and it would be more appropriate to use upper indices for $Z'$), and the form factor does not vanish only if  the product of all $Z$ and $Z'$ is a singlet of $\su(N)$ which is a natural requirement. 

\section{Traces of vertex operators}\la{traces}
\subsection*{General formula}
We want to compute traces of products of vertex operators 
defined as
\be
V(\t)=
 :\exp(i \phi(\t)):\,,
 \ee
where $\phi(\t)$ is a linear combination of the independent oscillators $a_i(t)$
\be
\phi(\t) = \int^\infty_{-\infty}{dt\over i t}\,c_i(t)\,a_i(t)e^{i\theta t}= \int^\infty_{0}dt \,\bar\a_i(t)\,a_i(t) + \int^\infty_{0}dt\,\,a_i^\dagger(t)\b_i(t)\,,
\ee
and 
\be
\bar\a_i(t)= {1\over i t}\,c_i(t)e^{i\theta t}\,,\quad \b_i(t)= -{1\over i t}\,c_i(-t)e^{-i\theta t}\,,\quad a_i^\dagger(t) = a_i(-t)\,.
\ee
It is sufficient to understand how to compute 
\be\la{trV}
\mbox{Tr}_F\left(\exp(2\pi i K) V(\t)\right)\,,\quad K = iH=i\int^{\infty}_{0}dt \, \sum^{N-1}_{i, j=1} h_{ij}(t) a_{i}^\dagger(t) a_{j}(t)\,,~~~~
\ee
where $F$ is the Fock space where $a_i(t)$ act. 
It is not difficult to show that if one has one set of oscillators $a, a^\dagger$ such that
\be
[a,a^\dagger]=1\,,\quad K = i\, h\, a^\dagger a\,,\quad V(\t)=
 :\exp(i \phi):\,,\quad \phi=\bar\a\,a + a^\dagger\b\,,
\ee
then 
\be\la{trV0} 
\mbox{Tr}_F \ e^{2\pi i K}  :\exp(i \phi):\, ={1
 \ov  1-e^{-2\pi h} }\exp\left({\bar\alpha\beta\ov 1-e^{2\pi h}}\right)\,.
\ee
We want to generalise the formula to the case of several coupled oscillators, and we can drop the $t$-dependence because the commutation relations are ultra-local in $t$. So we consider
\be
[a_i,a_j^\dagger]=f_{ij}\,,\quad K = i\, h_{ij}\, a_i^\dagger a_j\,,\quad V(\t)=
 :\exp(i \phi):\,,\quad \phi=\bar\a_i\,a_i + a_i^\dagger\b_i\,,
\ee
where $\bar f_{ij}=f_{ji}$ and $\bar h_{ij}=h_{ji}$, that is the matrices $f$ and $h$ are hermitian. Since $f$ is hermitian it can be diagonalized with a unitary matrix $U$
\be
U f U^\dagger = D^2\,,\quad D_{ij}=d_i\delta_{ij}\,,\quad d_i>0\,,
\ee
and the new oscillators $b=Ua\,,\ b^\dagger = a^\dagger U^\dagger$ satisfy the relations
\be
[b_i,b_j^\dagger]=d_i^2\delta_{ij}\,,\quad K = i\, b_i^\dagger (UhU^\dagger)_{ij}\, b_j\,,\quad  \phi=(\bar\a U^\dagger)_i\,b_i + b_i^\dagger(U\b)_i\,.
\ee
We then rescale $b_i$ to get the canonical commutation relations
\be
b_i = d_i c_i\,,\quad b_i^\dagger = d_ic_i^\dagger\,,\quad [c_i,c_j^\dagger]=\delta_{ij}\,,
\ee
and
 \be
K = i\, c_i^\dagger (DUhU^\dagger D)_{ij}\, c_j\,,\quad  \phi=(\bar\a U^\dagger D)_i\,c_i + c_i^\dagger(DU\b)_i\,.
\ee
The matrix $DUhU^\dagger D$ is obviously hermitian and can be diagonalized with a unitary matrix $W$
\be
DUhU^\dagger D = W^\dagger Q W\,,\quad Q_{ij}=q_i\delta_{ij}\,,
\ee
and introducing the new oscillators $A=Wc\,,\ A^\dagger = c^\dagger W^\dagger$ one gets
\be
[A_i,A_j^\dagger]=\delta_{ij}\,,\quad K = i\, q_iA_i^\dagger A_i\,,\quad  \phi =(\bar\a U^\dagger D W^\dagger)_i\,A_i + A_i^\dagger(WDU\b)_i\,,
\ee
and therefore
\be\la{trV1} 
\mbox{Tr}_F \ e^{2\pi i K}  :\exp(i \phi):\, =\prod_i{1
 \ov  1-e^{-2\pi q_i} }\exp\left({\bar\alpha'_i\beta'_i\ov 1-e^{2\pi q_i}}\right)\,,
\ee
where 
\be
\bar\alpha'_i = (\bar\a U^\dagger D W^\dagger)_i\,,\quad \beta'_i=(WDU\b)_i\,.
\ee
Formula \eqref{trV1} can be brought to the form
\be\la{trV2} 
\mbox{Tr}_F \ e^{2\pi i K}  :\exp(i \phi):\, ={1
 \ov  \det(1-e^{-2\pi hf}) }\exp\left(\bar\alpha_i \left( f{1\ov 1-e^{2\pi hf}}\right)_{ij}\beta_j\right)\,.
\ee
Notice the identity 
\be
f{1\ov 1-e^{2\pi hf}} ={1\ov 1-e^{2\pi fh}}f\,.
\ee
Fortunately, thanks to \eqref{fh}, $fh = t$, where we take into account that the actual commutation relations are $[a_i(t),a_j^\dagger(t')]=tf_{ij}\delta(t-t')$. Thus introducing the integral over $t$ one gets
\be\la{trVf}
{\mbox{Tr}_F\left(\exp(2\pi i K) V(\t)\right) \ov \mbox{Tr}_F\left(\exp(2\pi i K)\right) }= \exp\left(\int_0^\infty\, dt\,  {\bar\alpha_i(t)\,tf_{ij}(t)\,\beta_j(t)\ov 1-e^{2\pi t}}\right)\,.
\ee
This formula agrees with the prescription in \cite{FL}.
To show this let's consider
\be\la{trVV0}
\mbox{Tr}_F\left(\exp(2\pi i K) V_1V_2\right) \,,
\ee
where
\be
V_k=
 :\exp(i \phi_k):\,,\quad
\phi_k = \int^\infty_{0}dt \,\bar\a_i^{(k)}(t)\,a_i(t) + \int^\infty_{0}dt\,\,a_i^\dagger(t)\b_i^{(k)}(t)\,.
\ee
Then one gets
\be
V_1V_2 = g_{12}:V_1V_2:\,,\quad g_{12} =\exp\left( - \int_0^\infty\, dt\, \bar\alpha_i^{(1)}(t)\,tf_{ij}(t)\,\beta_j^{(2)}(t)\right)\,,
\ee
and
\be\la{trVV1}
{\mbox{Tr}_F\left(\exp(2\pi i K) V_1V_2\right) \ov \mbox{Tr}_F\left(\exp(2\pi i K)\right) }= \exp\left(  \int_0^\infty\, dt\, \left(-\bar\alpha_i^{(1)}(t)\,tf_{ij}(t)\,\beta_j^{(2)}(t) +{\bar\alpha_i(t)\,tf_{ij}(t)\,\beta_j(t)\ov 1-e^{2\pi t}}\right)\right)\,,
\ee
where
\be
\bar\alpha_i=\bar\alpha_i^{(1)}+\bar\alpha_i^{(2)}\,,\quad \beta_j(t)=\beta_j^{(1)}+\beta_j^{(2)}\,.
\ee
Thus,
\be\la{trVV2}
{\mbox{Tr}_F\left(\exp(2\pi i K) V_1V_2\right) \ov \mbox{Tr}_F\left(\exp(2\pi i K)\right) }= C_1 C_2 G_{12}\,,
\ee
where 
\be
C_k={\mbox{Tr}_F\left(\exp(2\pi i K) V_k\right) \ov \mbox{Tr}_F\left(\exp(2\pi i K)\right) }\,,
\ee
and 
\be\la{G12}
G_{12}= \exp\left(-  \int_0^\infty\, dt\, \left({\bar\alpha_i^{(1)}(t)\,tf_{ij}(t)\,\beta_j^{(2)}(t)\ov 1-e^{-2\pi t}}-{\bar\alpha_i^{(2)}(t)\,tf_{ij}(t)\,\beta_j^{(1)}(t)\ov 1-e^{2\pi t}}\right)\right)\,,
\ee
Introducing 
\bea\begin{aligned}
\langle\langle a_j(t)a_k(t')\rangle\rangle&={tf_{jk}(t) \ov 1- e^{-2 \pi t}}\delta(t+t')\, ,
\end{aligned}\eea
one finds
\bea\begin{aligned}
\langle\langle\phi_1\phi_2\rangle\rangle&=-\log G_{12}\, .
\end{aligned}\eea

\subsection*{Traces of single $V$'s}

To compute the traces of $V_\mu$ and $V'_\mu$ we use that
\be
\bar\a_\mu(t)=\bar\a'_\mu(t)=-\frac{i e^{i \t t}}{t}\,,\quad \b_\mu(t)=\b'_\mu(t)=\frac{i e^{-i \t t}}{t}\,,
\ee
and therefore from \eqref{trVf} one gets
\be\la{trVf1}
C_\mu\equiv {\mbox{Tr}_F\left(\exp(2\pi i K) V_\mu(\t)\right)\ov \mbox{Tr}_F\left(\exp(2\pi i K) \right)} = \exp\left(\int_0^\infty\, {dt\ov t}\,  {f_{\mu\mu}(t)\ov 1-e^{2\pi t}}\right)\,,
\ee
and similar formula for $V'$ with the obvious replacement $f\to f'$.

Computing the integrals one gets the constants
\be\la{cj}
\log C_j = -\text{log$\Gamma
   $}\left(1-\frac{1}{N}\right)+\gamma 
   \left(\frac{1}{N}-1\right)+\frac{\log (2 \pi )}{N}\,,\quad C_j=\frac{e^{\gamma  \left(\frac{1}{N}-1\right)} (2 \pi
   )^{\frac{1}{N}}}{\Gamma \left(\frac{N-1}{N}\right)}
\ee
\bea\nonumber
\log C_0 =\log C_N &=&1-\frac{3}{2
   N}+\frac{1}{2 N^2}-\frac{\gamma  (N-1)^2}{2
   N^2}+\frac{1-N}{N}\text{log$\Gamma
   $}\left(2-\frac{1}{N}\right)~~~~~~~~~~\\
   &+&\left(-\frac{1}{2
   N^2}+\frac{1}{N}-1\right) \log (2 \pi )+\psi ^{(-2)}\left(2-\frac{1}{N}\right)\,,~~~~~~~
\eea
where $\psi ^{(-2)}\left(z\right)$ is given by
\be
\psi ^{(-2)}\left(z\right) = \int_0^z\, dt\,  \text{log$\Gamma
   $}\left(t\right)\,.
\ee
and
 \be\la{cjp}
 C'_j = \frac{e^{\gamma  \left(-\frac{1}{N}-1\right)} (2 \pi )^{-1/N}}{\Gamma \left(\frac{N+1}{N}\right)}
\ee
\bea\nonumber
\log C'_0 &=&-\frac{\gamma  \left(N^2-1\right)}{2 N^2}-\frac{1}{2 N^2}+\left(\frac{1}{2 N^2}+\frac{1}{2}\right) \log
   (2 \pi )-\frac{1}{2 N}\\
   &&-\psi ^{(-2)}\left(1+\frac{1}{N}\right)+\frac{\log \left(\Gamma
   \left(1+\frac{1}{N}\right)\right)}{N}\,.~~~~~~~
\eea

\subsection*{Traces of $V_\mu V_\nu$ and functions $G_{\mu\nu}$}

To compute the traces we use \eqref{trVV2} and 
\eqref{G12} which takes the form
\be
G_{\mu \nu}(\b_2-\b_1)=\exp\big(-\langle\langle \p_\mu(\b_1) \p_\nu(\b_2)\rangle\rangle\big)\, ,
\ee
and therefore
\be
G_{\mu \nu}(\b)=\exp\Big(-\int^{\infty}_{0}{dt \ov t}{f_{\mu \nu}(t) \ov  \sinh \pi t}\cos\left(\b+ i \pi\right)t\Big)\, .
\ee
These satisfy the relations
\be
G_{\mu \nu}(\b-2\pi i)=G_{\mu \nu}(-\b)\, ,\quad S_{\mu \nu}(\b)={G_{\mu \nu}(-\b)\ov G_{\mu \nu}(\b)}\, ,
\ee
which are necessary to satisfy the form factors axioms.

\medskip

Computing the functions we get
the following representations for $G_{00}$
\be
G_{0 0}(\b)=G_{NN}(\b)=C_{00}\exp\Big(-2\int^{\infty}_{0}{dt \ov t}{\sinh{N-1 \ov N}\pi t \ov  \sinh^2 \pi t}e^{{\pi t\ov N}}\sinh^2\left({ \pi t\ov 2}-{i \b t\ov 2}\right)\Big)\, ,
\ee
\bea
C_{0 0}&=&\exp\Big(-\int^{\infty}_{0}{dt \ov t}{\sinh{N-1 \ov N}\pi t \ov  \sinh^2 \pi t}e^{{\pi t\ov N}}\Big) \\\nonumber
&=&
\frac{2^{\frac{N-1}{N^2}-\frac{5}{12}} \pi
   ^{\frac{N-1}{N^2}-1}}{A^3} \Gamma
   \left(\frac{3}{2}-\frac{1}{N}\right)^{\frac{2}{N}-1} e^{\frac{(N-1) (N+\gamma -1)}{N^2}+2 \psi
   ^{(-2)}\left(\frac{3}{2}-\frac{1}{N}\right)}
\, ,
\eea
where the integral representation for $G_{00}(\b)$ is well-defined for 
\be
G_{00}(\b):\quad -2 \pi <{\rm Im}(\beta )<0\quad {\rm and}\quad N>1\,.
\ee
Then, one can get the following representations for $G_{0N}$
\be
G_{0 N}(\b)=C_{0N}\exp\Big(-2\int^{\infty}_{0}{dt \ov t}{\sinh{\pi t\ov N} \ov  \sinh^2 \pi t}e^{{\pi t\ov N}}\sinh^2\left({ \pi t\ov 2}-{i \b t\ov 2}\right)\Big)\, ,
\ee
\bea
C_{0 N}&=&\exp\Big(-\int^{\infty}_{0}{dt \ov t}{\sinh{\pi t \ov N}\ov  \sinh^2 \pi t}e^{{\pi t\ov N}}\Big) \\\nonumber
&=&
(2 \pi )^{\frac{1}{N^2}+1} \Gamma
   \left(\frac{N-1}{N}\right)^{-2/N} e^{\frac{N+\gamma
   -1}{N^2}-2 \psi ^{(-2)}\left(\frac{N-1}{N}\right)}
\, ,
\eea
where the integral representation for $G_{0N}(\b)$ is well-defined for 
\be
G_{0N}(\b):\quad
\frac{2 \pi }{N}-3 \pi <{\rm Im}(\beta )<\pi -\frac{2
   \pi }{N}
 \,.
\ee
Then
\be\la{G00}
G_{jj}(\b)=\frac{i\, 4^{\frac{1}{N}+1} e^{\frac{2 \gamma }{N}} \pi
   ^{\frac{N+2}{N}} \sinh \left(\frac{\beta
   }{2}\right)}{\Gamma \left(-\frac{i \beta }{2 \pi
   }-\frac{1}{N}+1\right) \Gamma \left(\frac{i \beta }{2 \pi
   }-\frac{1}{N}\right)}\,,
\ee
where the integral representation for $G_{jj}(\b)$ is well-defined for 
\be G_{jj}(\b):\quad\frac{2 \pi }{N}-2 \pi <{\rm Im}(\b)<-\frac{2 \pi }{N}\quad {\rm and}\quad N>2\,.
\ee
Finally
\be\la{G01}
G_{j,j+1}(\b)=
e^{-\frac{\gamma }{N}} (2 \pi )^{-\frac{N+1}{N}} \Gamma
   \left(-\frac{i \beta }{2 \pi }-\frac{1}{2 N}+1\right)
   \Gamma \left(\frac{i \beta }{2 \pi }-\frac{1}{2 N}\right)\,,
\ee
where the integral representation for $G_{j,j+1}(\b)$ is well-defined for 
\be
G_{j,j+1}(\b):\quad\frac{\pi }{N}-2 \pi <{\rm Im}(\beta )<-\frac{\pi }{N}\quad {\rm and}\quad N>1\,.
\ee

For the functions $G_{\mu\nu}''(\b)$ one gets
\be
G''_{0 0}(\b)=G''_{NN}(\b)=C''_{00}\exp\Big(-2\int^{\infty}_{0}{dt \ov t}{\sinh{N-1 \ov N}\pi t \ov  \sinh^2 \pi t}e^{-{\pi t\ov N}}\sinh^2\left({ \pi t\ov 2}-{i \b t\ov 2}\right)\Big)\, ,
\ee
\bea
C''_{0 0}&=&\exp\Big(-\int^{\infty}_{0}{dt \ov t}{\sinh{N-1 \ov N}\pi t \ov  \sinh^2 \pi t}e^{-{\pi t\ov N}}\Big) \\\nonumber
&=&
A^3 2^{\frac{1}{N^2}-\frac{1}{N}+\frac{17}{12}} \pi
   ^{\frac{1}{N^2}-\frac{1}{N}+1} \Gamma
   \left(\frac{1}{2}+\frac{1}{N}\right)^{\frac{2}{N}-1}e^{-\frac{\gamma 
   (N-1)+1}{N^2}-2 \psi
   ^{(-2)}\left(\frac{1}{2}+\frac{1}{N}\right)}
\, ,
\eea
where the integral representation for $G''_{00}(\b)$ is well-defined for 
\be
G''_{00}(\b):\quad -\frac{2\pi }{N}-2 \pi <{\rm Im}(\beta )<\frac{2\pi }{N}\quad {\rm and}\quad N>1\,.
\ee
We also find that 
\be
G''_{0 N}(\b)=C''_{0N}\exp\Big(-2\int^{\infty}_{0}{dt \ov t}{\sinh{\pi t \ov N} \ov  \sinh^2 \pi t}e^{-{\pi t\ov N}}\sinh^2\left({ \pi t\ov 2}-{i \b t\ov 2}\right)\Big)\, ,
\ee
\bea
C''_{0 N}&=&\exp\Big(-\int^{\infty}_{0}{dt \ov t}{\sinh{\pi t \ov N}\ov  \sinh^2 \pi t}e^{-{\pi t\ov N}}\Big) \\\nonumber
&=&
(2 \pi )^{-\frac{1}{N^2}-1} \Gamma \left(1+\frac{1}{N}\right)^{-2/N} e^{\frac{N-\gamma
   +1}{N^2}+2 \psi ^{(-2)}\left(1+\frac{1}{N}\right)}\, ,
\eea
where the integral representation for $G''_{0N}(\b)$ is well-defined for 
\be
G''_{0N}(\b):\quad -3 \pi <{\rm Im}(\beta )<\pi\quad {\rm and}\quad N>1\,.
\ee
Then
\be\la{Gpp00}
G''_{jj}(\b)=
\frac{i\, 2 e^{-\frac{2 \gamma }{N}} (2\pi)
   ^{\frac{N-2}{N}} \sinh \left(\frac{\beta
   }{2}\right)}{\Gamma \left(-\frac{i \beta }{2 \pi
   }+\frac{1}{N}+1\right) \Gamma \left(\frac{i \beta }{2 \pi
   }+\frac{1}{N}\right)}
\,,
\ee
where the integral representation for $G''_{jj}(\b)$ is well-defined for 
\be  G''_{jj}(\b):\quad-2 \pi <{\rm Im}(\b)<0 \quad {\rm and}\quad N>0\,.
\ee
Finally
\be\la{Gpp01}
G''_{j,j+1}(\b)=
e^{\gamma /N} (2 \pi )^{\frac{1}{N}-1} \Gamma \left(-\frac{i
   \beta }{2 \pi }+\frac{1}{2 N}+1\right) \Gamma
   \left(\frac{i \beta }{2 \pi }+\frac{1}{2 N}\right)
\,,
\ee
where the integral representation for $G''_{j,j+1}(\b)$ is well-defined for 
\be
G''_{j,j+1}(\b):\quad -\frac{\pi }{N}-2 \pi <{\rm Im}(\beta )<\frac{\pi }{N}\quad {\rm and}\quad N>0\,.
\ee

Finally the functions $G_{\mu\nu}'(\b)$ are given by 
\be
G'_{0 0}(\b)=G'_{NN}(\b)=C'_{00}\exp\left(2\int^{\infty}_{0}{dt \ov t}{\sinh{N-1 \ov N}\pi t \ov  \sinh^2 \pi t}\sinh^2\Big({ \pi t\ov 2}-{i \b t\ov 2}\Big)\right)\, ,
\ee
\bea
C'_{0 0}&=&\exp\Big(\int^{\infty}_{0}{dt \ov t}{\sinh{N-1 \ov N}\pi t \ov  \sinh^2 \pi t}\Big) \\\nonumber
&=&
\left(\frac{\pi }{2}\right)^{1-\frac{1}{N}}
   \left(\frac{(N-1) \sec \left(\frac{\pi }{2
   N}\right)}{N}\right)^{1-\frac{1}{N}}e^{\frac{1-N}{N}-2 \psi
   ^{(-2)}\left(\frac{3}{2}-\frac{1}{2 N}\right)+2 \psi
   ^{(-2)}\left(\frac{N+1}{2 N}\right)}\, .
\eea
where the integral representation for $G'_{00}(\b)$ is well-defined for 
\be
G'_{00}(\b):\quad -\frac{2\pi }{N}-2 \pi <{\rm Im}(\beta )<\frac{\pi }{N}\quad {\rm and}\quad N>1\,.
\ee
Then 
\be
G'_{0 N}(\b)=C'_{0N}\exp\left(2\int^{\infty}_{0}{dt \ov t}{\sinh{\pi t \ov N} \ov  \sinh^2 \pi t}\sinh^2\Big({ \pi t\ov 2}-{i \b t\ov 2}\Big)\right)\, ,
\ee
with
\bea\begin{aligned}
C'_{0N}&=\exp\Big(\int^{\infty}_{0}{dt \ov t}{\sinh{\pi t \ov N} \ov  \sinh^2 \pi t}\Big)\\
&=\left(\frac{\pi\csc
   \left(\frac{\pi }{2 N}\right)}{2N}\right)^{\frac{1}{N}}
   e^{-\frac{1}{N}+2 \psi ^{(-2)}\left(1-\frac{1}{2
   N}\right)-2 \psi ^{(-2)}\left(1+\frac{1}{2
   N}\right)}\, ,
\end{aligned}\eea
where the integral representation for $G'_{0N}(\b)$ is well-defined for 
\be
G'_{0N}(\b):\quad -3 \pi+{\pi \ov N} <{\rm Im}(\beta )<\pi-{\pi \ov N}\quad {\rm and}\quad N>1\,.
\ee
Then
\be\la{Gp11}
G'_{jj}(\b)=
\frac{1}{2 \left(\cos \left(\frac{\pi }{N}\right)-\cosh
   (\beta )\right)}
   \,,
\ee
where the integral representation for $G'_{jj}(\b)$ is well-defined for 
\be
G'_{jj}(\b):\quad\frac{\pi }{N}-2 \pi <{\rm Im}(\beta )<-\frac{\pi }{N}\quad {\rm and}\quad N>1\,.
\ee
Finally
\be\la{Gp01}
G'_{j,j+1}(\b)=
2 i \sinh \left(\frac{\beta }{2}\right)
  \,,
\ee
where the integral representation for $G'_{j,j+1}(\b)$ is well-defined for 
\be  G'_{j,j+1}(\b):\quad-2 \pi <{\rm Im}(\b)<0 \quad {\rm and}\quad N>0\,.
\ee

The functions satisfy the following identities
\be\la{idG1}
G_{00}(\b-i \u_1)G_{0N}(\b+i \u_{N-1})={1\ov G_{01}(\b)}=G_{00}(\b+i \u_1)G_{0N}(\b-i \u_{N-1})\,,
\ee
\be\la{idG2}
G_{00}(\b+i \u_{N-1})G_{0N}(\b-i \u_{1})={1\ov G_{01}(\b+i \u_{N-2})}\,,
\ee
\be\la{idG3}
G_{00}(\b-i \u_{N-1})G_{0N}(\b+i \u_{1})={1\ov G_{01}(\b-i \u_{N-2})}\,,
\ee
and $G'$ and $G''$ functions obey the same identities too.

\subsection*{Traces of $V_{1\ldots r} V_{1\ldots s}$ and functions $G_{1\ldots r;1\ldots s}$}

To compute the traces we use \eqref{trVV2} and 
\eqref{G12} which takes the form
\be
G_{1\ldots r;1\ldots s}(\b_2-\b_1)=\exp\big(-\langle\langle \p_{1\ldots r}(\b_1) \p_{1\ldots s}(\b_2)\rangle\rangle\big)\, ,
\ee
and therefore
\be
G_{1\ldots r;1\ldots s}(\b)=\exp\Big(-\int^{\infty}_{0}{dt \ov t}{f_{1\ldots r;1\ldots s}(t) \ov  \sinh \pi t}\cos\left(\b+ i \pi\right)t\Big)\, .
\ee
 These can be computed in analogy with the previous sections so we simply state the results
    \bea\begin{aligned}
  &G_{1\ldots r;1\ldots s}(\b)&=&\exp\left(-\int^{\infty}_{0}{dt \ov t}{\sinh {\pi t r\ov N}\sinh {\pi t (N-s)\ov N}\ov \sinh^2\pi t \sinh{\pi t\ov N}}e^{{\pi t\ov N}}\cos(\b+i \pi)t\right)\, ,\\
   &G'_{1\ldots r;1\ldots s}(\b)&=&\exp\left(\int^{\infty}_{0}{dt \ov t}{\sinh {\pi t r\ov N}\sinh {\pi t (N-s)\ov N}\ov \sinh^2\pi t \sinh{\pi t\ov N}}\cos(\b+i \pi)t\right)\, ,\\
   &G''_{1\ldots r;1\ldots s}(\b)&=&\exp\left(-\int^{\infty}_{0}{dt \ov t}{\sinh {\pi t r\ov N}\sinh {\pi t (N-s)\ov N}\ov \sinh^2\pi t \sinh{\pi t\ov N}}e^{-{\pi t\ov N}}\cos(\b+i \pi)t\right)\, ,\\
  & \mbox{for}\quad s>r\, ,&
   \label{2FusedFF}
   \end{aligned}\eea
   and
      \bea\begin{aligned}
  C_{1\ldots r}&=\exp\left(-\int^{\infty}_{0}{dt \ov t}{\sinh {\pi t r\ov N}\sinh {\pi t (N-r)\ov N}\ov 2 \sinh^2\pi t \sinh{\pi t\ov N}}e^{{\pi t\ov N}} \, e^{- \pi t}\right)\, ,\\
       C'_{1\ldots r}&=\exp\left(-\int^{\infty}_{0}{dt \ov t}{\sinh {\pi t r\ov N}\sinh {\pi t (N-r)\ov N}\ov 2 \sinh^2\pi t \sinh{\pi t\ov N}}e^{-{\pi t\ov N}} \, e^{- \pi t}\right)\, .
   \end{aligned}\eea
   The ranges for which these integral representation are well-defined is as follows ($N>1$):
   \bea\begin{aligned}
   G_{1\ldots r;1\ldots s}(\b):&\quad-2 \pi-{(r-s) \pi\ov N} <{\rm Im}(\b)<{(r-s) \pi\ov N} \,,\\
   G'_{1\ldots r;1\ldots s}(\b):&\quad-2 \pi-{\pi \ov N}-{(r-s) \pi\ov N}  <{\rm Im}(\b)<{\pi \ov N}+{(r-s) \pi\ov N}\,,\\
   G''_{1\ldots r;1\ldots s}(\b):&\quad-2 \pi-{2\pi \ov N}-{(r-s) \pi\ov N}  <{\rm Im}(\b)<{2\pi \ov N}+{(r-s) \pi\ov N}\,.
   \end{aligned}\eea
       Note that these formulae include the cases $r=1$ and $s=1$, which gives the highest weight particle of rank 1. Some particular cases of interest are
       \bea\begin{aligned}
  G_{0; 1\ldots s}(\b)&=\exp\left(-\int^{\infty}_{0}{dt \ov t}{\sinh {\pi t (N-s)\ov N}\ov \sinh^2\pi t}e^{{\pi t\ov N}}\cos(\b+i \pi)t\right)\, ,\\
   G'_{0;1\ldots s}(\b)&=\exp\left(\int^{\infty}_{0}{dt \ov t}{\sinh {\pi t (N-s)\ov N}\ov \sinh^2\pi t }\cos(\b+i \pi)t\right)\, ,\\
   G''_{0;1\ldots s}(\b)&=\exp\left(-\int^{\infty}_{0}{dt \ov t}{\sinh {\pi t (N-s)\ov N}\ov \sinh^2\pi t }e^{-{\pi t\ov N}}\cos(\b+i \pi)t\right)\, ,
   \end{aligned}\eea
   for $r=1$ which gives the elementary particle with index $0$ and 
    \bea\begin{aligned}
  G_{1\ldots r;N}(\b)&=\exp\left(-\int^{\infty}_{0}{dt \ov t}{\sinh {\pi t r \ov N}\ov \sinh^2\pi t}e^{{\pi t\ov N}}\cos(\b+i \pi)t\right)\, ,\\
   G'_{1\ldots r;N}(\b)&=\exp\left(\int^{\infty}_{0}{dt \ov t}{\sinh {\pi t r \ov N}\ov \sinh^2\pi t }\cos(\b+i \pi)t\right)\, ,\\
   G''_{1\ldots r;N}(\b)&=\exp\left(-\int^{\infty}_{0}{dt \ov t}{\sinh {\pi t r \ov N}\ov \sinh^2\pi t }e^{-{\pi t\ov N}}\cos(\b+i \pi)t\right)\, ,
   \end{aligned}\eea
   for $s=N-1$ which gives the the fused rank $N-1$ particle, which takes index $N$.
     For the remaining traces of fused particles with $V_j$, we find
   \bea\begin{aligned}
  G_{r; 1\ldots r}(\b)&=\exp\left(\int^{\infty}_{0}{dt \ov t}{1\ov \sinh\pi t}e^{{\pi t\ov N}}\cos(\b+i \pi)t\right)\, ,\\
   G'_{r; 1\ldots r}(\b)&=\exp\left(-\int^{\infty}_{0}{dt \ov t}{1\ov \sinh\pi t}\cos(\b+i \pi)t\right)\, ,\\
   G''_{r; 1\ldots r}(\b)&=\exp\left(\int^{\infty}_{0}{dt \ov t}{1\ov \sinh\pi t}e^{-{\pi t\ov N}}\cos(\b+i \pi)t\right)\, .
   \end{aligned}\eea
   Remarkably, we find
   \bea\begin{aligned}
  G_{r; 1\ldots r}(\b)&=G_{r; r+1}(\b)\, ,\\
   G'_{r; 1\ldots r}(\b)&=G'_{r; r+1}(\b)\, ,\\
   G''_{r; 1\ldots r}(\b)&=G''_{r; r+1}(\b)\, .
   \end{aligned}\eea


\end{document}